\documentclass[10pt,journal,compsoc]{IEEEtran}

\usepackage[nocompress]{cite}
\usepackage[colorlinks,linkcolor=blue,citecolor=blue]{hyperref}

\usepackage{amsmath,amsthm,amsfonts,amssymb,amscd}
\usepackage{mathrsfs}
\usepackage{latexsym}
\usepackage{graphicx}
\usepackage{tikz}
\usepackage{subfig}
\usepackage{float}
\usepackage{listings}
\usepackage{textcomp}
\usepackage{xcolor}
\usepackage{algorithm}
\usepackage{algorithmicx}
\usepackage{algpseudocode}
\usepackage{booktabs}
\usepackage{multirow}
\usepackage{tabularx}
\usepackage{bm}

\newif\iftonsplit \tonsplitfalse

\theoremstyle{definition}
\newtheorem{theorem}{Theorem}[section]
\newtheorem{lemma}[theorem]{Lemma}
\newtheorem{corollary}[theorem]{Corollary}
\newtheorem{proposition}[theorem]{Proposition}

\newtheorem{remark}[theorem]{Remark}
\newtheorem{conjecture}[theorem]{Conjecture}
\newtheorem{definition}{Definition}[section]

\numberwithin{equation}{section}

\def\BibTeX{{\rm B\kern-.05em{\sc i\kern-.025em b}\kern-.08em
    T\kern-.1667em\lower.7ex\hbox{E}\kern-.125emX}}

\begin{document}

\title{Geometry, Reliability, and Spacetime Dynamics of Hyper-Simplex Fractal Networks}

\author{Hao~Xu,~Kaiwen~Yang,~Jiacheng~Qian,~Yinuo~Liu,~and~Chih-Lin~I
\thanks{H. Xu and J. Qian are with the College of Electronics and Information Engineering, Tongji University, Shanghai 201804, China (e-mail: hxu@tongji.edu.cn).}
\thanks{K. Yang is with the School of Mathematical Sciences, Tongji University, Shanghai 200092, China (e-mail: yangkaiwen@tongji.edu.cn).}
\thanks{Y. Liu is with Shanghai Customs University, Shanghai 201204, China (corresponding author, e-mail: liuyinuo@shcc.edu.cn).}
\thanks{C.-L. I is with the China Mobile Research Institute, Beijing 100053, China.}
\thanks{This work was supported by the Fundamental Research Funds for the Central Universities, Tongji University, under Project No. 22120240564.}
\thanks{A companion Lean~4 formalization verifying the combinatorial, latency, threshold-map, dual-homing, registration-bound, message-count, and reconfiguration-calculus core ($49$ modules, $853$ distinct kernel-audited statements, zero \texttt{sorry}, commit \texttt{664b9d7}) is linked at \url{https://github.com/Tongji708A/hsfn-lean}.}
\thanks{This work has been submitted to the IEEE for possible publication. Copyright may be transferred without notice, after which this version may no longer be accessible.}
}

\markboth{Preprint}
{Xu \MakeLowercase{\textit{et al.}}: Geometry, Reliability, and Spacetime Dynamics of Hyper-Simplex Fractal Networks}

\IEEEtitleabstractindextext{%
\begin{abstract}
Large-scale, delay-tolerant peer-to-peer networks that need strong consistency usually separate naming, routing, consensus, and ordering into independent layers. To bring deterministic network properties into complex peer-to-peer structures, we unify all four on a single recursive topology. Modeling mutual equality among $N$ peers in Euclidean space requires dimension at least $N{-}1$, which prevents flat equal groups from scaling at fixed dimension. The hyper-simplex fractal network $G(N,m)$ escapes this limit by recursion. Each non-leaf node anchors a half-scale $N$-clique. From the resulting digit-word addresses alone, nodes decide adjacency without routing tables, forward greedily along exact shortest paths, and converge geometrically to a self-similar attractor of dimension $\log_2 N$. Running hierarchical PBFT on the same prefix tree achieves linear message load and logarithmic fault-free latency, with an explicit reliability threshold under independent faults and analytical recursions for anchor exposure, dual homing, and event registration. Reversible topology edits generate the Higman--Thompson group $V_N$ on the address boundary, block order characterizes service-topology hosting beyond treewidth, and root-anchored block heights yield persistent spacetime coordinates with compact light-client proofs.
\end{abstract}

\begin{IEEEkeywords}
Hyper-simplex fractal network, address algebra, Byzantine fault tolerance, reliability thresholds, Higman--Thompson groups.
\end{IEEEkeywords}

}

\maketitle
\IEEEdisplaynontitleabstractindextext
\IEEEpeerreviewmaketitle

\section{Introduction}
\label{sec:intro}

Large decentralized networks rely on four coupled functions, naming nodes, routing messages, reaching Byzantine agreement, and ordering updates. Conventional stacks separate these functions across naming hierarchies, table-driven routing~\cite{jin2021dnsonchain}, and externally configured consensus committees~\cite{castro1999practical,androulaki2018hyperledger,hearn2019corda,mazzoni2022performance}. Each layer maintains its own state, failure boundary, and notion of time.

Of the four, Byzantine agreement is the hardest to scale. Flat PBFT incurs $\Theta(n^2)$ messages per decision~\cite{castro1999practical}, HotStuff concentrates traffic at a single leader~\cite{yin2019hotstuff}, and practical flat BFT rarely scales past a few hundred replicas~\cite{vukolic2016quest}. Hierarchical protocols such as Steward~\cite{amir2010steward}, multi-layer PBFT~\cite{li2020scalable}, GeoBFT~\cite{gupta2020resilientdb}, Kauri~\cite{neiheiser2021kauri}, TierFlow~\cite{yu2024tierflow}, and ORION~\cite{yahyaoui2024orion} relieve quadratic fanout through multi-tier committees, yet treat committee topology as an administrative overlay detached from naming and routing. Simultaneously bounding node degree, routing diameter, and per-node consensus fanout therefore requires one shared structural substrate.

We build that substrate from a geometric peer-equality principle. Representing equality among $N$ peers as mutual equidistance in Euclidean space requires ambient dimension at least $N{-}1$, yielding a regular $(N{-}1)$-simplex whose $1$-skeleton is the complete graph $K_N$. Flat equal groups therefore cannot grow at fixed dimension. The hyper-simplex fractal network (HSFN) overcomes this dimensional bound by recursion. Every non-leaf node anchors a half-scale $N$-clique, generating an $N$-ary hierarchy of peer groups called \emph{cells}, addressed by nonempty digit words of length at most $m$ over $\{0,\ldots,N{-}1\}$. Building on this single prefix tree, we unify network topology, table-free routing, consensus committees, and dynamic reconfiguration.

\subsection{Contributions and Organization}
\label{subsec:scope}

\begin{enumerate}
    \item \textbf{Geometry, addressing, and routing} (Sections~\ref{sec:model} and~\ref{sec:space}). We construct the sparse graph $G(N,m)$ and establish its counting laws, table-free two-clause adjacency, exact shortest-path greedy routing at degree at most $2N$ and diameter $2m{-}1$, and anchor cut-vertex separation. Under midpoint embedding, the self-similar limit set has Hausdorff dimension $\log_2 N$, and the family $\{F(N,m)\}$ nests isometrically in $(N,m)$.
    \item \textbf{Agreement and reliability theory} (Sections~\ref{sec:time} and~\ref{sec:fault}). We specify certified hierarchical PBFT with $f{+}1$ senders, fallback reporting, and a view-change lock, proving cell-local safety, exact linear message counts, exact ideal latencies ($N_c \geq 4$), and a general committee-tree proposition. Under independent crash faults, sufficient waiting, and timely report inclusion, we establish a depth-uniform overlay root threshold alongside recursions (exact for $3 \nmid N$, conservative certified bounds when $3 \mid N$) for single-anchor exposure, dual homing, and leaf-path registration.
    \item \textbf{Reconfiguration algebra, hosting, and $V_N$} (Section~\ref{sec:reconfig}). We formalize eight cell-atomic structural edits forming an inverse monoid, prove that subtree migration acts as a prefix substitution cutting and creating $N$ uplinks, characterize hostable topologies via biconnected block order and descent weight, identify the invertible units of the boundary join closure with the Higman--Thompson group $V_N$, and establish a Stay/Climb lifting dichotomy for local edits.
    \item \textbf{Spacetime coordinates and protocol validation} (Sections~\ref{sec:spacetime} and~\ref{sec:eval}). We pair addresses with root-anchored block heights into persistent spacetime coordinates with compact Ed25519 light-client proofs, and validate message counts, reliability recursions, safety, and live migrations through discrete-event protocol simulation and real signature verification.
\end{enumerate}

Earlier work sketched a digital-spacetime vision~\cite{xu2024web} and studied a dense facet-subdivision preprint~\cite{yang2024hsfnpart1} using $2N{-}2$ branching and $K_{N,N}$ bundles. This paper supersedes that earlier Part~I preprint with a unified theory of the sparse node-anchored network $G(N,m)$, adding anchor cut-vertex and dual-homed crash recursions, leaf-path registration, disciplined block--cut hosting, and reconfiguration algebra. Section~\ref{sec:related} situates the work in the literature, Sections~\ref{sec:model}--\ref{sec:eval} develop the theory and numerical validation, Section~\ref{sec:discussion} examines tradeoffs and open problems, and Section~\ref{sec:conclusion} concludes. \iftonsplit Appendices~\ref{app:space}--\ref{app:verification} in the supplemental material provide complete proofs, explicit combinatorial recursions, and verification tables.\else Appendices~\ref{app:space}--\ref{app:verification} provide complete proofs, explicit combinatorial recursions, and verification tables.\fi

\section{Related Work}
\label{sec:related}

\emph{Hierarchical Byzantine agreement.} Scaling BFT past a few hundred replicas decomposes replicas or traffic across geographic sites (Steward~\cite{amir2010steward}, GeoBFT~\cite{gupta2020resilientdb}), protocol layers (multi-layer PBFT~\cite{li2020scalable}, MBFT~\cite{du2020mbft}, HiBFT~\cite{thai2019hibft}, Saguaro~\cite{amiri2021saguaro}, TierFlow~\cite{yu2024tierflow}, ORION~\cite{yahyaoui2024orion}, TEE-assisted designs~\cite{chen2024parallel}), linear-view leaders and dissemination trees (flat leader-based HotStuff~\cite{yin2019hotstuff} and tree-pipelined Kauri~\cite{neiheiser2021kauri}), or random shards (Elastico~\cite{luu2016elastico}, OmniLedger~\cite{kokoris2018omniledger}, RapidChain~\cite{zamani2018rapidchain}). None of these systems derives committees, report senders, and shortest-path physical routes jointly from one positional address grammar. The HSFN derives committees, report senders, and routes directly from node addresses.

\emph{Structured overlays and recursive interconnects.} Prefix routing~\cite{plaxton1997accessing} drives Pastry~\cite{rowstron2001pastry} and Tapestry~\cite{zhao2004tapestry}, while Chord~\cite{stoica2003chord}, Kademlia~\cite{maymounkov2002kademlia}, and CAN~\cite{ratnasamy2001can} maintain soft-state tables via periodic stabilization. Data-center fabrics such as DCell~\cite{guo2008dcell}, BCube~\cite{guo2009bcube}, and fat-trees~\cite{alfares2008fattree} pair recursion with positional names. The HSFN decides adjacency by two prefix clauses without soft-state tables and routes along exact shortest paths (Appendix~\ref{app:space}).

\emph{Tree reliability and structural resilience.} Depth-uniform survival follows von Neumann's multiplexed majority organs~\cite{vonneumann1956probabilistic}, Moore--Shannon relays~\cite{moore1956reliable}, and tree percolation~\cite{lyons1990random}, while cut-vertex exposure links to targeted node removal~\cite{motter2002cascade}. Section~\ref{sec:fault} unites both lines for committee trees, pairing the explicit bound $q_0(b)$ with exact single-anchor and dual-homed recursions.

\emph{Logical time, fractal geometry, and Thompson groups.} Logical and vector clocks~\cite{lamport1978time,fidge1991logical} order distributed events without physical clocks, while temporal networks timestamp evolving edges~\cite{holme2012temporal}. Unlike Sierpi\'{n}ski graphs $S(m,N)$~\cite{klavzar1997sierpinski,hinz2017survey}, $G(N,m)$ retains every intermediate tier as a live anchor in a Harary block graph~\cite{harary1963block} whose midpoint attractor obeys Hutchinson's IFS theory~\cite{hutchinson1981fractals,falconer2014fractal}. Service hosting connects to virtual network embedding~\cite{fischer2013virtual,yu2008rethinking} and treewidth~\cite{robertson1986graph}, while invertible prefix rearrangements realize the full Higman--Thompson group $V_N$~\cite{higman1974finitely,lawson2007polycyclic,lawson2020revisited} rather than glued-space rearrangement subgroups~\cite{belk2019rearrangement}. Table~\ref{tab:related-comp} contrasts these architectural capabilities against representative baselines.

\begin{table*}[t]
\centering
\caption{Architectural Comparison Across Representative Baselines.}
\label{tab:related-comp}
\footnotesize
\setlength{\tabcolsep}{4.5pt}
\renewcommand{\arraystretch}{1.1}
\begin{tabular}{@{}lccccc@{}}
\toprule
System family & Table-free & Address-derived & Exact reliability & Reconfiguration & Spacetime \\
 & routing & committees & recursions & algebra & proofs \\
\midrule
Flat PBFT / HotStuff~\cite{castro1999practical,yin2019hotstuff} & No & No & No & No & Log index \\
GeoBFT / Kauri~\cite{gupta2020resilientdb,neiheiser2021kauri} & No & No & No & No & Log index \\
Steward / Multi-layer PBFT~\cite{amir2010steward,li2020scalable} & No & No & Partial & No & Log index \\
Pastry / Chord~\cite{rowstron2001pastry,stoica2003chord} & Soft state & No & No & Local repair & No \\
DCell / BCube~\cite{guo2008dcell,guo2009bcube} & Positional & No & No & No & No \\
\textbf{HSFN (this work)} & \textbf{Exact shortest path} & \textbf{Yes} & \textbf{Exact ($3 \nmid N$)} & \textbf{Boundary units ($V_N$)} & \textbf{Yes ($<5$ KB at $(8,6)$)} \\
\bottomrule
\end{tabular}
\end{table*}

\section{Design Principle and Network Model}
\label{sec:model}
\label{sec:construction}
\label{sec:prelim}

\subsection{Equality and the Dimensional Bound}
\label{subsec:equidistance}

An egalitarian peer group places every participant in the same geometric relationship to every other. Modeling equality among $N$ peers as mutual equidistance in Euclidean space imposes an exact dimensional lower bound.

\begin{lemma}[Equidistance prices dimension]
\label{lem:equidistance}
In $\mathbb{R}^{d}$ at most $d{+}1$ points have pairwise equal distances. Hence $N$ pairwise-equidistant points exist in $\mathbb{R}^{d}$ if and only if $d \geq N{-}1$, and for $d = N{-}1$ they are, up to similarity, the vertices of a regular simplex whose $1$-skeleton is the complete graph $K_N$.
\end{lemma}

The proof via Schoenberg's Gram criterion appears in Appendix~\ref{app:geometry}, and Fig.~\ref{fig:equidistance-spawning}(a)--(b) illustrates how each additional equidistant peer forces a lift into a new spatial axis. A flat egalitarian network of $V$ peers therefore requires ambient dimension $V{-}1$. The HSFN avoids this linear cost by fixing peer-group size at $N \geq 3$ and growing recursively, with each non-leaf node anchoring a half-scale $N$-clique (Fig.~\ref{fig:equidistance-spawning}(c)--(d)). One integer $N$ then governs the ambient dimension ($N{-}1$), the degree ceiling ($2N$), the Byzantine committee size, and the structural width of hosted topologies, while total population scales exponentially in depth $m$.

\begin{figure*}[!t]
\centering
\includegraphics[width=0.82\textwidth]{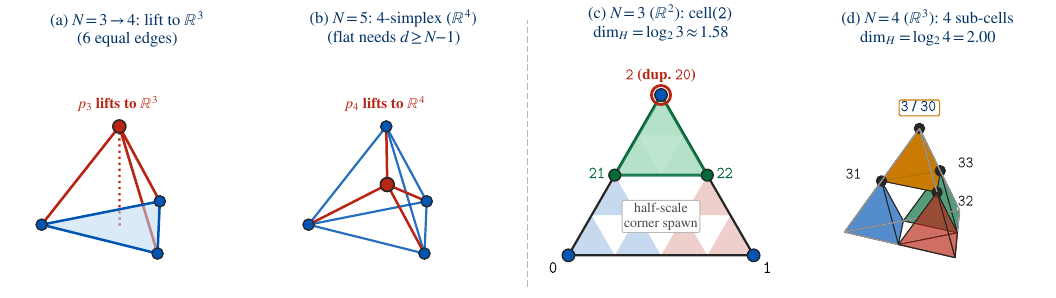}
\caption{Dimensional wall (Lemma~\ref{lem:equidistance}) and recursive escape (Section~\ref{subsec:geom-main}). (a), (b)~Adding a fourth or fifth mutually equidistant peer lifts out of $\mathbb{R}^2$ into $\mathbb{R}^3$ and out of $\mathbb{R}^3$ into $\mathbb{R}^4$ ($d \geq N{-}1$). (c), (d)~Fixing $N$ and spawning half-scale corner $N$-cliques grows population exponentially at fixed ambient dimension ($N = 3$ in $\mathbb{R}^2$ with marked duplicate $\mathtt{20}$ at anchor $\mathtt{2}$, and $N = 4$ in $\mathbb{R}^3$ with four corner sub-tetrahedra around a hollow octahedron). Higher-dimensional projections appear in Fig.~\ref{fig:hd-projections}.}
\label{fig:equidistance-spawning}
\end{figure*}

\subsection{Address Grammar and the Graph $G(N,m)$}
\label{subsec:generating-rule}

Although geometry motivates the construction, the network's combinatorics is defined purely on digit words before any Euclidean coordinates are assigned. An address works like a variable-length postal code. Dropping the last digit moves one tier up to the unique parent anchor, while changing only the last digit moves sideways to a peer in the same $N$-node cell.

\begin{definition}[Hyper-simplex fractal network]
\label{def:event}
\label{def:scheme}
\label{def:address-II}
Fix $N \geq 3$, $m \geq 1$, and alphabet $\Sigma_N = \{0, 1, \ldots, N{-}1\}$.
\begin{itemize}
    \item \emph{Addresses and prefixes.} The \emph{address set} $\mathcal{W}_{N,m}$ contains all nonempty words over $\Sigma_N$ of length at most $m$. The \emph{tier} of $a$ is $|a|$, $\operatorname{pre}(a)$ drops its final digit, and $\operatorname{pre}_s(a)$ takes its length-$s$ prefix ($\operatorname{pre}_0(a) = \varepsilon$, a virtual root naming no node).
    \item \emph{Cells and anchors.} Each $w$ with $|w| \leq m{-}1$ defines an $N$-node \emph{cell} $\operatorname{cell}(w) = \{w0, \ldots, w(N{-}1)\}$. For $|w| \geq 1$, node $w$ is its \emph{anchor} and $w0$ is its \emph{marked duplicate}, a distinct logical node co-located geometrically with $w$ (Section~\ref{subsec:geom-main}).
    \item \emph{Edges and marks.} The graph $G(N,m)$ joins all distinct members of each cell by \emph{sibling links} and links each $wd \in \operatorname{cell}(w)$ ($1 \leq |w| \leq m{-}1$) to anchor $w$ by an \emph{uplink} (the seed $\operatorname{cell}(\varepsilon)$ has sibling links only). With duplicate set $M = \{a0 : 1 \leq |a| \leq m{-}1\}$, the marked network is $F(N,m) = (\mathcal{W}_{N,m}, E, M)$.
\end{itemize}
\end{definition}

Every cell is an $N$-clique, every node at tiers $1, \ldots, m{-}1$ anchors one cell, and $\operatorname{cell}(\varepsilon)$ is the seed. The cells therefore form the complete $N$-ary tree of $m$ levels, read directly off address prefixes.

\begin{proposition}[Node and cell counts]
\label{thm:unified-count}
For every $1 \leq t \leq m$, tier $t$ holds $N^{t}$ nodes in $N^{t-1}$ cells. The network $G(N,m)$ contains
\begin{equation}
\label{eq:unified-V}
    V(N,m) = \sum_{t=1}^{m} N^{t} = \frac{N(N^{m}-1)}{N-1}
\end{equation}
nodes in $V(N,m)/N$ cells, and the cells form the complete $N$-ary tree with $m$ levels.
\end{proposition}

Grouping length-$t$ words by their length-$(t{-}1)$ prefix yields $N^{t-1}$ cells of $N$ members each, and summing over $t = 1, \ldots, m$ gives~\eqref{eq:unified-V}. The next result translates the two link classes into local predicates on digit words, removing all routing-table state.

\begin{proposition}[Two-clause adjacency law]
\label{thm:adjacency-II}
Distinct nodes $u, v$ of $G(N,m)$ are adjacent if and only if either
\begin{enumerate}
    \item[(S)] $|u| = |v| = t$ and $\operatorname{pre}_{t-1}(u) = \operatorname{pre}_{t-1}(v)$ (siblings in one cell), or
    \item[(P)] one address extends the other by a single digit (anchor--offspring uplink).
\end{enumerate}
\end{proposition}

Adjacency is therefore decided by two prefix comparisons without any routing table (Fig.~\ref{fig:addr-adj}, proof in Appendix~\ref{app:space}). In our running example ($N = 5$, $m = 3$), tiers $1$, $2$, and $3$ hold $5$, $25$, and $125$ nodes, totaling $155$ nodes in $31$ cells. Node $42$ anchors $\operatorname{cell}(42) = \{420, \ldots, 424\}$, with $420$ its marked duplicate.

\begin{figure}[t]
\centering
\includegraphics[width=0.76\linewidth]{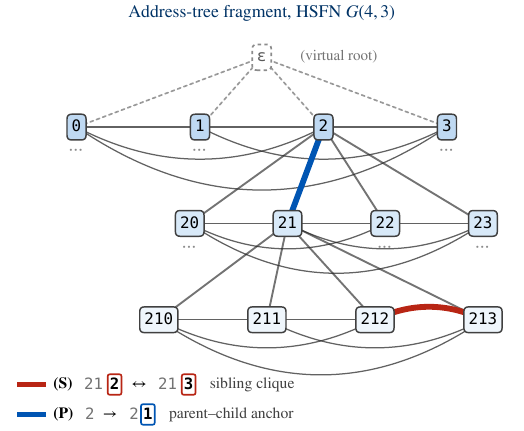}
\caption{Address-based adjacency in $G(4,3)$ across tiers $1$ to $3$. Clause~(S) of Proposition~\ref{thm:adjacency-II} generates sibling cliques and clause~(P) generates anchor uplinks. Digit $0$ designates the marked duplicate.}
\label{fig:addr-adj}
\end{figure}

\subsection{Transports, Fault Model, and Timing Assumptions}
\label{subsec:overlay}
\label{subsec:faults}

The edges of $G(N,m)$ represent the sparse physical links provisioned by a deployment. Hierarchical consensus also induces a logical \emph{protocol overlay} joining sibling members within each cell and every member of a child cell to every member of its parent cell. On $G(N,m)$, cross-cell messages transit the child's anchor in one hop if the anchor is an endpoint and two hops otherwise.

\begin{remark}[Two transports and the redundancy spectrum]
\label{rem:comm-abstraction}
We analyze two benchmark transports (Fig.~\ref{fig:net-compare}, Appendix~\ref{app:space}). Under the \emph{overlay} transport, parent--child cell links are direct ($1$ hop) and no single crash severs a child cell, realized at the upper extreme by complete-bipartite $K_{N,N}$ bundles ($G_{\mathrm{dense}}$, connectivity $N$, degree $\Theta(N^2)$). Under the \emph{physical} transport, all traffic traverses $G(N,m)$ at degree at most $2N$, making each anchor a single-node cut for its child cell. Dual homing (Section~\ref{subsec:anchor-redundancy}) bridges these extremes at degree at most $3N{+}1$. Logical node identities remain distinct even when an anchor and its duplicate share a geometric coordinate. Independent node faults $P_f$ model separate physical hosts (so an anchor and its geometrically co-located duplicate fail independently across committees), whereas $p_s$ models independent whole-cell site shocks rather than single-host co-location of cross-committee pairs.
\end{remark}

\emph{Fault and timing model.} Each $N$-member cell tolerates $f = \lfloor (N{-}1)/3 \rfloor$ Byzantine faults (so nontrivial Byzantine tolerance begins at $N \geq 4$) with quorum $Q = \lceil (N{+}f{+}1)/2 \rceil$ satisfying $2Q \geq N{+}f{+}1$. Messages are digitally signed. Consensus safety requires no timing assumptions, liveness follows standard PBFT partial synchrony (with physical non-null root progress requiring a majority of surviving child branches), and ideal latency uses fixed per-hop delay $\delta$. Reliability recursions assume independent node failure probability $P_f$ and optional independent whole-cell site outage $p_s$. Membership is fixed within each round, and reconfigurations apply between rounds. Table~\ref{tab:notation} summarizes the core notation.

\begin{table}[t]
\centering
\caption{Summary of Core Notation.}
\label{tab:notation}
\footnotesize
\setlength{\tabcolsep}{3.5pt}
\renewcommand{\arraystretch}{1.1}
\begin{tabular}{@{}lp{0.68\columnwidth}@{}}
\toprule
Symbol & Meaning \\
\midrule
$N, m$ & Cell size ($N \geq 3$) and network depth ($m \geq 1$) \\
$\Sigma_N, \mathcal{W}_{N,m}$ & Digit alphabet $\{0,\ldots,N{-}1\}$ and addresses of length $1 \ldots m$ \\
$G(N,m), F(N,m)$ & Sparse HSFN graph and duplicate-marked network \\
$\operatorname{cell}(w), \operatorname{pre}(a)$ & $N$-clique anchored at $w$; parent prefix of address $a$ \\
$V(N,m), L(N,m)$ & Total nodes $N(N^m{-}1)/(N{-}1)$ and total physical links \\
$f, Q, s$ & Per-cell Byzantine budget $\lfloor(N{-}1)/3\rfloor$, quorum, senders ($f{+}1$) \\
$P_f, p_s, q$ & Node fault prob., site outage prob., internal cell failure prob. \\
$f_b, g_N, h_N$ & Overlay, single-anchor physical, and dual-homed recursions \\
$q_0(b), q_c(b)$ & Certified depth-uniform threshold and saddle-node fold \\
$A, A^{\mathrm{ph}}$ & Leaf-to-root registration prob.\ (overlay / physical) \\
$x=(D,\mu,\zeta)$ & Configuration (deployed cells, occupied slots, disabled slots) \\
$\beta(H), \operatorname{dw}(H)$ & Max block order and block--cut descent weight of guest $H$ \\
$V_N, \chi=(a,h)$ & Higman--Thompson group and spacetime coordinate \\
\bottomrule
\end{tabular}
\end{table}

\section{Spatial Structure: Addressing, Routing, and Geometry}
\label{sec:space}
\label{sec:representation}

Every spatial metric and routing decision of the HSFN follows directly from the address grammar of Section~\ref{sec:model}, while pairing it with half-scale midpoint simplices fixes its Hausdorff dimension.

\subsection{Locators, Exact Metric, and Degree Profile}
\label{subsec:metric}

\begin{proposition}[Length-prefixed locator]
\label{prop:locator-II}
Within a fixed $(N,m)$ family context, encoding the zero-based tier $t{-}1$ in $\lceil \log_2 m \rceil$ bits (zero bits at $m = 1$) and $t$ base-$N$ digits in $t\lceil \log_2 N \rceil$ bits, together with a $\lceil \log_2 N \rceil$-bit parameter tag, gives a locator budget of $(t{+}1)\lceil \log_2 N \rceil + \lceil \log_2 m \rceil$ bits. As $m \to \infty$, its maximum length approaches $\lceil \log_2 N \rceil / \log_2 N$ times the information-theoretic lower bound $\log_2 V(N,m)$, reaching optimality precisely when $N$ is a power of two.
\end{proposition}

\begin{theorem}[Exact distance, greedy routing, and diameter]
\label{thm:distance-II}
Let $u \neq v$ be nodes of $G(N,m)$ at tiers $t_u$ and $t_v$, with longest common prefix length $\ell$. Call $u$ and $v$ \emph{prefix-comparable} if one is a prefix of the other.
\begin{enumerate}
    \item[(i)] The shortest-path distance is $d(u,v) = |t_u - t_v|$ when $u, v$ are prefix-comparable, and $d(u,v) = t_u + t_v - 2\ell - 1$ otherwise.
    \item[(ii)] \emph{Greedy locator routing} is exact. At any current node $z \neq v$ with common-prefix length $\lambda$, forward to child $\operatorname{pre}_{|z|+1}(v)$ if $z$ is a proper prefix of $v$, to sibling $\operatorname{pre}_{\lambda+1}(v)$ if $z$ and $v$ are incomparable with $|z| = \lambda{+}1$, and to anchor $\operatorname{pre}(z)$ otherwise. Every hop is computed from the two locators alone and reduces $d(\cdot,v)$ by exactly one.
    \item[(iii)] For every $m \geq 2$, the diameter of $G(N,m)$ is exactly $2m{-}1$.
\end{enumerate}
\end{theorem}

\begin{figure}[t]
\centering
\includegraphics[width=0.82\linewidth]{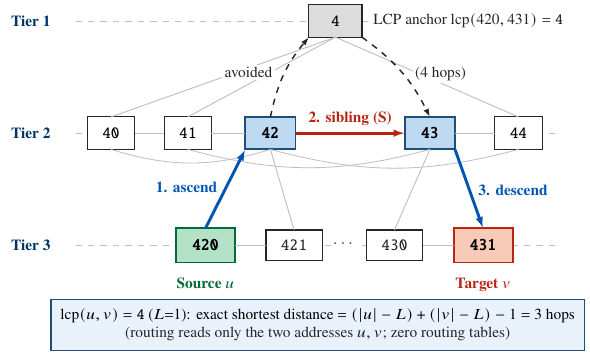}
\caption{Table-free greedy locator routing in $G(5,3)$ from $u = 420$ to $v = 431$ (Theorem~\ref{thm:distance-II}). The packet ascends to the diverging sibling cell, takes one sibling shortcut~(S) from $42$ to $43$ instead of climbing to anchor~$4$, and descends to $431$ in exactly three hops.}
\label{fig:greedy-routing}
\end{figure}

The greedy walk climbs toward the common prefix, crosses one sibling shortcut in the cell where the addresses diverge, and descends (Fig.~\ref{fig:greedy-routing}). In the running example ($N = 5, m = 3$), nodes $420$ and $431$ have $\ell = 1$ and distance $3$ along $420 \to 42 \to 43 \to 431$, whereas $000$ and $444$ achieve the diameter $5$.

\begin{proposition}[Degrees and links]
\label{thm:deg-II}
For $m \geq 2$, a tier-$t$ node of $G(N,m)$ has degree $2N{-}1$ at $t = 1$, $2N$ at $2 \leq t \leq m{-}1$, and $N$ at $t = m$. The average degree $\bar{d} = 2L(N,m)/V(N,m)$ increases monotonically with $m$ toward $N{+}1$ (closed-form expressions~\eqref{eq:links-II}--\eqref{eq:avgdeg-II} in Appendix~\ref{app:space}).
\end{proposition}

\begin{proposition}[Connectivity price]
\label{prop:conn-price}
Let $m \geq 2$. (i)~In $G(N,m)$, every anchor of tier $t \leq m{-}1$ is a cut vertex separating its $(N^{m-t+1}{-}N)/(N{-}1)$ descendants from the rest of the network, so $\kappa(G(N,m)) = 1$. (ii)~Realizing every parent--child cell edge by a complete bipartite bundle $K_{N,N}$ raises vertex connectivity to $N$ at maximum degree $\Theta(N^2)$.
\end{proposition}

Every node therefore maintains at most $2N$ links independently of total population $V(N,m)$. Appendix~\ref{app:space} supplies all proofs alongside overlay comparisons (Table~\ref{tab:overlays}), recursive interconnect benchmarks (Table~\ref{tab:interconnect}, covering DCell, BCube, fat-tree, and hypercube), and Sierpi\'{n}ski-graph contrasts. Without lateral subtree bridges, mean leaf-to-leaf distance stays within one hop of the diameter. Section~\ref{subsec:phys-recursion} quantifies the cut-vertex exposure of Proposition~\ref{prop:conn-price} and resolves it through dual homing.

\subsection{Midpoint Realization, Hausdorff Dimension, and Family Nesting}
\label{subsec:geom-main}

To connect the discrete address tree back to Lemma~\ref{lem:equidistance}, realize the seed cell on the vertices $\{p_1,\ldots,p_N\}$ of a regular $(N{-}1)$-simplex $\Delta_N \subset \mathbb{R}^{N-1}$ with $N$ half-scale corner similitudes $f_i(x) = p_i + \tfrac{1}{2}(x - p_i)$. The \emph{midpoint realization} $\iota \colon \mathcal{W}_{N,m} \to \mathbb{R}^{N-1}$ places each spawned cell as the half-scale corner sub-simplex at its anchor's vertex, mapping the marked duplicate $a0$ onto $\iota(a)$ and the remaining $N{-}1$ offspring onto the midpoints of edges leaving $\iota(a)$ (Definition~\ref{def:embedding-maps}, Fig.~\ref{fig:equidistance-spawning}(c)--(d)). Geometrically coincident nodes remain distinct logical addresses, and coincidence never creates edges.

\begin{theorem}[Hausdorff dimension of the HSFN limit set]
\label{thm:dim-II}
Under the midpoint realization $\iota$, the tier-$t$ cell simplexes are precisely the images $f_{i_1} \circ \cdots \circ f_{i_{t-1}}(\Delta_N)$ over $\{1,\ldots,N\}^{t-1}$, in bijection with the $N^{t-1}$ cells of Proposition~\ref{thm:unified-count}. The closure of the union of realized node positions across all depths is the unique compact attractor $\mathcal{S}\Delta_N$ of the iterated function system $\{f_1,\ldots,f_N\}$, with exact Hausdorff and similarity dimension
\begin{equation}
\label{eq:dim-main}
    \dim_{\mathrm{H}} \mathcal{S}\Delta_N = \dim_{\mathrm{S}} \mathcal{S}\Delta_N = \log_2 N.
\end{equation}
\end{theorem}

\begin{figure*}[!t]
\centering
\includegraphics[width=0.78\textwidth]{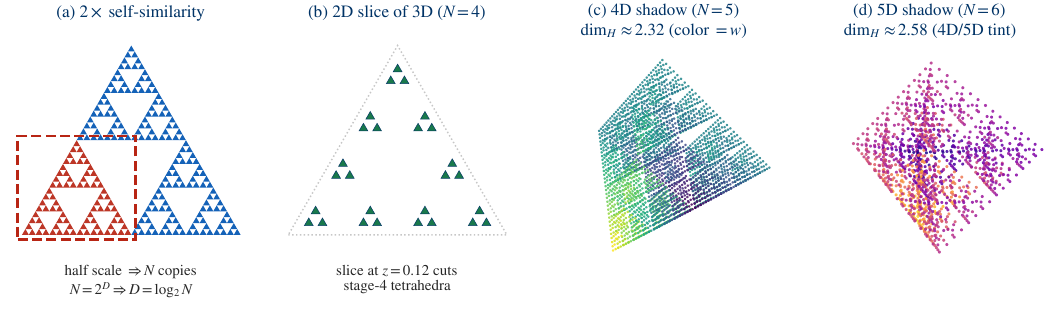}
\caption{Visualizing higher-dimensional HSFN attractors $\mathcal{S}\Delta_N$. (a)~Zooming $2\times$ into one corner reproduces the entire attractor, giving $N = 2^D$ and $D = \log_2 N$. (b)~A horizontal $2\text{D}$ slice ($z = 0.12$) through the fourth finite approximant of the $3\text{D}$ attractor at $N = 4$ cuts it into triangular fragments. (c), (d)~Three-dimensional orthogonal shadows of the $4\text{D}$ attractor ($N = 5$, colored by fourth coordinate $w$) and $5\text{D}$ attractor ($N = 6$, tinted by higher coordinates), showing voids persisting beyond three dimensions.}
\label{fig:hd-projections}
\end{figure*}

In barycentric coordinates, each image $f_i(\Delta_N)$ occupies the corner half-space $\lambda_i \geq \tfrac{1}{2}$, so open interiors never overlap and the Moran--Hutchinson theorem yields~\eqref{eq:dim-main} (Appendix~\ref{app:geometry}, Fig.~\ref{fig:barycentric-ifs}). For fixed $N$, ambient dimension stays $N{-}1$ across all depths, whereas increasing $N$ raises both $N{-}1$ and $\log_2 N$ without bound. Halving the spatial scale reveals $N$ self-similar copies of the whole (Fig.~\ref{fig:hd-projections}(a)). The attractor dimension rises smoothly from $1.58$ at $N = 3$ to planar scaling ($2.0$) at $N = 4$, lifts into higher Euclidean spaces at $N = 5$ and $N = 6$ (Fig.~\ref{fig:hd-projections}(b)--(d)), and reaches volumetric scaling ($3.0$) at $N = 8$.

At finite depth $m$, $G(N,m)$ has logarithmic diameter $\Theta(\log_N V)$, whereas $\log_2 N$ measures the self-similar geometric attractor $\mathcal{S}\Delta_N$. On a unit-edge seed simplex, the sibling links of all tier-$m$ leaf cells carry total Euclidean length $\binom{N}{2}(N/2)^{m-1}$, which diverges inside the bounded seed hull for every $N \geq 3$ like the unbounded frontier of a fractal snowflake. Keeping geometrically coincident boundary points logically unglued preserves table-free shortest paths, Harary block structure, and the boundary closure generating $V_N$ (Section~\ref{subsec:vn-main}). Moreover, the family $\{F(N,m)\}$ embeds into larger parameters without renumbering existing addresses.

\begin{lemma}[Family monotonicity]
\label{lem:family-mono}
For all $3 \leq N \leq N'$ and $1 \leq m \leq m'$, the inclusion $\mathcal{W}_{N,m} \subseteq \mathcal{W}_{N',m'}$ embeds $F(N,m)$ as an induced, isometric, mark-preserving subnetwork of $F(N',m')$, and these inclusions compose into a directed system.
\end{lemma}

Expanding depth $m$ adds new leaf cells directly, while expanding $N$ preserves all existing addresses alongside a committee-size upgrade (Appendix~\ref{app:geometry}).

\section{Agreement Layer: Certified Hierarchical PBFT}
\label{sec:time}
\label{sec:consensus}

\begin{figure*}[!t]
\centering
\includegraphics[width=0.78\textwidth]{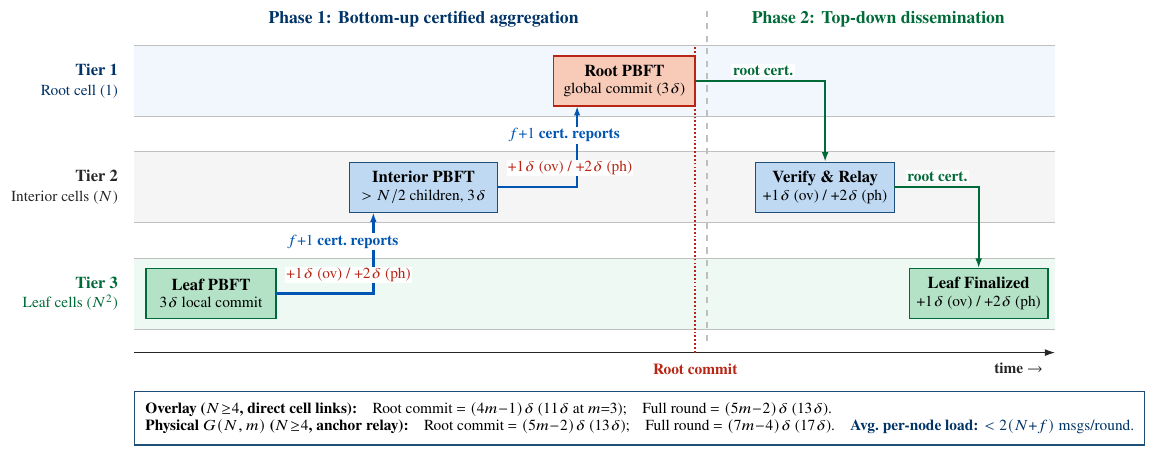}
\caption{Single-round timeline of certified hierarchical PBFT for depth $m = 3$ and $N \geq 4$ (fault-free, timer-free round). Phase~1 aggregates local $3\delta$ cell commits upward via $f{+}1$ certified reports per child cell, and Phase~2 disseminates the signed root decision downward. Direct overlay links commit at the root in $(4m{-}1)\delta$ and finish in $(5m{-}2)\delta$, whereas anchor-relayed physical paths on $G(N,m)$ take $(5m{-}2)\delta$ and $(7m{-}4)\delta$ (Proposition~\ref{prop:latency}).}
\label{fig:proto-timeline}
\end{figure*}

The prefix tree of addresses doubles as the consensus hierarchy. Each $N$-node cell runs PBFT locally and passes a signed receipt upward, keeping network-average per-node traffic $O(N)$ (and total logical messages $\Theta(V)$ for fixed $N$) while total population grows exponentially in depth $m$ (Fig.~\ref{fig:proto-timeline}).

\subsection{Consensus Tree and Protocol Specification}
\label{subsec:consensus-tree}
\label{subsec:multi-pbft}

Each cell $\operatorname{cell}(w)$ ($|w| \leq m{-}1$) forms one PBFT committee of $N$ members with children $\operatorname{cell}(w0), \ldots, \operatorname{cell}(w(N{-}1))$ when $|w| \leq m{-}2$. Tier-$m$ cells are \emph{leaf cells}, and $\operatorname{cell}(\varepsilon)$ is the \emph{root cell}.

Execution proceeds in numbered rounds $r = 1, 2, \ldots$ sequentially, each round producing at most one block per cell. Every signed message binds its type, round, cell identifier, view, and digest, preventing cross-round or cross-cell replay. A \emph{prepared certificate} locks a local quorum onto one block, and a \emph{commit certificate} serves as a signed receipt verified offline by the parent cell.
\begin{enumerate}
    \item \textbf{Intra-cell agreement.} Each cell runs PBFT (pre-prepare, prepare, commit) at quorum $Q$, with view-$v$ primary $v \bmod N$. A prepared certificate pairs the primary's pre-prepare with $Q{-}1$ matching non-primary prepares, and a commit certificate gathers $Q$ matching commits. A member sending a view change for view $w$ ceases participation below $w$ and attaches its highest prepared certificate. A new-view message collects $Q$ view changes and proposes the highest prepared digest among them, or a valid fresh proposal if none exists.
    \item \textbf{Validity and in-order resolution.} A leaf cell proposes its local input. An interior cell proposes an \emph{aggregate} of $(\text{child}, \text{digest})$ pairs for round~$r$, each backed by that child's valid commit certificate, covering at least $\lfloor b/2 \rfloor + 1$ of its $b$ children ($b = N$ in a full HSFN). Rounds resolve strictly in order by either a committed block or a certified null block produced via a closing view change. A correct member signs no round-$r$ message until all earlier rounds resolve, and every block header hashes the previous resolution, creating a hash-linked cell chain whose round numbers define block height in Section~\ref{sec:spacetime}.
    \item \textbf{Certified report.} Upon committing, a cell's $s = f{+}1$ \emph{senders} (the committing view's primary and the next $f$ members in digit order) send the digest and commit certificate to every member of the parent cell. Any other committed correct member sends a fallback report if no root decision arrives after half its tier budget. An interior member proposes an aggregate including every verified child report received so far once it holds reports from all children, or from a majority after grace period $T_g$. Without a majority by its tier deadline, the cell closes the round with a certified null block.
    \item \textbf{Certified dissemination.} Once the root commits, its $f{+}1$ senders forward the root decision and commit certificate to all members of child cells, which verify, record, and forward downward through their own senders.
\end{enumerate}

\begin{remark}[Report senders and withholding resilience]
\label{rem:reporters}
Relying solely on the primary ($s = 1$) to report upward lets a single Byzantine primary complete intra-cell agreement and then silently withhold the report. Designating $s = f{+}1$ senders guarantees at least one correct sender whenever a cell stays within its fault budget $f$. If Byzantine members selectively help only non-sender correct replicas commit, the step-3 fallback report still delivers the certified block upward.
\end{remark}

\subsection{Safety and Validity}
\label{subsec:safety}

\begin{proposition}[Agreement and validity]
\label{prop:safety}
Suppose a cell has at most $f$ Byzantine members.
\begin{enumerate}
    \item[(i)] No two correct members of the cell commit different blocks in the same round.
    \item[(ii)] For any child cell that also has at most $f$ Byzantine members, every valid commit certificate from that child names its unique certified block for that round (matching the block committed by every correct member of that child that finishes the round). Consequently, when every included child cell stays within budget $f$, every committed aggregate contains only legitimately certified child blocks.
\end{enumerate}
In particular, if every cell along the path from a cell to the root stays within its fault budget $f$, any block of that cell included in the root decision through certified aggregates is that cell's unique certified block for the round.
\end{proposition}

Because $2Q \geq N{+}f{+}1$, any two prepared quorums share at least one correct replica, and any $Q$ view changes intersect the $Q{-}f$ correct committers of an earlier view. Committed digests are therefore preserved across views and certified aggregates up to the root (full proof in Appendix~\ref{app:protocol}). Omitting the step-1 view lock admits safety violations under heavy-tailed delays (Section~\ref{sec:eval}), whereas liveness follows standard PBFT partial synchrony (with non-null physical root progress requiring a majority of surviving child branches, as in Section~\ref{subsec:faults}) and step-3 fallback overcoming withholding.

\subsection{Exact Message Count, Committee Trees, and Latency}
\label{subsec:complexity}
\label{subsec:delay}

\begin{proposition}[Executed message count]
\label{prop:exec-count}
Consider one fault-free round on a fully populated $G(N,m)$ of $V = V(N,m)$ nodes with $s$ senders per cell ($s = f{+}1$ by default), starting with all earlier rounds resolved and no timer firing. The round sends exactly
\begin{equation}
\label{eq:exec-count}
    C_{\mathrm{exec}} = 2(N{-}1)V + 2s(V - N)
\end{equation}
messages, traversing $2(N{-}1)V + 2s(2N{-}1)(V/N - 1)$ physical links on $G(N,m)$.
\end{proposition}

Each cell executes $2N(N{-}1)$ intra-cell messages, and each non-root cell exchanges $sN$ reports upward and $sN$ disseminations downward, averaging $2N{-}1$ physical hops per child cell per sender via its anchor ($2N{-}1$ upward, and $N$ directly or $2N$ via a sibling relay downward, proof in Appendix~\ref{app:protocol}). Averaged across the network for fixed $N$, a node sends fewer than $2(N{+}f)$ messages per round (with designated senders carrying the additional $N$ cross-tier messages per child or parent edge), whereas flat PBFT sends $2(V{-}1)$ messages per replica across a single global committee of all $V$ nodes. Because both message counts and tier delays are strictly local to each committee and parent--child edge, the accounting generalizes immediately to arbitrary committee trees.

\begin{proposition}[Committee trees]
\label{thm:committee-tree}
Run the certified protocol on any rooted tree of PBFT committees where committee $c$ has $N_c$ members, $s_c$ senders, and parent $p(c)$, with longest root-to-leaf path of $L_{\max}$ tiers. A fault-free round with non-firing timers sends exactly $\sum_{c} 2N_c(N_c{-}1) + \sum_{c \neq \mathrm{root}} (s_c N_{p(c)} + s_{p(c)} N_c)$ messages. When every committee has $N_c \geq 4$ (nontrivial Byzantine tolerance, $Q_c - 1 \geq 2$) under fixed per-hop overlay delay $\delta$ and non-firing timers, the root commits at $(4L_{\max}{-}1)\delta$ and every node holds the decision by $(5L_{\max}{-}2)\delta$.
\end{proposition}

\begin{proposition}[Round latency]
\label{prop:latency}
Under the fault-free fixed-delay assumptions of Proposition~\ref{thm:committee-tree} on $G(N,m)$ with $N \geq 4$, the overlay transport commits at the root in $t_{\mathrm{root}} = (4m{-}1)\delta$ and completes dissemination at $t_{\mathrm{conf}} = (5m{-}2)\delta$. On the physical graph $G(N,m)$, anchor relaying adds one hop per cross-tier crossing, yielding $t_{\mathrm{root}} = (5m{-}2)\delta$ and $t_{\mathrm{conf}} = (7m{-}4)\delta$ (compared with $3\delta$ for all-to-all flat PBFT at $N \geq 4$; at $N = 3$, where $f = 0$ and $Q = 2$, local commits complete in $2\delta$, reducing each tier by $\delta$).
\end{proposition}

Proofs of Proposition~\ref{thm:committee-tree} and Proposition~\ref{prop:latency}, together with cross-protocol comparisons against HotStuff, Kauri, multi-layer PBFT, and Steward (Table~\ref{tab:hierbft}), appear in Appendix~\ref{app:protocol}. Round latency grows linearly in depth $m = \log_N V + O(1)$, trading tree height against intra-cell verification cost as $N$ varies (Section~\ref{sec:eval}).

\section{Reliability Theory: Thresholds, Anchors, and Path Registration}
\label{sec:fault}
\label{sec:security}

\begin{figure*}[!t]
\centering
\includegraphics[width=0.76\textwidth]{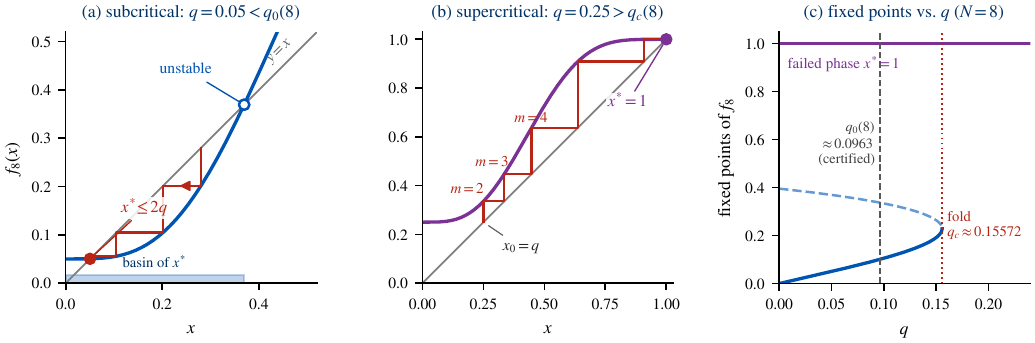}
\caption{HSFN overlay reliability recursion $f_8(x) = 1 - (1{-}q)\Pr[\mathrm{Bin}(8,x) \leq 3]$ as a one-dimensional dynamical system ($b = N = 8$). (a)~Subcritical cobweb at $q = 0.05 < q_0(8)$ starting from $x_0 = 0.28$ inside the basin of attraction and converging into the certified invariant interval $[q, 2q] = [0.05, 0.10]$ of Theorem~\ref{prop:threshold-b}. (b)~Supercritical cobweb at $q = 0.25 > q_c(8)$ climbing toward root collapse $x^* = 1$. (c)~Bifurcation diagram placing the certified bound $q_0(8) \approx 0.0963$ below the saddle-node fold $q_c(8) \approx 0.15572$ (six-digit value $0.155719$).}
\label{fig:rg-cobweb}
\end{figure*}

\subsection{Overlay Recursion and Depth-Uniform Threshold}
\label{subsec:fpd}
\label{subsec:threshold}

Because each interior cell takes a majority vote over its $N$ child branches, local failure probabilities undergo a threshold transition as depth $m$ grows (Fig.~\ref{fig:rg-cobweb}). At $N = 8$, for example, a cell tolerates up to $f = 2$ member faults and succeeds whenever at most $3$ of its $8$ child branches fail. Below the certified threshold $q_0(N)$, majority voting traps root failure below $2q$ at every depth. Above the fold $q_c(N)$, faults amplify up the tree until the root collapses.

Assume each node fails independently with probability $P_f$. A cell fails internally when more than $f = \lfloor (N{-}1)/3 \rfloor$ of its $N$ members fail, occurring with probability
\begin{equation}
\label{eq:P-leaf}
    q = P_m = 1 - \sum_{i=0}^{f} \binom{N}{i} (1 - P_f)^{N-i} P_f^{\,i}.
\end{equation}
Let $S$ denote the recursive event that the root stays within budget $f$ and strictly more than half of the children of every such cell satisfy $S$. Every cell on this majority spine stays inside budget $f$ and carries authentic digests by Proposition~\ref{prop:safety}. Under crash faults (or silent over-budget cells) with sufficient grace periods, $S$ coincides with non-null root commitment when $3 \nmid N$ and lower-bounds commitment when $3 \mid N$.

\begin{theorem}[Failure recursion]
\label{thm:security}
Consider a depth-$m$ tree of $N$-member cells where level-$k$ cells have $b_k$ children ($k = 1$ is the root). Define $S_b(x) = \Pr[\mathrm{Bin}(b, x) \leq \lceil b/2 \rceil - 1]$ and $f_b(x) = 1 - (1-q)S_b(x)$. Then level failure probabilities satisfy $P_m = q$ and $P_k = f_{b_k}(P_{k+1})$ for $1 \leq k < m$, with $\Pr[S] = 1 - P_1$. On the uniform $N$-ary tree of $G(N,m)$,
\begin{equation}
\label{eq:P-hsfn}
    P_1(N,m) = f_N^{\,m-1}(q).
\end{equation}
On any heterogeneous committee tree where cell $w$ has internal failure rate $q_w$ and $b_w$ children with failure rates $p_i$, interior cells satisfy the same majority-tail recursion using independent child indicators $X_i \sim \mathrm{Bernoulli}(p_i)$.
\end{theorem}

\begin{theorem}[Depth-uniform reliability threshold]
\label{prop:threshold-b}
Let $b \geq 3$ and write $c = \lceil b/2 \rceil$. The iterates $f_b^{\,h}(q)$ are nondecreasing in $h$ and converge to the least fixed point $x_b^*(q)$ of $f_b$ in $[q,1]$. If
\begin{equation}
\label{eq:q0-def}
    q < q_0(b) := \left[2^{c}\binom{b}{c}\right]^{-1/(c-1)},
\end{equation}
then $f_b$ maps $[q, 2q]$ into itself, certifying $f_b^{\,h}(q) \leq 2q$ for every $h \geq 0$. Moreover, there exists a critical threshold $q_c(b) \in \bigl[q_0(b),\, 1 - \tbinom{b}{\lfloor b/2 \rfloor + 1}^{-1}\bigr]$ such that $x_b^*(q) < 1$ for $q < q_c(b)$ and $x_b^*(q) = 1$ for $q > q_c(b)$.
\end{theorem}

\begin{corollary}[Depth-uniform overlay root certificate]
\label{cor:threshold-II}
Under the overlay recursion of Theorem~\ref{thm:security}, if $q < q_0(N)$, then overlay root failure satisfies $P_1(N,m) \leq 2q$ for every depth $m \geq 1$.
\end{corollary}

\begin{figure*}[!t]
\centering
\includegraphics[width=0.78\textwidth]{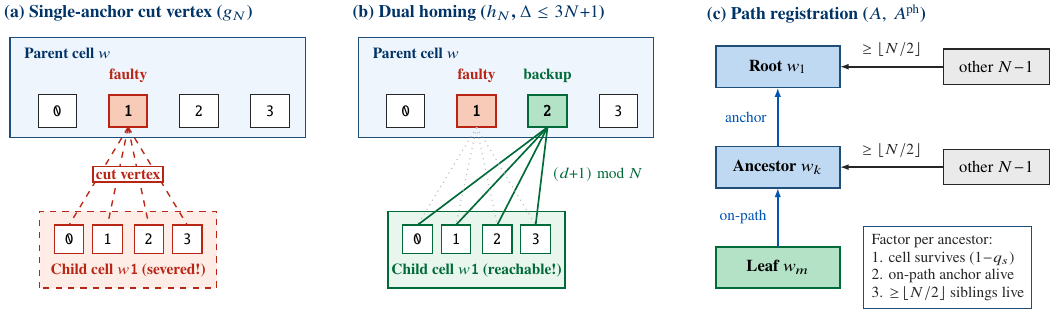}
\caption{Anchor reliability mechanisms on $G(N,m)$. (a)~A faulty parent member severs its anchored child cell, driving the physical recursion $g_N$ (Proposition~\ref{prop:phys-recursion}). (b)~Dual homing links each child cell to backup anchor $(d{+}1) \bmod N$ at maximum degree $3N{+}1$, eliminating every single-node cut (Proposition~\ref{prop:dual}). (c)~Leaf-to-root path registration multiplies one survival-and-majority factor per ancestor along the unique path (Proposition~\ref{prop:registration}).}
\label{fig:anchor-dual-reg}
\end{figure*}

Complete proofs appear in Appendix~\ref{app:protocol}, and Fig.~\ref{fig:rg-cobweb} illustrates the cobweb dynamics and bifurcation portrait at $N = 8$. The certified threshold $q_0(b)$ depends only on the branching factor $b$, while committee size $N$ enters through the leaf failure rate $q(P_f)$. Because every tier of the HSFN branches uniformly at $b = N$, Corollary~\ref{cor:threshold-II} holds without an additive root correction. Across $N = 4, 5, 6, 8$, the bound evaluates to $q_0(N) = 0.0417$, $0.1118$, $0.0791$, and $0.0963$, oscillating with parity because $\lceil N/2 \rceil$ rises only at odd $N$.

Independent whole-cell \emph{site outages} with probability $p_s$ enter by replacing $q$ with $q_s = p_s + (1 - p_s)q$ on the overlay and scaling per-cell survival by $1 - p_s$ on the physical graph.

\subsection{Anchor Exposure, Dual Homing, and Path Registration}
\label{subsec:phys-recursion}
\label{subsec:anchor-redundancy}
\label{subsec:reg-theory}

On the physical graph $G(N,m)$, each member of a parent cell anchors one child cell, so a parent crash silences that entire child branch (Proposition~\ref{prop:conn-price}, Fig.~\ref{fig:anchor-dual-reg}(a)). Conditioning on faulty parent members yields an exact crash recursion $g_N$ coupling internal faults to severed subtrees. Conversely, \emph{dual-homing} each child cell $\operatorname{cell}(wd)$ (anchored at parent member $wd$ of digit $d$) with $N$ backup uplinks from all $N$ members of $\operatorname{cell}(wd)$ to backup parent member $w((d{+}1) \bmod N)$ raises maximum degree only to $3N{+}1$ and eliminates every single-node cut (Fig.~\ref{fig:anchor-dual-reg}(b), analogous to multi-parent links in RPL~\cite{winter2012rpl} and backup pointers in Tapestry~\cite{zhao2004tapestry}).

\begin{proposition}[Failure recursion on the physical graph]
\label{prop:phys-recursion}
Run the protocol on $G(N,m)$ under independent crash faults with all traffic routed along $G(N,m)$, assuming timers admit every completing child subtree and every in-budget cell reports. Letting $r = \lceil N/2 \rceil - 1$, conditioning on $j \leq \min(f,r)$ faulty parent members severing $j$ distinct child subtrees defines the anchor-coupled recursion
\begin{equation}
\label{eq:gN-main}
\begin{split}
    g_N(x) = 1 - \sum_{j=0}^{\min(f, r)} &\tbinom{N}{j} P_f^{\,j} (1-P_f)^{N-j} \\
    &\cdot \Pr\bigl[\mathrm{Bin}(N{-}j, x) \leq r - j\bigr].
\end{split}
\end{equation}
If $3 \nmid N$, then $P_1^{\mathrm{phys}} = g_N^{\,m-1}(q)$ is exact; if $3 \mid N$, it upper-bounds root failure. Moreover, $g_N(x) \geq f_N(x)$ for all $x \in [0,1]$, hence $g_N^{\,m-1}(q) \geq f_N^{\,m-1}(q)$ at every depth.
\end{proposition}

\begin{proposition}[Dual-homed anchor recursion]
\label{prop:dual}
Under the hypotheses of Proposition~\ref{prop:phys-recursion} with dual-homed cells, a child cell is severed only when both of its anchors $(d, (d{+}1)\bmod N)$ lie in the faulty parent subset $J$. Writing $a(J)$ for the number of such faulty consecutive pairs in $J$ and conditioning on $J$ (with $|J| \leq f, a(J) \leq r$) defines the dual-homed recursion $h_N(x)$ (explicit sum~\eqref{eq:hN-def} in Appendix~\ref{app:protocol}), giving $P_1^{\mathrm{dual}} = h_N^{\,m-1}(q)$ exactly for $3 \nmid N$ and an upper bound for $3 \mid N$. Whenever $f = 1$, every admissible fault set has $|J| \leq 1$, forcing $a(J) = 0$ and collapsing $h_N \equiv f_N$ identically.
\end{proposition}

Table~\ref{tab:collapse-thresh} compares the resulting collapse fault-rate thresholds across transports. Across the listed rows ($N \in \{5, 8, 16\}$), single-anchor exposure reduces the collapse threshold $P_f$ by $23\%$--$28\%$ (and by nearly half at $N = 4$, from $0.098$ to $0.051$). Consequently, the overlay certificate of Corollary~\ref{cor:threshold-II} does not transfer automatically to the single-anchor physical graph or to designated leaf registration. At $N = 4$, $P_f = 0.08$, and $m = 10$, $q = 0.0344 < q_0(4)$ keeps overlay root failure at $0.046$, whereas single-anchor physical root failure reaches $0.989$. Even deep inside the subcritical regime ($N = 8$, $P_f = 0.05$, $m = 50$), where overlay and single-anchor root failure rates nearly coincide ($5.79 \times 10^{-3}$ versus $5.82 \times 10^{-3}$), designated leaf-path registration satisfies $A \leq (1{-}q_s)^m$ and drops to $A = 0.748$ and $A^{\mathrm{ph}} = 0.067$. Dual homing restores overlay root reliability identically at $N \in \{4, 5\}$ and nearly in full at larger $N$, while preventing any single targeted node removal~\cite{motter2002cascade} from severing a subtree (versus $19.9\%$ detached by one tier-$1$ anchor on single-anchor $G(5,4)$).

\begin{table}[t]
\centering
\caption{Collapse Fault-Rate Thresholds ($P_f$) Across Transports.}
\label{tab:collapse-thresh}
\scriptsize
\setlength{\tabcolsep}{3.5pt}
\begin{tabular}{@{}ccccc@{}}
\toprule
$N$ ($f$) & Deg.\ (phys./dual) & Overlay $f_N$ & Phys.\ $g_N$ & Dual $h_N$ \\
\midrule
$5$ ($1$) & $10$ / $16$ & $0.150$ & $0.115$ & $0.150$ \\
$8$ ($2$) & $16$ / $25$ & $0.177$ & $0.127$ & $0.173$ \\
$16$ ($5$) & $32$ / $49$ & $0.269$ & $0.206$ & $0.258$ \\
\bottomrule
\end{tabular}
\end{table}

Finally, a leaf event is \emph{registered} in a round when its leaf cell commits and every ancestor along its unique root path includes its certified report (Fig.~\ref{fig:anchor-dual-reg}(c)).

\begin{proposition}[Path registration probability]
\label{prop:registration}
Under independent node faults $P_f$, cell site outages $p_s$ with $q_s = p_s + (1{-}p_s)q$, and the crash-fault hypotheses of Proposition~\ref{prop:phys-recursion} for $3 \nmid N$, the probability that a tier-$m$ leaf block is anchored by the root factors across the leaf and its $m{-}1$ ancestors into $A$ on the overlay (and dual-homed $G(N,m)$ for $f = 1$) and $A^{\mathrm{ph}}$ on single-anchor $G(N,m)$. Each ancestor contributes a survival factor requiring freedom from internal and site failure, an intact on-path anchor in the physical case, and at least $\lfloor N/2 \rfloor$ off-path child reports, as given explicitly in~\eqref{eq:reg-overlay}--\eqref{eq:reg-phys} and proved in Appendix~\ref{app:protocol}.
\end{proposition}

\section{Dynamics and Algebra: Reconfiguration, Hosting, and $V_N$}
\label{sec:reconfig}

\subsection{Configuration Space and Operations Calculus}
\label{subsec:calculus-main}

\begin{figure*}[!t]
\centering
\includegraphics[width=0.72\textwidth]{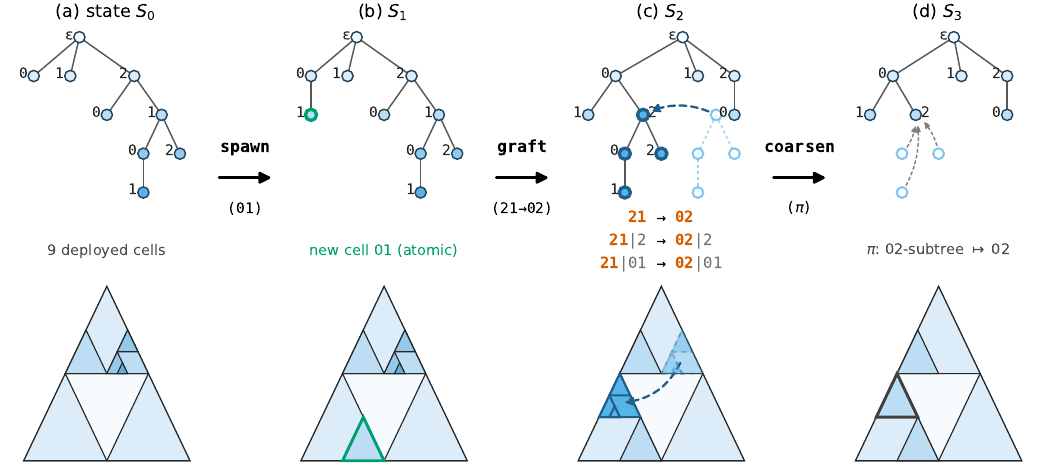}
\caption{The operations calculus at $N = 3$, showing the deployed cell tree (top) over similarity-scaled patches of the midpoint realization (bottom). (a)~A configuration with nine deployed cells. (b)~$\mathsf{spawn}(01)$ deploys one whole cell atomically. (c)~$\mathsf{graft}(21 \to 02)$ acts as a prefix substitution leaving suffixes untouched (Proposition~\ref{thm:graft-iso}), with ghost outlines marking the vacated source. (d)~Derived macro $\mathsf{coarsen}$ collapses the subtree below $02$ back into $02$.}
\label{fig:ops-filmstrip}
\end{figure*}

Eight reversible cell-atomic edits govern network growth, shrinkage, and load balancing (Table~\ref{tab:typing}, Fig.~\ref{fig:ops-filmstrip}). They open or close a cell ($\mathsf{spawn}$, $\mathsf{retract}$), admit or remove a member ($\mathsf{join}$, $\mathsf{leave}$), reorder members within a cell ($\mathsf{permute}$), relocate a subtree to a new parent ($\mathsf{graft}$), or retire and restore a vacant slot ($\mathsf{disable}$, $\mathsf{enable}$). Each edit collects root-anchored approvals from every affected committee against the active configuration version and activates in round $r{+}1$ after the final required approval is anchored in round~$r$ (treating atomic before/after states as a static-epoch assumption), while in-order resolution brings lagging cells up to round~$r$ before they advance.

\begin{definition}[Configuration]
\label{def:calc-state}
A \emph{configuration} of $F(N,m)$ is a triple $x = (D, \mu, \zeta)$, where $D \subseteq \Sigma_N^{\leq m-1}$ is a finite prefix-closed set of atomically deployed cells containing $\varepsilon$, yielding slot set $\operatorname{Sl}(x) = \bigcup_{w \in D} \operatorname{cell}(w)$; $\mu \subseteq \operatorname{Sl}(x)$ marks occupied slots; and $\zeta \subseteq \operatorname{Sl}(x)$ disjoint from $\mu \cup (D \setminus \{\varepsilon\})$ marks decommissioned slots. Every deployed cell carries a stable identifier, block chain, and commit certificates that travel with the cell under migration (retracted identifiers are retired).
\end{definition}

Table~\ref{tab:typing} lists the eight typed generators acting on the structural state $(D, \mu, \zeta)$ (under the forgetful map stripping retired cell identifiers and historical chains, Appendix~\ref{app:calculus}). Every generator preserves the structural configuration invariants and admits a two-sided partial inverse inside the table ($\mathsf{retract} = \mathsf{spawn}^{-1}$, $\mathsf{leave} = \mathsf{join}^{-1}$, $\mathsf{enable} = \mathsf{disable}^{-1}$, $\mathsf{permute}(w,\tau)^{-1} = \mathsf{permute}(w,\tau^{-1})$, and $\mathsf{graft}(a \to b)^{-1} = \mathsf{graft}(b \to a)$ subject to depth fit). In $\mathsf{spawn}(a)$, marking $a0$ designates the duplicate mark $M$ rather than occupying the slot. Words of generators therefore form an inverse monoid $\mathcal{M}(N,m)$ of partial bijections on structural configurations (Theorem~\ref{thm:rep-tree}).

\begin{table}[t]
\centering
\caption{The Eight Reconfiguration Generators. Footprints include guard reads, and $h_x(a)$ is the height of the deployed cell subtree below $a$.}
\label{tab:typing}
\footnotesize
\setlength{\tabcolsep}{2.0pt}
\renewcommand{\arraystretch}{1.2}
\begin{tabular}{@{}l>{\raggedright}p{0.29\columnwidth}>{\raggedright}p{0.24\columnwidth}>{\raggedright\arraybackslash}p{0.21\columnwidth}@{}}
\toprule
Operation & Guards & Effect & Footprint \\
\midrule
$\mathsf{spawn}(a)$ & $a \in \operatorname{Sl}(x) \setminus (D \cup \zeta)$, $|a| < m$ & deploy $\operatorname{cell}(a)$ vacant, $a0$ marked & $\{\operatorname{pre}(a), a\} \cup \operatorname{cell}(a)$ \\
$\mathsf{retract}(a)$ & $a \in D \setminus \{\varepsilon\}$, $\operatorname{cell}(a) \cap (D \cup \mu \cup \zeta) = \emptyset$ & remove $\operatorname{cell}(a)$ & $\{a\} \cup \operatorname{cell}(a)$ \\
$\mathsf{join}(s)$ & $s \in \operatorname{Sl}(x) \setminus (\mu \cup \zeta)$ & $\mu \mathbin{{+}{=}} \{s\}$ & $\{\operatorname{pre}(s), s\}$ \\
$\mathsf{leave}(s)$ & $s \in \mu$ & $\mu \mathbin{{-}{=}} \{s\}$ & $\{s\}$ \\
$\mathsf{permute}(w, \tau)$ & $w \in D$, $\tau \in S_N$, $\tau(0) = 0$ if $w \neq \varepsilon$ & recode digit $|w|{+}1$ below $w$ by $\tau$ & $\operatorname{cone}(w)$ \\
$\mathsf{graft}(a \to b)$ & $a \in D$, $a, b \in \mu$, $b \notin D$, $\operatorname{pre}(b) \in D$, $a \not\preceq b$, $|b| + h_x(a) \leq m$ & substitute prefix $a$ by $b$ below $a$ & $\operatorname{cone}(a) \cup \operatorname{cone}(b) \cup \{\operatorname{pre}(b)\}$ \\
$\mathsf{disable}(s)$ & $s \in \operatorname{Sl}(x) \setminus (\mu \cup D \cup \zeta)$ & $\zeta \mathbin{{+}{=}} \{s\}$ & $\{\operatorname{pre}(s), s\}$ \\
$\mathsf{enable}(s)$ & $s \in \zeta$ & $\zeta \mathbin{{-}{=}} \{s\}$ & $\{s\}$ \\
\bottomrule
\end{tabular}
\end{table}

Subtree migration ($\mathsf{graft}$) carries the structural weight of the calculus and touches only its boundary links.

\begin{proposition}[Migration is a prefix substitution]
\label{thm:graft-iso}
Let $\mathsf{graft}(a \to b)$ be applicable at configuration $x$ with result $x'$.
\begin{enumerate}
    \item[(i)] Moved slots re-address via $aw \mapsto bw$ for nonempty $w$, leaving every suffix digit unchanged and carrying duplicate, occupancy, and decommission marks with their cells.
    \item[(ii)] The map $aw \mapsto bw$ is a graph isomorphism from the induced subgraph on the moved slots onto its image.
    \item[(iii)] If $|a| = |b|$, slot count, participant count, provisioned link count, and the occupancy-histogram message proxies of Proposition~\ref{prop:kpi-affine} stay invariant, while exactly $N$ uplinks are cut at $a$ and created at $b$. (Under heterogeneous live committee sizes in Proposition~\ref{thm:committee-tree}, changing parent from size $N_{\mathrm{old}}$ to $N_{\mathrm{new}}$ with $s_{\mathrm{old}}, s_{\mathrm{new}}$ senders alters cross-edge messages by $s_c(N_{\mathrm{new}} - N_{\mathrm{old}}) + k(s_{\mathrm{new}} - s_{\mathrm{old}})$ for a moved root cell of size $k$ and $s_c$ senders.)
\end{enumerate}
\end{proposition}

Appendix~\ref{app:calculus} proves Proposition~\ref{thm:graft-iso} alongside three companion laws. Operations with disjoint footprints (guard reads included) commute as partial maps on structural configurations, whereas write-set disjointness alone fails on $\mathsf{graft}(a \to b)$ and $\mathsf{leave}(b)$ (Proposition~\ref{prop:commute}). Slot, participant, link, and proxy-load totals are affine functionals of per-tier histograms maintainable in $O(1)$ arithmetic per fixed-tier edit (Proposition~\ref{prop:kpi-affine}). By contrast, under the fixed-threshold ragged model of Proposition~\ref{prop:rel-exact}, exact root reliability $R(x)$ is \emph{not} determined by per-tier histograms (balanced versus skewed deployments with identical tier histograms separate at $0.6226$ versus $0.6202$ for $N = 3$ and $0.6737$ versus $0.6691$ for $N = 4$), yet updates incrementally along affected anchor paths in $O(N^2 m)$ direct convolutions (or $O(Nm)$ with cached per-cell convolution polynomials).

\subsection{Hosting Service Topologies}
\label{subsec:hosting-main}

To characterize which connected communication graphs $H$ fit inside $F(N,m)$ without trivializing cell capacity, we restrict reductions to \emph{disciplined realizations} $\mathrm{Real}(S,\mathcal{C})$ (Definition~\ref{def:realization}), selecting a subgraph $S \subseteq F(N,m)$ and contracting only vertex-disjoint chains of marked digit-$0$ duplicates $w, w0, w00, \ldots$ before taking the simple quotient.

\begin{figure*}[!t]
\centering
\includegraphics[width=0.70\textwidth]{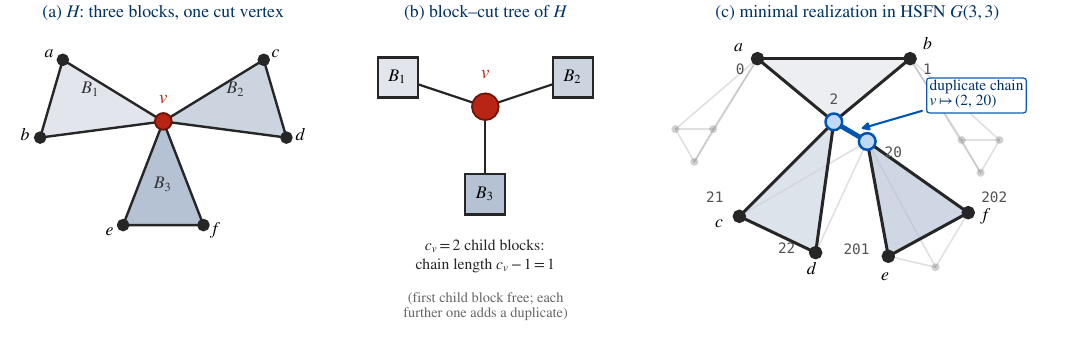}
\caption{Tight boundary case of Theorem~\ref{thm:blockcut} at $N = 3$. Hosting three triangles sharing one cut vertex extends a duplicate chain by one step and contracts $v \sim v0$, requiring $m \geq 3$.}
\label{fig:blockcut}
\end{figure*}

Decompose $H$ into its biconnected components (\emph{blocks}) meeting at cut vertices, as in the three-petal flower of Fig.~\ref{fig:blockcut}. Write $\beta(H)$ for the largest block order in $H$, call a block of order $N{+}1$ \emph{full}, and let $\operatorname{dw}(H)$ be the maximum \emph{descent weight} (sum of child-block counts along a root-to-leaf path of the block--cut tree $\mathcal{T}(H)$). Because $F(N,m)$ is a Harary block graph whose maximal cliques have order $N$ (tier~$1$) or $N{+}1$ (anchored cliques $\{v\} \cup \operatorname{cell}(v)$, Lemma~\ref{lem:maxclique}), cell size $N$ bounds the largest tightly knit cluster in $H$, while depth $m$ accommodates branching at shared cut vertices.

\begin{theorem}[Hosting characterization]
\label{thm:blockcut}
Let $H$ be a finite connected graph.
\begin{enumerate}
    \item[(i)] If $\beta(H) \leq N$, then $H$ is realized by $F(N,m)$ for every $m \geq 1 + \operatorname{dw}(H)$.
    \item[(ii)] If $H$ is realized by some $F(N,m)$, then $\beta(H) \leq N{+}1$ and every cut vertex of $H$ lies in at most two full blocks.
    \item[(iii)] If no cut vertex of $H$ lies in two full blocks, then $H$ is realized by some $F(N,m)$ if and only if $\beta(H) \leq N{+}1$.
\end{enumerate}
\end{theorem}

\begin{lemma}[Treewidth bound]
\label{lem:tw-nec}
For every $N \geq 3$, $\mathrm{tw}(F(N,1)) = N{-}1$ and $\mathrm{tw}(F(N,m)) = N$ for all $m \geq 2$. Hence $\mathrm{tw}(H) \leq N$ for every graph $H$ realized by $F(N,m)$.
\end{lemma}

Cell size $N$ therefore fixes the structural width ceiling, whereas depth $m$ grows only with branching at shared cut vertices (Fig.~\ref{fig:blockcut}). Crucially, disciplined hosting is strictly finer than treewidth $\mathrm{tw}(H) \leq N$. Because an interior chain clique identifies its consecutive pair $(x0^i, x0^{i+1})$ and drops to order at most $N$, full blocks of order $N{+}1$ can sit only at the two ends of a cut vertex's duplicate chain (birth clique or terminal anchored clique, Proposition~\ref{prop:blockcut-caveat}). This two-slot placement law immediately forbids three full blocks at one cut vertex and yields the stronger local obstruction of Proposition~\ref{prop:two-full-sep} in Appendix~\ref{app:hosting}, refusing any non-full block holding two doubly-full cut vertices at \emph{any} depth $m$ despite $\mathrm{tw}(H) \leq N$. Sufficiency of the resulting global orientation condition remains open (Section~\ref{sec:discussion}).

\subsection{The Higman--Thompson Group $V_N$ and Operation Lifting}
\label{subsec:vn-main}

\begin{figure*}[!t]
\centering
\includegraphics[width=0.68\textwidth]{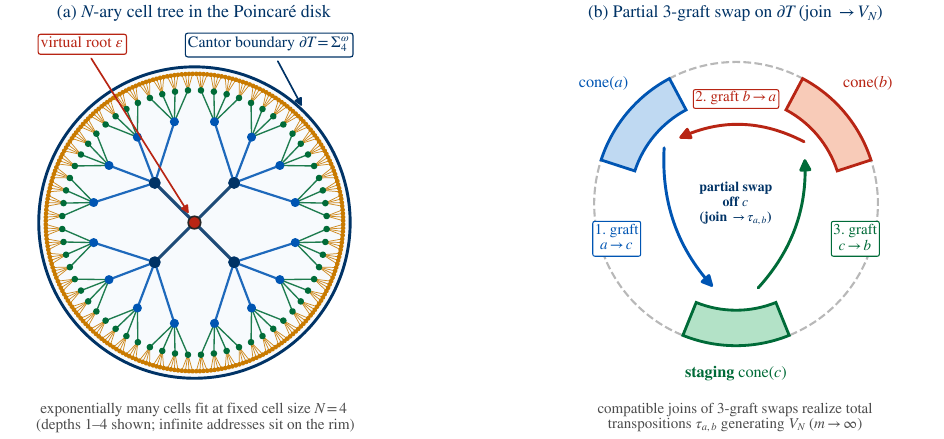}
\caption{Hyperbolic atlas of the address tree and the unit group $V_N$ of its boundary closure. (a)~The $4$-ary cell tree packed into the Poincar\'{e} disk (depths $1$ to $4$), with the rim illustrating the Cantor boundary $\partial T = \Sigma_4^{\omega}$ schematically. (b)~Three grafts through a staging cone $c$ exchange $\operatorname{cone}(a)$ and $\operatorname{cone}(b)$ outside $\operatorname{cone}(c)$; joining two such swaps over disjoint staging cones yields the total cone transpositions $\tau_{a,b}$ that generate $V_N$ (Theorem~\ref{thm:vn-identification}).}
\label{fig:poincare-vn}
\end{figure*}

Two incomparable subtrees $a$ and $b$ exchange places via an occupied, cell-free staging slot $c \in \mu \setminus D$ through the three-graft word $\mathsf{swap}(a,b;c) = \mathsf{graft}(c \to b) \circ \mathsf{graft}(b \to a) \circ \mathsf{graft}(a \to c)$, which interchanges $\operatorname{cone}(a)$ and $\operatorname{cone}(b)$ verbatim outside $\operatorname{cone}(c)$ (Fig.~\ref{fig:swap}, Appendix~\ref{app:calculus}). Because a single swap leaves $\operatorname{cone}(c)$ outside its partial-shadow domain, taking the compatible join of two swaps using disjoint staging slots recovers a total bijection on the infinite address boundary $\partial T = \Sigma_N^{\omega}$ (Lemma~\ref{lem:bridge}). Fig.~\ref{fig:poincare-vn} embeds the $N$-ary cell tree inside a Poincar\'{e} disk whose rim illustrates the Cantor boundary $\partial T$ schematically, showing how three grafts cyclically permute two boundary cones through a staging cone. Formally, rigid calculus words (without $\mathsf{permute}$) act as partial prefix-substitution bijections whose boundary shadows generate Lawson's Cuntz inverse monoid $C_N$~\cite{lawson2007polycyclic,lawson2020revisited} under composition and finite compatible joins (Theorem~\ref{thm:rep-tree}, Appendix~\ref{app:calculus}).

\begin{theorem}[The boundary join closure has unit group $V_N$]
\label{thm:vn-identification}
Fix $N \geq 3$ and take the inductive limit over arbitrary depths $m \to \infty$ (with each finite $\mathcal{M}(N,m)$ realizing depth-$\leq m$ fragments).
\begin{enumerate}
    \item[(i)] The partial boundary shadows of rigid calculus words generate the Cuntz inverse monoid $C_N$ under composition and finite compatible joins.
    \item[(ii)] A bijection of $\partial T$ is a finite compatible join of partial shadows of rigid words if and only if it belongs to the Higman--Thompson group $V_N = G_{N,1}$, the group of units of $C_N$.
    \item[(iii)] Under the two-stage compatible join of Lemma~\ref{lem:bridge}, pairs of derived three-graft swaps realize the total cone transpositions $\tau_{a,b}$, which generate $V_N$.
\end{enumerate}
\end{theorem}

Unlike Belk--Forrest rearrangement groups on glued fractal limit spaces~\cite{belk2019rearrangement}, keeping geometrically coincident addresses logically distinct recovers the \emph{full} group $V_N$ as the unit group of the boundary join closure $C_N$. Even though duplicate marking restricts cell permutations to $\operatorname{Stab}_{S_N}(0) \cong S_{N-1}$, compatible joins of three-graft swaps built solely from grafts generate all of $V_N$ (Appendix~\ref{app:calculus}).

Combining hosting (Theorem~\ref{thm:blockcut}) with the calculus yields a \emph{lifting dichotomy} for elementary edits of a hosted graph $H$ (Fig.~\ref{fig:lifting-gate}, proved in full as Theorem~\ref{thm:lifting} in Appendix~\ref{app:calculus}). Every edit of a tidy realization lifts to a vertex-tracking calculus word in some $F(N',m')$ (taking $N' = N + |V(H)| + 1$ supplies fresh vacant, enabled, cell-free digits in every deployed cell, including the tier-$1$ seed cell, with $\mathsf{enable}$ clearing any disabled source duplicate). Deletions always Stay local at $N' = N$. On tidy, block-separated, slot-clean realizations (as delivered by Theorem~\ref{thm:blockcut}), additions, merges, and block-preserving or clique-fitting splits likewise Stay local at $N' = N$ whenever $|B'| \leq N$, or when $|B'| = N{+}1$ with Theorem~\ref{thm:blockcut}(iii) satisfied at a free terminal anchor or cell-free tier-$1$ anchor (with block-fracturing splits behaving like deletions on tidiness and larger split unions handled by climb or re-embedding), leaving the crossing-chain and two-full sufficiency cases open (Conjecture~\ref{conj:crossing-transport}, Remark~\ref{rem:two-full}), and Climb to $N' > N$ whenever any hosting obstruction fails at $N$.

\begin{figure}[t]
\centering
\includegraphics[width=0.84\linewidth]{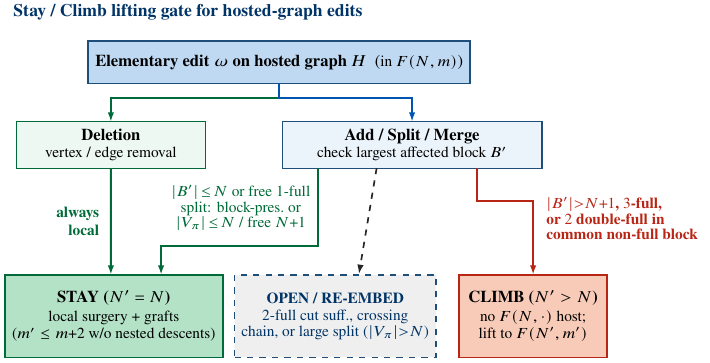}
\caption{Stay/Climb operation-lifting gate for elementary edits of a hosted topology $H$ (Theorem~\ref{thm:lifting}, Appendix~\ref{app:calculus}). Cell size $N$ governs whether an edit stays local or forces a climb to $N' > N$.}
\label{fig:lifting-gate}
\end{figure}

\section{Spacetime Coordinates and Certified Event Registration}
\label{sec:spacetime}
\label{subsec:spacetime}

Without a trusted global clock, distant participants require a shared reference for both location and logical time. Combining spatial addresses (Section~\ref{sec:space}), certified block heights (Section~\ref{sec:time}), registration probabilities (Section~\ref{sec:fault}), and reconfiguration dynamics (Section~\ref{sec:reconfig}) yields a unified spacetime coordinate system.

\subsection{Anchored Time and Global Spacetime Order}
\label{subsec:rotation}
\label{sec:update}

The address string $a$ supplies the spatial location, and the cell's committed block number $h$ supplies discrete logical time. Because every root block digests certified headers from its reporting descendants, the root chain acts as a tamper-evident global metronome synchronizing certified snapshots across all anchored cells.

\begin{definition}[Spacetime coordinate]
\label{def:spacetime-coordinate}
Index each cell's chain by round numbers. A block committed by a cell in round $r$ has \emph{height} $h = r$, while a null resolution marks a gap. The \emph{spacetime coordinate} of node $s$ is $\chi(s) = (a(s), h(s))$, pairing its address with its cell's latest block height, and an event at index $i$ inside cell $a$'s round-$h$ block carries the locator $(a, h, i)$. A block $B$ of cell $a$ is \emph{anchored} if the round-$r$ root block includes $B$ through a chain of certified aggregates, making the root block a certified snapshot of all anchored cells at time $r$. The \emph{worldline} of an entity is the sequence of coordinates it traces across rounds.
\end{definition}

\begin{proposition}[Anchored time]
\label{prop:anchored-time}
Suppose every cell on the path from cell $a$ to the root has at most $f$ Byzantine members in every round.
\begin{enumerate}
    \item[(i)] Cell $a$ commits at most one block per round, and all correct members hold prefixes of one strictly ordered resolution chain.
    \item[(ii)] For any round-$r$ block of cell $a$ included by the root decision, that entry equals cell $a$'s unique certified block for round $r$ and agrees with every correct member of $a$ that decides in round $r$.
    \item[(iii)] Every correct node verifying the root decision of round $r$ and the aggregate bodies below it recovers the exact same snapshot of anchored cells and blocks.
\end{enumerate}
\end{proposition}

Across cells that commit in parallel within one round, a deterministic depth-first traversal of the deployed cell tree provides a canonical tie-breaking order (proofs of Proposition~\ref{prop:anchored-time} and Proposition~\ref{thm:update-main} in Appendix~\ref{app:calculus}).

\begin{proposition}[Cyclic update order]
\label{thm:update-main}
For each cell $\operatorname{cell}(a)$, define the depth-first slot list $\mathcal{D}(a)$ recursively as $a0, \ldots, a(N{-}1)$ followed by $\mathcal{D}(a0), \ldots, \mathcal{D}(a(N{-}1))$, omitting child lists when $|a| = m{-}1$. The wrap-around successor map of $\mathcal{D}(\varepsilon)$ forms a single cycle visiting all $V(N,m)$ node slots, determined purely by the address grammar.
\end{proposition}

Ordering anchored blocks first by height $h$, then by cell position in the digit-ordered depth-first traversal of the deployed cell tree starting at the root cell $\varepsilon$ (which includes $\varepsilon$ itself and tolerates vacant anchors on ragged deployments), and finally by event index $i$ inside a block defines the global \emph{spacetime order}, computable locally from the certified root snapshot, aggregate bodies, and authenticated address-migration history. Four primitives follow without an external directory.
\begin{itemize}
    \item \emph{Locate.} Map a historical coordinate $(a, h)$ through root-anchored migrations after $h$ using stable cell identifiers and route greedily along intact physical paths (Theorem~\ref{thm:distance-II}).
    \item \emph{Order.} Compare heights inside a cell and spacetime order across anchored cells.
    \item \emph{Audit.} Check a block's commit certificate against the authenticated committee keys recorded for that round.
    \item \emph{Reconcile.} Align lagging subtrees against certified root snapshots.
\end{itemize}

\subsection{Coordinate Persistence Under Dynamics}
\label{subsec:coord-persist}

Applying a reconfiguration word $w$ transports surviving addresses without resetting block heights.

\begin{proposition}[Coordinate persistence]
\label{prop:coordinate-persistence}
Let $w$ be a word of reconfiguration operations from Table~\ref{tab:typing} applicable at configuration $x$, and let $s$ be an occupied slot surviving $w$.
\begin{enumerate}
    \item[(i)] The finite address of $s$ transforms by the ordered composite $\widehat{\Phi}(w)$ of prefix substitutions ($\mathsf{graft}$, fixing both finite anchors and moving strict descendants) and cell relabelings ($\mathsf{permute}$), preserving every suffix below a substituted prefix.
    \item[(ii)] The cell identifier, block chain, and commit certificates travel with the cell, so $h(s)$ never resets and historical certificates remain verifiable given each round's authenticated committee keys and configuration record.
    \item[(iii)] On a deployed cell-tree traversal (and its block-expanded node cycle $\mathcal{D}(\varepsilon)$), each single graft splices the cyclic update order with at most three successor pointer updates under stable cell identities.
    \item[(iv)] Consequently, $\chi(w(s)) = (\widehat{\Phi}(w)(a(s)),\, h(s))$. For any rigid word representative $w$, $\widehat{\Phi}(w)$ coincides on surviving finite slots with the finite-slot cocycle $\Phi(w) \in B_N$ (Theorem~\ref{thm:rep-tree}), whereas the induced boundary shadow of $w$ on $\partial T = \Sigma_N^{\omega}$ lies in $C_N$ and any total bijective compatible join completion defines a unit $g \in V_N$ (Theorem~\ref{thm:vn-identification}) agreeing with that partial boundary shadow on its domain (and, as a sufficient condition, matching uniform prefix replacement by $\widehat{\Phi}(w)(s)$ on $\operatorname{cone}(s)$ whenever $\operatorname{cone}(s)$ remains inside a single substitution piece of $w$ and no letter of $w$ fixes an intermediate finite image of $s$ as a graft anchor).
\end{enumerate}
\end{proposition}

\begin{figure}[t]
\centering
\includegraphics[width=0.80\linewidth]{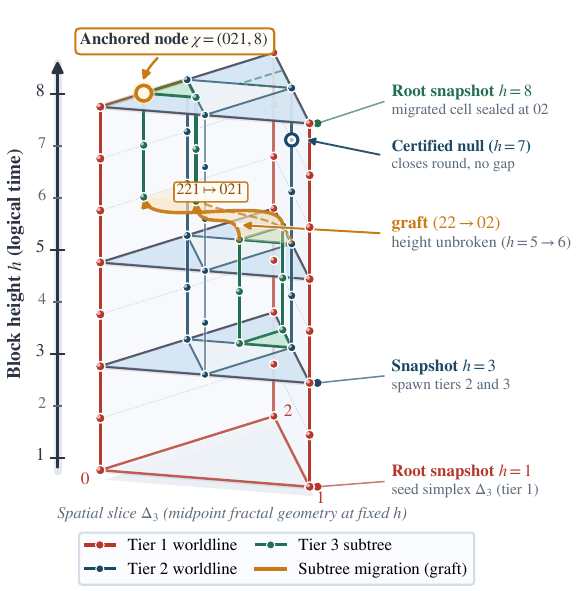}
\caption{Illustrative spacetime prism stacking the midpoint simplex geometry $\Delta_3$ along discrete block height $h$ (logical time) on a partially deployed configuration $D = \{\varepsilon, 0, 1, 2, 22\}$ with occupied anchor $22$ and $02 \in \mu \setminus D$. Hash-linked vertical worldlines carry discrete round ticks (the hollow marker at $h = 7$ shows a certified null resolution on surviving member $22$ in its tier-$2$ committee), while $\mathsf{graft}(22 \to 02)$ relocates the tier-$3$ subtree ($221 \mapsto 021$, $222 \mapsto 022$) between rounds $5$ and $6$ without resetting height or cell history (Proposition~\ref{prop:coordinate-persistence}; events in $\operatorname{cell}(02)$ at height $8$ carry locator $(02, 8, i)$).}
\label{fig:worldlines}
\end{figure}

Together, these mechanisms organize discrete consensus ticks into an auditable discrete spacetime with persistent histories (proof in Appendix~\ref{app:calculus}). Sequential hash-linked resolutions (including certified null resolutions carrying the previous-round hash) preserve gap-free round order inside each cell, the deployed cell-tree traversal provides a total deterministic ordering across anchored cells while splicing in at most three pointer updates per graft, and worldlines relocate across the simplex geometry under structural reconfigurations without resetting cell identifiers or block heights (Fig.~\ref{fig:worldlines}).

\subsection{Light-Client Event Registration Proofs}
\label{subsec:service}

Spacetime coordinates directly enable certified event registration for resource-constrained verifiers. A client submits an event to a leaf cell. Once the root anchors the leaf block, the client receives the event locator $(a, h, i)$ (leaf cell address $a$, block height $h$, event index $i$) and a compact \emph{registration proof}. For each of the $m$ cells on the leaf-to-root path, the standard $Q$-signature proof packs the round's $81$-byte block header, a $9$-byte commit-certificate prefix (view and signer count, $90$ bytes combined), and its $Q$-signature commit certificate. Each interior cell contributes the aggregate body naming its child's digest, and the leaf cell supplies a domain-separated Merkle path for the event inside a leaf body committing to $E$ events~\cite{nakamoto2008bitcoin}.

Given external trust inputs comprising the authenticated round-$r$ committee public keys along the $m$-cell path, the expected round, the cell-path identifiers resolved from $(a, h)$ via root-anchored configuration history, and the trusted round-$r$ root committee keys authenticating the root header, the light verifier accepts certificates carrying at least $Q$ valid Ed25519 signatures (with an application optionally comparing the verified root header digest against an externally pinned checkpoint) and checks SHA-256 hash links without downloading any other branch of the tree (wire anatomy in Appendix~\ref{app:protocol}, Fig.~\ref{fig:reg-proof-packet}). In the concrete wire codec (8-bit signer count $Q \leq 255$, 16-bit aggregate child count, 32-bit event count and index, 64-bit cell identifiers and rounds), with $90$-byte header-plus-prefix records ($81$ B header + $9$ B certificate prefix), $72$-byte signer-signature entries, $40$-byte aggregate entries, and $32$-byte SHA-256 hashes, the exact-$Q$ serialized registration proof length is precisely
\begin{equation}
\label{eq:proof-size}
    8 + 32\lceil \log_2 E \rceil + m(90 + 72Q) + \sum_{k=1}^{m-1} (2 + 40 b_k)
\end{equation}
bytes for $b_k \leq N$ children per aggregate. Proof length grows linearly with depth $m = O(\log_N V)$, while registration probability is governed by Proposition~\ref{prop:registration}. At the $(8,6)$ design point ($V = 299\,592$, Table~\ref{tab:service}), a complete proof remains under $5$~KB.

\section{Protocol Evaluation and Numerical Validation}
\label{sec:eval}
\label{sec:results}

\begin{figure*}[!t]
\centering
\includegraphics[width=0.78\textwidth]{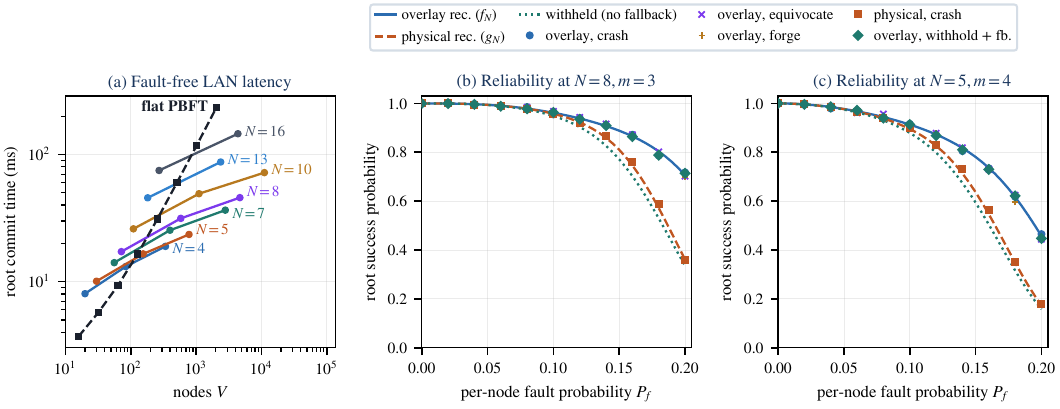}
\caption{Message-level protocol verification. (a)~Causal-relay simulated LAN root commit time versus population $V$ on $G(N,m)$ compared with flat PBFT (five-round means for HSFN, three-round means for flat PBFT through $V = 256$, and single runs at $V \in \{512, 1024, 2048\}$). (b), (c)~Executed root success rate versus $P_f$ at $(8,3)$ and $(5,4)$ (archived $2000$-round E2 campaign) plotted against the overlay recursion $f_N$, the physical recursion $g_N$, and the withheld single-sender recursion without fallback.}
\label{fig:pbft-exec}
\end{figure*}

We evaluate the protocol using a discrete-event protocol simulator executing multi-round hierarchical agreement, Byzantine faults, chain synchronization, and live reconfigurations across full and ragged deployments, paired with a real SHA-256 and Ed25519 registration-proof codec (exhaustive combinatorial checks and formal proof boundaries are cataloged in Appendix~\ref{app:verification}).

\subsection{Message-Level Protocol Verification}
\label{subsec:pbft-exec}

\emph{Simulator and Byzantine strategies.} The simulator executes Section~\ref{subsec:multi-pbft} message by message with an abstract signature registry (checking tuple membership and certificate quorums on every report and decision, while real Ed25519 signing and verification are benchmarked separately in Section~\ref{subsec:reg-eval}), proposal validity checks, and an agreement monitor flagging conflicting commits. Three Byzantine strategies are modeled, namely \emph{withholding} (drops reports and relays), \emph{equivocating} (splits pre-prepares and ignores view changes), and \emph{forging} (discards view-change certificates to propose conflicting aggregates). Faulty members inside over-budget cells ($> f$ faults) stay silent to isolate the certified-safety boundary while correct members still execute local rules.

\emph{Delay regimes.} Three delay models span analytical and LAN settings. The \emph{fixed-delay} model assigns exact hop delay $\delta$ with zero processing cost and non-firing timers. The \emph{ideal} model uses hop delay $1 + \mathrm{Exp}(0.5)$ units, view-change timeout $10$ units, and grace period $T_g = 3$. The \emph{LAN} model uses hop delay $0.5 + \mathrm{Exp}(0.25)$~ms, $100$~Mbit/s per-node serialization, $10$~$\mu$s message processing, and $50$~$\mu$s signature verification ($128$-byte headers, $64$-byte signatures, $1$-KiB payloads, $50$~ms view-change timeout raised to $150$~ms at $N = 16$). The LAN timings reported below represent simulated protocol latencies under these link and cryptographic parameters.

\emph{Exact counts and LAN scaling.} Across $100$ fault-free configurations spanning $25$ $(N,m)$ instances ($20 \leq V \leq 11\,110$, $N \geq 4$), both transports, and both cost models, every round matched Proposition~\ref{prop:exec-count} identically. Across $128$ fixed-delay runs ($N \geq 4$) and seven irregular committee trees (uniform $N_c \geq 4$ with ragged branching), latencies matched Proposition~\ref{prop:latency} and Proposition~\ref{thm:committee-tree} exactly.

Table~\ref{tab:eval-count} and Fig.~\ref{fig:pbft-exec}(a) compare simulated fault-free LAN rounds under the finite grace timer (where interior cells propose immediately once all children report, or upon expiry of the $T_g = 3$~ms grace timer started by the first majority subject to the tier deadline, anchoring for example $73.8\%$ of leaf reports at $(13,2)$ and omitting $115$ child entries across five rounds at $(8,4)$) against flat PBFT under identical signature and serialization parameters (cross-protocol complexities appear in Table~\ref{tab:hierbft}). The two architectures operate under distinct fault models. Flat PBFT runs one global committee tolerating up to $\lfloor(V{-}1)/3\rfloor$ arbitrarily placed Byzantine replicas at $\Theta(V^2)$ message cost, whereas uniform HSFN trades worst-case global adversary concentration for $\Theta(V)$ scaling under per-cell ($\leq f$) or independent faults and produces a root majority aggregate over cells (heterogeneous root enlargement below raises the targeted root corruption barrier at negligible network-wide message overhead). Flat PBFT is dominated by $2V$ signature checks per replica, rising from $3.7$~ms at $V = 16$ to $234$~ms at $V = 2048$ ($8.4 \times 10^6$ messages, $1.6$~GB per round). HSFN latency under causal relay scheduling grows linearly with depth $m$ (roughly $6.7$~ms per tier at $N = 5$ and $14.3$~ms at $N = 8$), crossing flat PBFT near $100$--$200$ nodes. At $V = 4680$ ($N = 8, m = 4$), a round commits in $45.7$~ms using $93\,552$ messages ($62.2$~MB), versus $4.4 \times 10^7$ messages for flat PBFT.

\begin{table*}[t]
\centering
\caption{Simulated Fault-Free Rounds Under the LAN Model on $G(N,m)$ (Causal Relay Scheduling). Message counts match Proposition~\ref{prop:exec-count} identically. Times report mean [minimum, maximum] over five rounds for HSFN and three rounds for flat PBFT ($m = 1$, $N = V$).}
\label{tab:eval-count}
\footnotesize
\setlength{\tabcolsep}{4pt}
\begin{tabular}{rrrrrrr}
\toprule
$N$ & $m$ & $V$ & messages & MB & $t_{\mathrm{root}}$ (ms) & $t_{\mathrm{conf}}$ (ms) \\
\midrule
4 & 4 & 340 & 3{,}384 & 2.6 & 18.9 [18.4, 19.4] & 28.7 [27.8, 29.2] \\
5 & 4 & 780 & 9{,}340 & 6.5 & 23.5 [23.4, 23.7] & 36.4 [36.1, 36.6] \\
8 & 3 & 584 & 11{,}632 & 7.7 & 31.4 [31.1, 31.7] & 51.0 [50.2, 51.6] \\
13 & 3 & 2{,}379 & 80{,}756 & 55.8 & 87.2 [86.5, 87.8] & 139.6 [138.2, 140.9] \\
7 & 4 & 2{,}800 & 50{,}358 & 34.9 & 36.5 [36.4, 36.5] & 58.2 [57.9, 58.5] \\
16 & 3 & 4{,}368 & 183{,}264 & 130.0 & 145.5 [144.9, 146.3] & 228.0 [227.3, 229.0] \\
8 & 4 & 4{,}680 & 93{,}552 & 62.2 & 45.7 [45.5, 46.0] & 74.1 [73.6, 74.9] \\
9 & 4 & 7{,}380 & 162{,}306 & 101.1 & 49.2 [49.1, 49.3] & 83.5 [82.8, 83.8] \\
10 & 4 & 11{,}110 & 288{,}780 & 197.5 & 71.9 [71.2, 72.6] & 115.4 [114.3, 116.4] \\
\midrule
128 & 1 & 128 & 32{,}512 & 6.4 & 16.5 [16.3, 16.7] & 18.8 [18.7, 18.9] \\
256 & 1 & 256 & 130{,}560 & 25.3 & 31.1 [30.9, 31.2] & 35.2 [35.1, 35.3] \\
\bottomrule
\end{tabular}
\end{table*}

\emph{Reliability validation.} In the archived E2 reliability campaign ($2000$ rounds per point across $P_f \in \{0, 0.02, \ldots, 0.20\}$ on $(5,4), (8,3), (6,3), (9,3), (4,5)$), crash runs track $f_N$ and $g_N$ on $3 \nmid N$ networks (while remaining above the certified lower bounds when $3 \mid N$), and Byzantine withholding tracks $f_N$ once fallback reporting is active (Fig.~\ref{fig:pbft-exec}(b), (c)). A separate $10$-point dual-homed sweep ($P_f \in \{0.02, \ldots, 0.20\}$ across $(4,5), (5,4), (8,3)$) confirms $h_N$ across all three topologies. At $(4,5), P_f = 0.08$, the analytical single-sender withholding recursion predicts $0.729$ without fallback, whereas executed leader-only reporting with step-3 fallback achieves $0.957$ (and default $f{+}1$ reporting with fallback achieves $0.962$). At $(4,5), P_f = 0.10$, single-anchor versus dual-homed crash recursions predict $0.572$ versus $0.914$, matching observed empirical proportions of $0.563$ versus $0.916$ respectively.

Bootstrap tests across $30$ points ($20\,000$ rounds each) on strict physical withholding and dual-homed crashes yield mean absolute deviations of $0.00115$ and $0.00167$ ($p = 0.9641$ and $p = 0.6082$, Appendix~\ref{app:verification}). On $3 \mid N$ deployments such as $(6,3)$ at $P_f = 0.20$, executed overlay crash empirical proportion ($0.886$) exceeds the overlay certified-safety bound $1 - f_6^{\,2}(q) = 0.147$ (Theorem~\ref{thm:security}), and executed physical crash empirical proportion ($0.598$) exceeds the physical bound $1 - g_6^{\,2}(q) = 0.050$ (Proposition~\ref{prop:phys-recursion}), because cells still reach quorum $Q$ with $f{+}1$ crashed members.

\emph{Adversarial timing safety and ablation.} Under heavy-tailed Pareto delays (shape $1.5$, capped at $200$ units) and all three Byzantine strategies at $P_f \in \{0.1, 0.2, 0.3\}$ across four networks ($1000$ independent trials per setting, $36\,000$ rounds pooled), zero agreement violations occurred (one-sided $95\%$ pooled upper bound $8.3 \times 10^{-5}$ per round under an independent-trial pooled risk estimand). Removing only the step-1 view-change lock produced $29 / 36\,000$ agreement violations across all networks and strategies. Targeted tests also confirmed rejection of equivocating self-prepares, replayed child certificates, invalid leaf proposals in interior cells, and selective-commit withholding.

\subsection{Anchored Time, Multi-Round Chains, and Migration}
\label{subsec:chains}

\begin{figure}[t]
\centering
\includegraphics[width=0.82\linewidth]{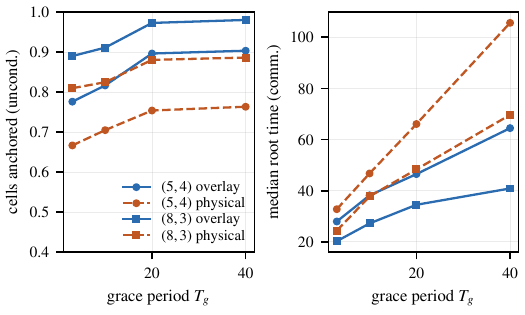}
\caption{Unconditional anchored cell fraction (left) and root-commit-conditioned median root commit time (right) versus grace period $T_g$ under crash faults ($P_f = 0.05$, $1000$ rounds per point, causal relay scheduling).}
\label{fig:anchoring}
\end{figure}

Fig.~\ref{fig:anchoring} measures the anchoring tradeoff of Section~\ref{sec:spacetime}. Fault-free rounds anchor $99.9\%$ of cells on the overlay and $98.5\%$--$99.8\%$ on $G(N,m)$. Under $P_f = 0.05$ crashes at $(5,4)$, lengthening $T_g$ from $3$ to $40$ raises the unconditional overlay anchored fraction from $0.78$ to $0.90$ (root-commit-conditioned median root time $28$ to $64$ ideal delay units), while the unconditional physical-graph fraction rises from $0.67$ and plateaus near $0.76$ (conditioned median $33$ to $106$ units) because cells beneath crashed anchors keep committing locally without reaching the root. Across $120$ fixed-fault observation windows of $50$ rounds each ($6000$ rounds total), detached subtrees kept committing locally with zero disagreements.

A chain-level simulator carries persistent cell chains, state transfer, in-order round closing, and Byzantine forgery injection across rounds. Over $960$ executions of $30$ rounds under heavy-tailed delays, Byzantine faults ($P_f \in \{0.1, 0.2\}$), and per-round offline probability ($0.05, 0.1$), state transfer moved $346\,941$ verified entries and rejected every forged or replayed block in in-budget cells (removing certificate checks across $240$ separate ablation runs admitted $252\,620$ forged entries). Dropping in-order round resolution caused $14$ member chains in $2$ of $540$ reconfiguration runs to skip committed blocks, whereas in-order closing preserved identical histories across all correct members. A reproducible causal-relay check binding previous-round logical digests into every certified null resolution ($30$ heavy-tailed runs across $(4,3), (5,3), (8,2)$, overlay and physical transports, five seeds $9001$--$9030$, $25$ rounds each, $P_f = 0.1$, offline $0.1$, totaling $750$ rounds and $2819$ null resolutions across the deployment) likewise recorded zero broken predecessor links, zero mismatches, and zero chain inconsistencies across all monitored in-budget cells.

Executing live reconfigurations (subtree migration at round $5$, leave at round $10$, join at round $15$) across $(4,4)$, $(5,3)$, and $(8,3)$ verified exact Proposition~\ref{thm:committee-tree} message counts before and after migration in all $15$ fault-free runs (which scheduled the graft alone). Across $540$ faulty runs ($25$ rounds each, $P_f = 0.1$, offline $0.05$--$0.1$), members fetched $931$ missed root decisions and rejected $2765$ forged ones. Of the $540$ faulty runs, $419$ activated the scheduled graft (with $189$ collecting committee approvals across two distinct rounds before activation). Following completed migrations, every post-migration coordinate lookup traced historical addresses through root-anchored migrations to the relocated cell ($2823/2823$ faulty, $105/105$ fault-free), and every historical certificate verified against its round's committee ($858$ times from newly joined nodes), with zero disagreements.

\subsection{Registration Budgets and Operations Calculus Simulation}
\label{subsec:reg-eval}

A SHA-256 and Ed25519 implementation of the registration proof in Section~\ref{subsec:service} matched~\eqref{eq:proof-size} identically across all rows of Table~\ref{tab:service} and rejected all nine forged proof variants. Across $27$ crash and site-outage points ($P_f \in \{0.02, 0.05, 0.10\}$, $p_s \in \{0, 0.01, 0.05\}$, $4000$ rounds per point on $(5,4), (8,3), (4,5)$), analytical $A$ fell inside the $95\%$ empirical interval at $26/27$ overlay points and $A^{\mathrm{ph}}$ at $25/27$ physical-graph points. Near threshold at $(4,5), P_f = 0.10, p_s = 0.05$, the majority factors lower path survival from $0.592$ to $A = 0.410$ against $0.409$ executed.

Table~\ref{tab:service} summarizes deployment budgets at $P_f = 0.01, p_s = 0.001$ up to $V = 299\,592$ nodes, pairing simulated or tier-fitted LAN commit times with analytical registration probabilities $A$ and $A^{\mathrm{ph}}$. Per-node messages and baseline signature checks stay asymptotically flat in population for fixed $N$, proof sizes stay under $5$~KB, and single-anchor exposure reduces registration probability by $0.020$--$0.049$, a drop fully restored by dual homing at $N \in \{4,5\}$. Finally, an event-driven simulation of $6 \times 10^4$ proposed operations ($52\,081$ executed after guards, including $3885$ grafts) across six family instances verified Lemma~\ref{lem:edge-cut} and the three-pointer cycle-repair bound on every graft. Incremental affine indicators and anchor-path reliability updates matched exact recomputation identically across all steps, yielding $2.7\times$--$33.5\times$ median speedups for reliability and $28.2\times$--$416\times$ for histograms (Appendix~\ref{app:verification}).

\begin{table}[t]
\centering
\caption{Certified Event Registration Budgets at $P_f = 0.01, p_s = 0.001$ ($E = 64$ Events per Leaf Block). Root times report five-round causal-relay LAN simulated means on $G(N,m)$, except linear tier-fitted rows marked $^{\dagger}$.}
\label{tab:service}
\scriptsize
\setlength{\tabcolsep}{1.6pt}
\begin{tabular}{@{}rrrrrrrrr@{}}
\toprule
$(N,m)$ & $V$ & msgs/n & sigs/n$^{\ast}$ & root (ms) & $A$ & $A^{\mathrm{ph}}$ & proof (B) & diam. \\
\midrule
$(5,4)$ & $780$ & $12.0$ & $23.9$ & $23.5$ & $0.9921$ & $0.9638$ & $2318$ & $7$ \\
$(8,3)$ & $584$ & $19.9$ & $49.5$ & $31.4$ & $0.9968$ & $0.9770$ & $2410$ & $5$ \\
$(8,4)$ & $4\,680$ & $20.0$ & $49.9$ & $45.7$ & $0.9958$ & $0.9663$ & $3254$ & $7$ \\
$(7,5)$ & $19\,607$ & $18.0$ & $42.0$ & $47.7^{\dagger}$ & $0.9948$ & $0.9557$ & $3578$ & $9$ \\
$(8,5)$ & $37\,448$ & $20.0$ & $50.0$ & $60.0^{\dagger}$ & $0.9947$ & $0.9556$ & $4098$ & $9$ \\
$(10,4)$ & $11\,110$ & $26.0$ & $73.9$ & $71.9$ & $0.9960$ & $0.9664$ & $3782$ & $7$ \\
$(8,6)$ & $299\,592$ & $20.0$ & $50.0$ & $74.3^{\dagger}$ & $0.9937$ & $0.9451$ & $4942$ & $11$ \\
\bottomrule
\multicolumn{9}{@{}p{0.96\columnwidth}@{}}{\rule{0pt}{2.2ex}$^{\ast}$Baseline count ($1$ check per intra-cell message plus $Q$ per cross-tier message; simulator timing adds $1$ sender check and bundled child checks, e.g., $57.6$ per node at $(8,3)$). $^{\dagger}$Linear per-tier fit to the causal E1 physical LAN series ($m \in \{2,3,4\}$); all other entries are closed-form or directly executed.}
\end{tabular}
\end{table}

\subsection{Analytical Comparison Against External Baselines}
\label{subsec:analytical-baselines}

Because our discrete-event simulator benchmarks HSFN directly only against flat PBFT, Table~\ref{tab:analytical-baselines} places HSFN alongside external consensus, structured-overlay, and recursive-interconnect baselines on exact analytical scaling laws and concrete values at $V = 4\,680$ ($G(8,4)$), complementing the extended reference tables in Appendices~\ref{app:space} and~\ref{app:protocol} (Tables~\ref{tab:overlays}, \ref{tab:interconnect}, and~\ref{tab:hierbft}).

\begin{table*}[t]
\centering
\caption{Analytical Scaling Comparison Against External Baselines (Population $V$, Evaluated at $V = 4\,680$, i.e., $G(8,4)$ with $N = 8, m = 4, f = 2$).}
\label{tab:analytical-baselines}
\scriptsize
\setlength{\tabcolsep}{3pt}
\renewcommand{\arraystretch}{1.12}
\begin{tabular}{@{}>{\raggedright\arraybackslash}p{0.15\textwidth}>{\raggedright\arraybackslash}p{0.22\textwidth}>{\raggedright\arraybackslash}p{0.17\textwidth}>{\raggedright\arraybackslash}p{0.17\textwidth}>{\raggedright\arraybackslash}p{0.24\textwidth}@{}}
\toprule
\multicolumn{5}{@{}l}{\textbf{Panel A. Flat, Pipelined, Layered, and Sharded BFT Protocols}} \\
\midrule
Protocol & Msgs / decision ($V = 4\,680$) & Max recv / node ($V = 4\,680$) & Ideal latency & Fault model and routing state \\
\midrule
Flat PBFT~\cite{castro1999practical} & $2V(V{-}1)$ ($4.38 \times 10^{7}$) & $2(V{-}1)$ ($9\,358$) & $O(1)$ ($3$ phases) & $f < V/3$ global, depth $1$; all-to-all $O(V)$ \\
HotStuff~\cite{yin2019hotstuff} & $2(V{-}1)$ steady ($9\,358$) & $V{-}1$ leader ($4\,679$) & $O(1)$ chained & $f < V/3$ global, depth $1$; star $O(V)$ \\
Kauri~\cite{neiheiser2021kauri} & $2(V{-}1)$ steady ($9\,358$) & $b{+}1$ ($9$ at $b = 8$) & $\Theta(\log_b V)$ ($4$ hops) & $f < V/3$ global, ext.\ tree ($b{+}1$) \\
Multi-layer PBFT~\cite{li2020scalable} & $(k{+}1)^2 + k(c{+}1)^2$ ($\geq 1.58 \times 10^{5}$)$^{\dagger}$ & $O(c + V/c)$ & $O(1)$ ($2$ layers) & $f_c < c/3$ per group, ext.\ mesh \\
OmniLedger / RapidChain~\cite{kokoris2018omniledger,zamani2018rapidchain} & $O(c^{2})$/shard, $c = \Theta(\log V)$ & $O(c)$ per shard & $O(\log V)$ cross-shard & Per-shard w.h.p.\ (epoch reshuffle), ext.\ DHT \\
\textbf{HSFN $G(N,m)$ (this work)} & $\mathbf{2(N{-}1)V + 2(f{+}1)(V{-}N)}$ ($\mathbf{93\,552}$) & $\mathbf{2(N{-}1)+(N{+}1)(f{+}1)}$ ($\mathbf{41}$) & $\mathbf{(4m{-}1)\delta}$ / $\mathbf{(5m{-}2)\delta}$ ($\mathbf{15}$/$\mathbf{18}$) & $f < N/3$ per cell + exact $q_0, f_N, g_N, h_N$; $\leq 2N$ \\
\midrule
\multicolumn{5}{@{}l}{\textbf{Panel B. Structured Overlays and Recursive Interconnect Topologies}} \\
\midrule
System / topology & Per-node routing state / degree & Path length / diameter & Vertex conn.\ $\kappa$ & Built-in agreement and churn model \\
\midrule
Chord / Pastry / Kademlia~\cite{stoica2003chord,rowstron2001pastry,maymounkov2002kademlia} & $O(\log V)$ soft-state ($\lceil \log_2 V \rceil = 13$) & $O(\log V)$ w.h.p.\ (non-exact) & Overlay-dependent & None; periodic stabilization / refresh \\
DCell$_k$ / BCube$_k$~\cite{guo2008dcell,guo2009bcube} & $k{+}1$ links (positional) & $\leq 2^{k+1}{-}1$ / $k{+}1$ & $\leq k{+}1$ / $k{+}1$ & None; data-center fabrics \\
\textbf{HSFN $G(N,m)$ (this work)} & $\mathbf{\leq 2N}$ ($\mathbf{16}$ at $N = 8$, $O(1)$ in $V$) & \textbf{Exact diam.}\ $\mathbf{2m{-}1}$ ($\mathbf{7}$); asymp.\ leaf mean $2m{-}1{-}\tfrac{2}{N-1}$$^{\ddagger}$ & $\mathbf{1}$ ($N$ on $K_{N,N}$) & \textbf{Table-free shortest path + BFT + proofs} \\
\bottomrule
\multicolumn{5}{@{}p{0.97\textwidth}@{}}{\rule{0pt}{2.0ex}$^{\dagger}$Two-layer PBFT~\cite{li2020scalable} ($k$ groups of $c$) scales at best as $\Theta(V^{4/3})$; at $V = 4\,680$, exact $kc = 4\,680$ minimizes at $(c,k)=(20,234)$ ($158\,419$ msgs), and $kc \geq 4\,680$ at $(21,223)$ ($158\,108$ msgs, vs.\ $390\,781$ at $c = 8$). $^{\ddagger}$Tier-$m$ leaf-pair mean holds up to $+O(m N^{-m})$ (App.~\ref{app:space}). HotStuff/Kauri use pipelined steady state; HSFN lists overlay / physical ideal hops.}
\end{tabular}
\end{table*}

Two analytical tradeoffs separate HSFN from linear-authenticator global BFT and sharded ledgers. Pipelined HotStuff and Kauri achieve fewer total messages per decision ($2(V{-}1)$ versus HSFN's $2(N{-}1)V + 2(f{+}1)(V{-}N)$) because they aggregate threshold votes for one global leader rather than running full PBFT inside every committee, while preserving a global $\lfloor(V{-}1)/3\rfloor$ fault bound. HotStuff incurs a linear $V{-}1$ leader bottleneck ($4\,679$ incoming messages at $V = 4\,680$), whereas Kauri removes that bottleneck through an externally configured fanout-$b$ tree ($b{+}1$ per node). Two-layer PBFT reaches at best $\Theta(V^{4/3})$ messages because the top committee grows with group count, while sharded protocols keep per-shard cost $O(c^2)$ only by scaling shard size as $c = \Theta(\log V)$ under periodic epoch reshuffling and separate cross-shard overlays. HSFN instead keeps peak per-node reception at $2(N{-}1)+(N{+}1)(f{+}1)$ (constant in $V$ for fixed $N$, growing as $\Theta(N^2)$ in $N$) and average per-node load at $O(N)$, keeps cell-local committees autonomous during partitions, and derives committees, fallback senders, and shortest physical routes from one address tree.

Against overlays and data-center interconnects (Panel~B), Chord, Pastry, and Kademlia store $O(\log V)$ soft-state routing entries maintained by periodic stabilization and give probabilistic $O(\log V)$ path lengths rather than exact shortest paths. DCell$_k$ and BCube$_k$ achieve higher physical vertex connectivity ($\kappa = k{+}1$) than single-anchor $G(N,m)$ ($\kappa = 1$) without providing Byzantine agreement or light-client proofs. HSFN trades physical cut-vertex exposure ($\kappa = 1$) for a constant degree cap $2N$, table-free exact shortest-path routing, explicit analytical pricing of cut-vertex failure ($g_N$), and dual-homed recovery ($h_N$).

\section{Discussion and Open Problems}
\label{sec:discussion}

\emph{Parameter tradeoffs and hardening.} Averaged per node for fixed $N$, the message load is $2(N{-}1) + 2(f{+}1)(1 - N/V) < 2(N{+}f)$ messages per round, while ideal delay scales linearly with $m = \log_N V + O(1)$ ($N \geq 4$). Under the modeled LAN serialization and verification costs, intra-cell delay grows with $N$, making $N \in [7,10]$ a balanced operating range in Tables~\ref{tab:eval-count} and~\ref{tab:service}. On the physical graph $G(N,m)$, single-anchor cut vertices reduce the collapse threshold by $23\%$--$28\%$ across $N \in \{5,8,16\}$ (and nearly half at $N = 4$), which dual homing restores completely at $N \in \{4,5\}$ and nearly in full at larger $N$. Under the heterogeneous committee-tree accounting of Proposition~\ref{thm:committee-tree} and the overlay recursion of Theorem~\ref{thm:security}, enlarging only the root committee to $N_{\mathrm{root}} > N$ while keeping its $N$ child branches and all deeper cells unchanged sharply strengthens root resilience at negligible network-wide message cost. At $(8,6)$ ($V = 299\,592$, $5.99 \times 10^6$ messages/round), raising $N_{\mathrm{root}}$ from $8$ to $16$, $31$, or $64$ increases the targeted root corruption threshold $f_{\mathrm{root}}{+}1$ from $3$ to $6$, $11$, or $22$ nodes and lowers root internal failure $q_{\mathrm{root}}(0.05)$ from $5.79 \times 10^{-3}$ to $8.09 \times 10^{-5}$, $1.62 \times 10^{-7}$, or $2.46 \times 10^{-13}$, while increasing total network messages by only $0.013\%$, $0.047\%$, or $0.18\%$ (and by $0.80\%$ or $3.01\%$ for $N_{\mathrm{root}} = 16$ or $31$ at $(8,4)$). Analyzing adaptive adversary migration or periodic root membership rotation requires separate adversary and reconfiguration models.

\emph{Role of the three mathematical layers.} Each mathematical layer answers a distinct networking question without claiming protocol equivalence where none exists. Simplex geometry provides the spatial carrier motivating fixed peer-group size $N$ and lossless address expansion (Lemma~\ref{lem:family-mono}). Block--cut hosting determines which application topologies fit inside a fixed-$N$ substrate strictly beyond treewidth (Theorem~\ref{thm:blockcut}). The boundary join closure $C_N$ and its unit group $V_N$ supply the asymptotic symmetry invariant of invertible clopen branch rearrangements on the Cantor address boundary $\partial T$ rather than a finite-round runtime primitive (Theorem~\ref{thm:vn-identification}). Multi-host wide-area deployment, cross-tier pipelining, and WAN benchmarking against GeoBFT, HotStuff, and Kauri form natural companion systems work.

\emph{Open mathematical problems.} Four open problems merit further study.
\begin{enumerate}
    \item Whether the global orientation condition at the two-full hosting boundary (Remark~\ref{rem:two-full}) is sufficient as well as necessary.
    \item Closing the crossing-chain transport case (Conjecture~\ref{conj:crossing-transport}) and designing a tidy-preserving deletion policy in Theorem~\ref{thm:lifting}.
    \item Constructing a confluent rewriting system for configuration-independent word equivalence in $\mathcal{M}(N,m)$.
    \item Defining explicit cross-subtree bridge rules on tier-$m$ leaves (with or without identifying geometrically coincident midpoints) to close the fractal frontier, and characterizing the resulting routing stretch, cut-vertex mitigation, and boundary rearrangement subgroup.
\end{enumerate}

\section{Conclusion}
\label{sec:conclusion}

The hyper-simplex fractal network shows that naming, routing, consensus, reconfiguration, and logical time need not live in separate protocol layers. Motivated by the Euclidean dimension bound of equal peers, recursive simplex spawning organizes a sparse $N$-ary hierarchy indexed by finite digit words. From this address grammar alone follow bounded-degree adjacency and exact shortest-path greedy routing, while pairing it with half-scale midpoint simplices yields a self-similar attractor of Hausdorff dimension $\log_2 N$.

Executing certified hierarchical PBFT on the same prefix tree aligns committee structure with physical connectivity. For fixed committee size $N \geq 4$, average per-node load stays independent of network size, ideal latency grows logarithmically, and under independent crash faults, sufficient grace waiting, and timely report inclusion, majority aggregation yields an explicit depth-uniform overlay root threshold alongside recursions (exact for $3 \nmid N$, conservative certified bounds when $3 \mid N$) separating physical single-anchor exposure, dual-homing recovery, and designated leaf-path registration. Over time, eight reversible structural edits form an inverse monoid whose rigid word shadows generate a boundary join closure with the Higman--Thompson group $V_N$ as its group of units. Biconnected block order simultaneously governs service-topology hosting and local edit lifting. Pairing addresses with root-anchored block heights yields persistent spacetime coordinates that survive migration and admit compact Ed25519 light-client proofs, validated by discrete-event protocol simulation and serialized proof verification.

\bibliographystyle{IEEEtran}
\bibliography{reference}

\clearpage
\appendices
\section{Addressing, Routing, and Interconnect Proofs}
\label{app:space}

This appendix supplies the proofs of Sections~\ref{sec:model} and~\ref{sec:space} and situates $G(N,m)$ among recursively defined interconnects.

\begin{corollary}[Per-tier and asymptotic growth]
\label{cor:count-II}
The tier-$t$ node count $N^{t}$ and cell count $N^{t-1}$ grow geometrically with ratio $N$. Moreover, $V(N,m) = \Theta(N^{m})$, with $V(N,m) = \frac{N}{N-1} N^{m} - \frac{N}{N-1}$, so $V(N,m) / N^{m} \to N/(N-1)$ as $m \to \infty$.
\end{corollary}

\begin{IEEEproof}
Immediate from~\eqref{eq:unified-V}, since $\frac{N(N^{m}-1)}{N-1} = \frac{N}{N-1}N^{m} - \frac{N}{N-1}$.
\end{IEEEproof}

\subsection{Digit-Word Addresses and Automorphisms}

Every node below tier $m$ spawns one full cell of $N$ nodes without exclusions, so addresses are unrestricted nonempty digit words over $\Sigma_N = \{0,\ldots,N{-}1\}$. Positional digit naming parallels recursive interconnects such as DCell and BCube~\cite{guo2008dcell,guo2009bcube}, while variable word length encodes tier directly.

\begin{proposition}[Address bijection]
\label{prop:addr-bij}
For every $t \in \{1, \ldots, m\}$, Definition~\ref{def:address-II} establishes a bijection between the tier-$t$ nodes of $G(N,m)$ and $\Sigma_N^{t}$. Consequently, tier $t$ holds $N^{t}$ nodes and $V(N,m) = \sum_{t=1}^{m} N^{t} = N(N^{m}-1)/(N-1)$.
\end{proposition}

\emph{Classical counterpart.} Positional digit tuples index servers in BCube and DCell~\cite{guo2009bcube,guo2008dcell}. Counting nodes as words carries over, whereas variable word length lets one alphabet index all tiers simultaneously.

Under this bijection, the \emph{edge inventory} of $G(N,m)$ partitions into $E = E_{\mathrm{sib}} \cup E_{\mathrm{up}}$, where
\begin{equation*}
\begin{aligned}
E_{\mathrm{sib}} &= \bigl\{\, \{wd,\, wd'\} \;:\; 0 \leq |w| \leq m{-}1,\; d \neq d' \,\bigr\}, \\
E_{\mathrm{up}}  &= \bigl\{\, \{w,\, wd\}  \;:\; 1 \leq |w| \leq m{-}1,\; d \in \Sigma_N \,\bigr\}.
\end{aligned}
\end{equation*}
Geometric coincidence under the midpoint realization $\iota$ never creates edges (Remark~\ref{rem:coincident}).

\begin{lemma}[Duplicate-digit convention]
\label{lem:dup-digit}
Let $\iota$ be the midpoint realization of Definition~\ref{def:embedding-maps} and let $M = \{\,a0 : 1 \leq |a| \leq m{-}1\,\}$ be the set of marked duplicates.
\begin{enumerate}
    \item[(i)] For every $a$ with $1 \leq |a| \leq m{-}1$, exactly one member of $\operatorname{cell}(a)$ is co-located with its anchor, namely $a0$ ($\iota(a0) = \iota(a)$, while $\iota(ad) \neq \iota(a)$ for $d \neq 0$). The tier-$1$ cell carries no duplicate, so $M$ is well defined.
    \item[(ii)] Every family $(\tau_w)$ of cell-wise digit permutations ($\tau_\varepsilon \in S_N$ and $\tau_w \in S_N$ for $1 \leq |w| < m$) induces an automorphism $\varphi$ of $G(N,m)$ via
    \begin{equation*}
        \varphi(a_1 a_2 \cdots a_t) = b_1 b_2 \cdots b_t,
        \qquad
        b_s = \tau_{\operatorname{pre}_{s-1}(a)}(a_s),
    \end{equation*}
    and $\varphi(M) = M$ if and only if $\tau_w \in \operatorname{Stab}_{S_N}(0) \cong S_{N-1}$ for all $w \neq \varepsilon$, with $\tau_\varepsilon \in S_N$ unconstrained.
\end{enumerate}
In particular, $M$ is imposed by the address convention rather than being an abstract graph invariant of $G(N,m)$.
\end{lemma}

\emph{Classical counterpart.} Part~(ii) is the rooted $N$-ary tree automorphism group, an iterated permutational wreath product of copies of $S_N$~\cite{nekrashevych2005self}. Marking cuts every non-root factor to $\operatorname{Stab}_{S_N}(0) \cong S_{N-1}$.

\emph{Running example ($N = 5$, $m = 3$).} The graph $G(5,3)$ has $5 + 25 + 125 = 155$ nodes. Nodes $u = 420$ and $v = 431$ have longest common prefix $4$ ($\ell = 1$), giving $d(u,v) = 3 + 3 - 2 - 1 = 3$ along $420 \to 42 \to 43 \to 431$. The sibling hop $42 \to 43$ saves one hop over the pure tree walk through anchor $4$. Nodes $000$ and $444$ achieve the diameter $5$. With $\lceil \log_2 5 \rceil = 3$ and $\lceil \log_2 3 \rceil = 2$, locator $\langle 5, 3, 431 \rangle$ occupies $14$ bits, and the asymptotic ratio of Proposition~\ref{prop:locator-II} is $3/\log_2 5 \approx 1.29$.

\subsection{Comparison with Structured Overlays and Recursive Interconnects}
\label{sec:graphs}

\begin{figure*}[!t]
\centering
\includegraphics[width=\textwidth]{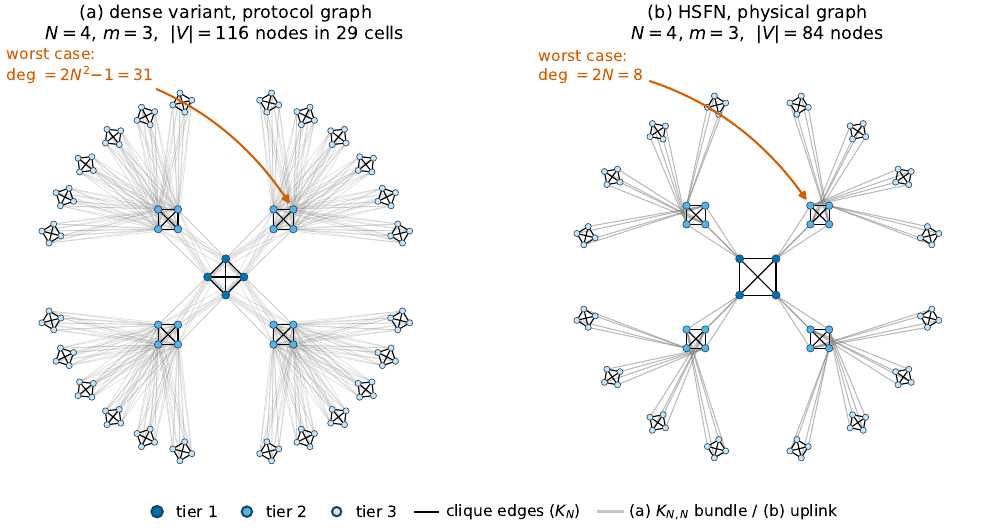}
\caption{Two ways of linking parent and child cells at $N = 4$, $m = 3$. (a)~Anchor-redundant complete-bipartite bundles $K_{4,4}$ in gray (as in the dense reference overlay of~\cite{yang2024hsfnpart1}), yielding vertex connectivity $N$ at $\Theta(N^2)$ degree. (b)~The sparse physical graph $G(4,3)$ with sibling cliques in black and single-anchor uplinks in gray, capping degree at $2N$ with every anchor a cut vertex.}
\label{fig:net-compare}
\end{figure*}

Table~\ref{tab:overlays} contrasts HSFN's grammar-driven addressing and shortest-path routing with classical structured overlays, while Table~\ref{tab:interconnect} and Fig.~\ref{fig:net-compare} compare $G(N,m)$ and its complete-bipartite overlay reference $G_{\mathrm{dense}}$ against DCell~\cite{guo2008dcell}, BCube~\cite{guo2009bcube}, the fat-tree~\cite{alfares2008fattree}, and the hypercube $Q_d$. On a depth-$m$ cell tree with parent--child $K_{N,N}$ bundles, each member of a root cell has degree $(N{-}1) + b_{\mathrm{root}} N$, an interior member has $(N{-}1) + (1 + b_{\mathrm{int}})N$, and a leaf member has $2N{-}1$. On the $N$-ary HSFN tree this overlay has maximum degree $N^2 + 2N - 1$. On the dense reference construction ($b_{\mathrm{int}} = 2N{-}2$)~\cite{yang2024hsfnpart1} it reaches $2N^2 - 1$. Both achieve vertex connectivity $N$ at $\Theta(N^2)$ degree, whereas $G(N,m)$ holds physical degree at most $2N$ at vertex connectivity $1$.

\begin{table*}[!t]
\centering
\caption{Addressing and Routing Compared with Classical Structured Overlays (Population $n$).}
\label{tab:overlays}
\footnotesize
\setlength{\tabcolsep}{3pt}
\renewcommand{\arraystretch}{1.15}
\begin{tabular}{@{}>{\raggedright}p{0.10\textwidth}>{\raggedright}p{0.17\textwidth}>{\raggedright}p{0.19\textwidth}>{\raggedright}p{0.19\textwidth}>{\raggedright}p{0.08\textwidth}>{\raggedright\arraybackslash}p{0.21\textwidth}@{}}
\toprule
Scheme & Identifier space & Per-node routing state & Routing hops & Hops from two IDs & Maintenance on churn \\
\midrule
Chord~\cite{stoica2003chord} & $m$-bit ring & finger table, $O(\log n)$ entries & $O(\log n)$ w.h.p. & no & periodic stabilization \\
Pastry~\cite{rowstron2001pastry} & $128$-bit, base $2^{b}$ & $\lceil \log_{2^{b}} n \rceil$ rows, leaf/neighborhood sets & $< \lceil \log_{2^{b}} n \rceil$ normal & no & $O(\log_{2^{b}} n)$ join messages \\
Kademlia~\cite{maymounkov2002kademlia} & $160$-bit, XOR metric & one $k$-bucket per bit & $\log n + c$ typical & no & idle bucket refresh \\
CAN~\cite{ratnasamy2001can} & $d$-dimensional torus & $2d$ neighbors & $(d/4)n^{1/d}$ average & no & zone split, periodic updates \\
HSFN & words of length $\leq m$ over $\Sigma_N$ & at most $2N$ links (Proposition~\ref{thm:deg-II}) & exact $d(u,v) \leq 2m{-}1$ (Theorem~\ref{thm:distance-II}) & yes & cell-atomic $\mathsf{spawn}$/$\mathsf{join}$ (Section~\ref{sec:reconfig}) \\
\bottomrule
\end{tabular}
\end{table*}

\begin{table}[t]
\centering
\caption{HSFN Among Recursive Interconnect Topologies. Degree and connectivity are per host.}
\label{tab:interconnect}
\scriptsize
\setlength{\tabcolsep}{1.6pt}
\begin{tabular}{lccccc}
\toprule
Topology & Deg. & Diam. & Mean & $\kappa$ & Nodes \\
\midrule
HSFN $G(N,m)$ & $\leq 2N$ & $2m{-}1$$^{\mathrm{a}}$ & $2m{-}1{-}\tfrac{2}{N-1}$$^{\mathrm{b}}$ & $1$ & $\Theta(N^{m})$ \\
Overlay $K_{N,N}$$^{\ast}$ & $\Theta(N^{2})$ & $\leq 2m$ & $\leq 2m$ & $N$ & $\Theta(N^{m})$ \\
DCell$_k$~\cite{guo2008dcell} & $k{+}1$ & $\leq 2^{k+1}{-}1$ & -- & $\leq k{+}1$ & $\gtrsim (n{+}\tfrac{1}{2})^{2^{k}}$ \\
BCube$_k$~\cite{guo2009bcube} & $k{+}1$ & $k{+}1$ & $(k{+}1)\tfrac{n-1}{n}$ & $k{+}1$ & $n^{k+1}$ \\
Fat-tree~\cite{alfares2008fattree} & $1$ & $6$ & $\leq 6$ & $1$ & $k^{3}/4$ \\
Hypercube $Q_d$ & $d$ & $d$ & $d/2$ & $d$ & $2^{d}$ \\
\bottomrule
\multicolumn{6}{@{}l@{}}{\rule{0pt}{2ex}$^{\ast}$Logical protocol overlay with $K_{N,N}$ parent--child bundles.} \\
\multicolumn{6}{@{}l@{}}{$^{\mathrm{a}}$Exact (Theorem~\ref{thm:distance-II}). \ $^{\mathrm{b}}$Tier-$m$ pairs up to $O(mN^{-m})$.} \\
\end{tabular}
\end{table}

\emph{Mean path length.} Draw $u, v$ independently and uniformly from tier $m$ (which holds an $(N{-}1)/N$ asymptotic share of $V(N,m)$). Their common prefix length $\ell$ satisfies $\Pr[\ell \geq j] = N^{-j}$ for $0 \leq j \leq m$, with $u = v$ iff $\ell = m$. Since distinct equal-tier nodes are never prefix-comparable, Theorem~\ref{thm:distance-II} gives $d(u,v) = 2m - 2\ell - 1$ whenever $u \neq v$. Conditioning on $\ell < m$ yields
\begin{equation*}
\begin{split}
    \mathbb{E}[\ell \mid u \neq v]
    &= \frac{\sum_{j=1}^{m} N^{-j} - m N^{-m}}{1 - N^{-m}} \\
    &= \frac{1}{N-1} - O\bigl(m N^{-m}\bigr),
\end{split}
\end{equation*}
so $\mathbb{E}[d(u,v)] = 2m - 1 - \tfrac{2}{N-1} + O(m N^{-m})$, with geometric concentration $\Pr[d(u,v) \leq 2m{-}1{-}2j] \leq N^{-j}$. Without lateral shortcuts across subtrees, the typical path sits within $2/(N{-}1) \leq 1$ of the diameter.

By contrast, Sierpi\'{n}ski graphs $S(m,N)$~\cite{klavzar1997sierpinski,hinz2017survey} keep only the $N^m$ deepest-stage words and connect sub-copies laterally via bridge edges, avoiding cut vertices. Retaining all $\sum_{t=1}^m N^t$ tiers as live vertical anchors produces both the three-class degree profile of Proposition~\ref{thm:deg-II} and the cut vertices of Proposition~\ref{prop:conn-price}.

\subsection{Proofs}
\label{app:space-proofs}

\begin{IEEEproof}[Proof of Proposition~\ref{prop:addr-bij}]
We induct on $t$. At $t = 1$, $\operatorname{cell}(\varepsilon) = \Sigma_N$ gives a bijection. Assuming the bijection at tier $t < m$, every tier-$t$ node spawns one disjoint cell of $N$ tier-$(t{+}1)$ nodes labeled by its one-digit extensions. Words of distinct cells already differ in their length-$t$ prefixes, and every word in $\Sigma_N^{t+1}$ extends a unique length-$t$ address. Summing gives $V(N,m)$.
\end{IEEEproof}

\begin{IEEEproof}[Proof of Lemma~\ref{lem:dup-digit}]
(i)~For $1 \leq |a| \leq m{-}1$, Definition~\ref{def:embedding-maps} places $\iota(a0) = \iota(a)$ and $\iota(ad)$ at the midpoint of a nondegenerate simplex edge from $\iota(a)$ to a distinct sibling for $d \neq 0$, so $\iota(ad) \neq \iota(a)$. Tier-$1$ vertices are affinely independent and unspawned, hence unmarked.

(ii)~The map $\varphi$ is bijective with inverse defined by $\tau'_{\varphi(w)} = \tau_w^{-1}$, preserves word lengths, and satisfies $\operatorname{pre}_s(\varphi(a)) = \varphi(\operatorname{pre}_s(a))$. Thus it maps cells to cells and anchors to anchors, preserving both $E_{\mathrm{sib}}$ and $E_{\mathrm{up}}$. Since $\varphi(w0) = \varphi(w)\,\tau_w(0)$, $\varphi(M) = M$ holds iff every non-root $\tau_w$ fixes $0$, leaving $\tau_\varepsilon$ unconstrained.
\end{IEEEproof}

\begin{IEEEproof}[Proof of Proposition~\ref{thm:adjacency-II}]
Identify nodes with addresses via Proposition~\ref{prop:addr-bij}. For $t \geq 2$, distinct members of $\operatorname{cell}(a)$ ($|a| = t{-}1$) are precisely pairs of distinct length-$t$ words sharing their $(t{-}1)$-prefix. At $t = 1$, all pairs in $\operatorname{cell}(\varepsilon)$ agree vacuously on their $0$-prefix. Thus $E_{\mathrm{sib}}$ is exactly clause~(S). Similarly, $\{w, wd\} \in E_{\mathrm{up}}$ holds iff one address extends the other by one digit, which is clause~(P). Clauses~(S) and~(P) are disjoint because (P) pairs have distinct lengths, and $E = E_{\mathrm{sib}} \cup E_{\mathrm{up}}$ is exhaustive.
\end{IEEEproof}

\begin{IEEEproof}[Proof of Proposition~\ref{prop:locator-II}]
By~\eqref{eq:unified-V}, $\log_2 V(N,m) = m \log_2 N + \log_2\frac{N}{N-1} + \log_2(1 - N^{-m}) = m \log_2 N + O(1)$. Within a fixed $(N,m)$ family context, encoding the zero-based tier $t{-}1$ takes $\lceil \log_2 m \rceil$ bits (zero bits when $m = 1$) and $t$ fixed-width base-$N$ digits take $t\lceil \log_2 N \rceil$ bits; budgeting one additional $\lceil \log_2 N \rceil$-bit tag for the parameter slot gives $(t{+}1)\lceil \log_2 N \rceil + \lceil \log_2 m \rceil$ bits, maximized at $t = m$. Dividing by $\log_2 V(N,m)$ and letting $m \to \infty$ leaves $\lceil \log_2 N \rceil / \log_2 N \in [1,\, 1 + 1/\log_2 N)$, equal to $1$ iff $N$ is a power of two.
\end{IEEEproof}

\begin{IEEEproof}[Proof of Theorem~\ref{thm:distance-II}]
By Proposition~\ref{thm:adjacency-II}, clause-(S) hops keep tier fixed and clause-(P) hops change tier by $\pm 1$.

\emph{Step 1 (prefix invariance).} Every node on a path staying at tier $\geq s \geq 1$ shares the same $(s{-}1)$-prefix, checked immediately on a single clause-(S) or clause-(P) edge.

\emph{Step 2 (upper bounds).} If $u$ is a prefix of $v$ ($t_u \leq t_v$), descending prefixes gives a clause-(P) path of length $t_v - t_u$. Otherwise $\ell \leq \min(t_u, t_v) - 1$, and ascending from $u$ to $\operatorname{pre}_{\ell+1}(u)$, taking one clause-(S) hop to $\operatorname{pre}_{\ell+1}(v)$, and descending to $v$ has length $t_u + t_v - 2\ell - 1$.

\emph{Step 3 (lower bounds).} Let $u = z_0 \sim \cdots \sim z_k = v$ be any path. If $u, v$ are prefix-comparable, tier changes by at most $1$ per hop, forcing $k \geq |t_u - t_v|$. Otherwise let $s^{*}$ be the minimum tier on the path. If $s^{*} \geq \ell + 2$, Step~1 forces $\operatorname{pre}_{\ell+1}(u) = \operatorname{pre}_{\ell+1}(v)$, contradicting maximality of $\ell$. If $s^{*} \leq \ell$, splitting at a tier-$s^{*}$ node gives $k \geq (t_u - s^{*}) + (t_v - s^{*}) \geq t_u + t_v - 2\ell > t_u + t_v - 2\ell - 1$. If $s^{*} = \ell + 1$, let $z_i$ and $z_j$ ($i \leq j$) be the first and last tier-$(\ell{+}1)$ nodes on the path. By Step~1, $z_i = \operatorname{pre}_{\ell+1}(u)$ and $z_j = \operatorname{pre}_{\ell+1}(v)$, which are distinct, so $i < j$ and the middle segment takes at least $1$ hop while the outer segments take at least $t_u - (\ell{+}1)$ and $t_v - (\ell{+}1)$ hops, proving~(i).

(ii)~For $z \neq v$, the three forwarding cases partition all pairs, every step traverses a valid clause-(P) or clause-(S) edge, and direct substitution into~(i) shows that each step reduces $d(\cdot,v)$ by exactly $1$. Arrival therefore occurs in exactly $d(u,v)$ hops.

(iii)~Prefix-comparable pairs have $d \leq m - 1$, whereas incomparable pairs satisfy $d \leq 2m - 1$, attained by $0^m$ and $1^m$ (using $N \geq 3$). Hence the diameter is $2m - 1$.
\end{IEEEproof}

\begin{IEEEproof}[Proof of Proposition~\ref{thm:deg-II}]
(i)~A tier-$t$ node has $N{-}1$ sibling links, $1$ uplink iff $t \geq 2$, and $N$ downlinks iff $t \leq m{-}1$, giving $2N{-}1$, $2N$, and $N$.

(ii)~There is one unanchored seed cell contributing $\binom{N}{2}$ links and $\sum_{t=1}^{m-1} N^t = (N^m - N)/(N-1)$ anchored cells each contributing $\binom{N}{2} + N = N(N{+}1)/2$ links, yielding the closed-form link count
\begin{equation}
\label{eq:links-II}
    L(N,m) = \frac{N(N-1)}{2} + \frac{N(N+1)(N^{m} - N)}{2(N-1)}.
\end{equation}

(iii)~Dividing twice~\eqref{eq:links-II} by $V(N,m)$ gives the exact average degree
\begin{equation}
\label{eq:avgdeg-II}
    \bar{d} = \frac{2L(N,m)}{V(N,m)} = (N{+}1) - \frac{2(N-1)}{N^{m}-1},
\end{equation}
which increases monotonically with $m$ toward $N{+}1$.
\end{IEEEproof}

\begin{IEEEproof}[Proof of Proposition~\ref{prop:conn-price}]
(i)~Following uplinks connects every node to the tier-$1$ clique, so $G(N,m)$ is connected. For any anchor $a$ ($1 \leq t = |a| \leq m{-}1$), its descendant set $D(a) = \{x : a \text{ is a proper prefix of } x\}$ has size $\sum_{j=1}^{m-t} N^j = (N^{m-t+1} - N)/(N-1)$. No sibling edge crosses between $D(a)$ and its complement because sibling addresses share their prefix, and the only uplinks leaving $D(a)$ land on $a$. Deleting $a$ therefore disconnects $D(a)$ from all tier-$1$ nodes differing from $\operatorname{pre}_1(a)$, giving $\kappa(G(N,m)) = 1$.

(ii)~On any depth-$m$ ($m \geq 2$) full $N$-ary cell tree with intra-cell $K_N$ cliques and parent--child $K_{N,N}$ bundles (or any cell tree of at least three cells), deleting any set $S$ of at most $N{-}1$ nodes leaves at least one survivor in every cell. Survivors within a cell are adjacent, and survivors in parent--child cells are adjacent, so paths follow the cell tree and the graph stays connected. Conversely, deleting all $N$ members of the parent of any leaf cell detaches that leaf cell from the remaining cells, proving $\kappa = N$. Since interior cells of branching $b = \Theta(N)$ attach to $b$ children via $K_{N,N}$, their members have degree $\Theta(N^2)$.
\end{IEEEproof}

\section{Geometric Realization Proofs and Views}
\label{app:geometry}

This appendix proves Lemma~\ref{lem:equidistance}, Theorem~\ref{thm:dim-II}, and Lemma~\ref{lem:family-mono}, formalizes the midpoint realization map $\iota$, and records its vertical and horizontal coincidences.

\subsection{Simplex Geometry and Equidistance}

\begin{definition}[Simplex]
\label{def:simplex}
An \emph{$(N{-}1)$-simplex} $\Delta_N$ is the convex hull of $N$ affinely independent points $\{p_1, p_2, \ldots, p_N\}$ in $\mathbb{R}^{N-1}$, called its \emph{vertices}. We use the shorthand ``$N$-simplex'' for a simplex on $N$ vertices.
\end{definition}

\begin{figure}[t]
\centering
\includegraphics[width=\linewidth]{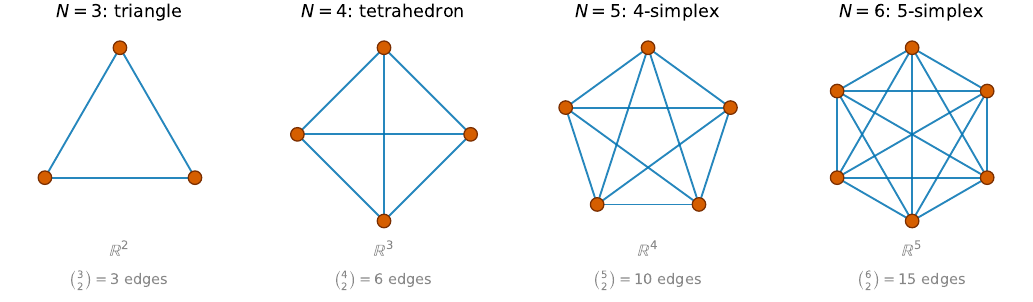}
\caption{Seed simplexes across $N = 3, \ldots, 6$. Each cell is an $N$-clique realized as the $1$-skeleton of a regular $(N{-}1)$-simplex.}
\label{fig:simplex-family}
\end{figure}

A \emph{facet} of $\Delta_N$ is the $(N{-}2)$-simplex formed by omitting one vertex, so $\Delta_N$ has $N$ vertices and $N$ facets~\cite{schneider2013convex,ewald1996combinatorial}. Fig.~\ref{fig:simplex-family} illustrates the seed $N$-clique across $N = 3,\ldots,6$.

\begin{IEEEproof}[Proof of Lemma~\ref{lem:equidistance}]
Let $p_1, \ldots, p_N \in \mathbb{R}^{d}$ satisfy $\|p_i - p_j\| = 1$ for all $i \neq j$, and set $v_i = p_i - p_N$ for $i < N$. Then $\|v_i\| = 1$ and $2\langle v_i, v_j\rangle = \|v_i\|^2 + \|v_j\|^2 - \|v_i - v_j\|^2 = 1$ for $i \neq j$. The Gram matrix of $v_1, \ldots, v_{N-1}$ is $G = \tfrac{1}{2}(I + J)$, where $J$ is the all-ones matrix. Its eigenvalues are $\tfrac{1}{2}$ (multiplicity $N{-}2$) and $N/2$, all strictly positive. Thus $G$ is positive definite, $v_1, \ldots, v_{N-1}$ are linearly independent, and $d \geq N{-}1$. The regular simplex attains $d = N{-}1$ and is unique up to similarity because the Gram matrix determines pairwise inner products up to isometry.
\end{IEEEproof}

\emph{Classical counterpart.} This is the equilateral specialization of Schoenberg's Gram criterion for Euclidean embeddability~\cite{schoenberg1935remarks}, read here as the ambient dimension cost per peer.

\subsection{Iterated Function System and Dimension Proof}
\label{subsec:dimension}

An \emph{iterated function system} (IFS)~\cite{barnsley1985iterated,hutchinson1981fractals,falconer2014fractal,edgar2008measure,mandelbrot1982fractal} is a finite collection of contractions on a complete metric space possessing a unique nonempty compact attractor.

\begin{IEEEproof}[Proof of Theorem~\ref{thm:dim-II}]
\emph{Cells are IFS images.} We induct on $t$. At $t = 1$, the seed cell is $\Delta_N = f_\varepsilon(\Delta_N)$. Suppose a tier-$t$ cell realizes as $S = f_w(\Delta_N)$ with $|w| = t{-}1$ and vertices $f_w(p_1), \ldots, f_w(p_N)$. Spawning a corner sub-simplex at $f_w(p_i)$ applies the ratio-$\tfrac{1}{2}$ homothety $h_{f_w(p_i)}$ centered at $f_w(p_i)$ to $S$. Since $f_w$ is affine, it intertwines corner homotheties via $h_{f_w(p_i)} \circ f_w = f_w \circ h_{p_i} = f_w \circ f_i$, so this sub-simplex equals $f_w(f_i(\Delta_N)) = f_{wi}(\Delta_N)$. Thus tier-$(t{+}1)$ cells are precisely the $N^t$ word images of length $t$.

\emph{Closure is the attractor.} Each $p_j$ fixes $f_j$, placing $p_j \in \mathcal{S}\Delta_N$. Invariance then puts every realized vertex $f_w(p_j)$ in $\mathcal{S}\Delta_N$. Conversely, by Hutchinson's theorem~\cite{hutchinson1981fractals}, the iterates $\bigcup_{|w|=t} f_w(K)$ for $K = \{p_1, \ldots, p_N\}$ converge to $\mathcal{S}\Delta_N$ in the Hausdorff metric, and the $t$-th iterate is precisely the tier-$(t{+}1)$ realized vertex set.

\emph{Dimension.} In barycentric coordinates $\lambda = (\lambda_1, \ldots, \lambda_N)$ with $\sum_k \lambda_k = 1$, $f_i$ acts by $\lambda \mapsto \tfrac{1}{2}(\lambda + e_i)$, so $f_i(\Delta_N) = \{\lambda \in \Delta_N : \lambda_i \geq \tfrac{1}{2}\}$. For the open interior $U = \operatorname{int}\Delta_N$, $f_i(U) \subset U$ and $f_i(U) \cap f_j(U) = \varnothing$ for $i \neq j$, because $\lambda_i > \tfrac{1}{2}$ and $\lambda_j > \tfrac{1}{2}$ contradict $\sum_k \lambda_k = 1$. The open set condition holds for every $N \geq 3$. By the Moran--Hutchinson theorem~\cite{hutchinson1981fractals,falconer2014fractal,edgar2008measure}, $\dim_{\mathrm{H}} \mathcal{S}\Delta_N = \dim_{\mathrm{S}} \mathcal{S}\Delta_N = s$, where $N (1/2)^s = 1$, giving $s = \log_2 N$.
\end{IEEEproof}

\emph{Classical counterpart.} The limit set is the $N$-vertex simplex gasket, whereas the HSFN keeps touching corner vertices logically distinct rather than glued.

\begin{figure}[t]
\centering
\includegraphics[width=0.95\linewidth]{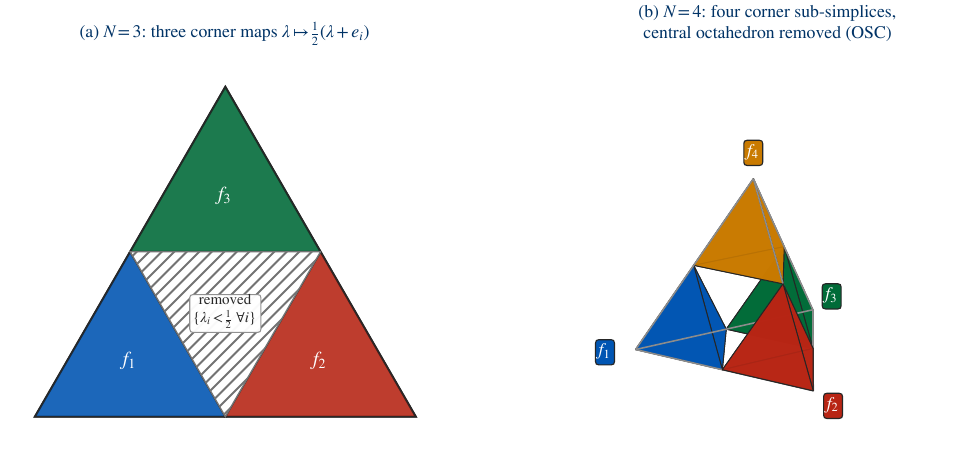}
\caption{Barycentric open set condition behind Theorem~\ref{thm:dim-II}. Each similitude $f_i$ maps $\Delta_N$ to the corner half $\{\lambda_i \geq \tfrac{1}{2}\}$, so interiors never overlap in any dimension $N{-}1$.}
\label{fig:barycentric-ifs}
\end{figure}

\begin{figure*}[t]
\centering
\includegraphics[width=0.78\textwidth]{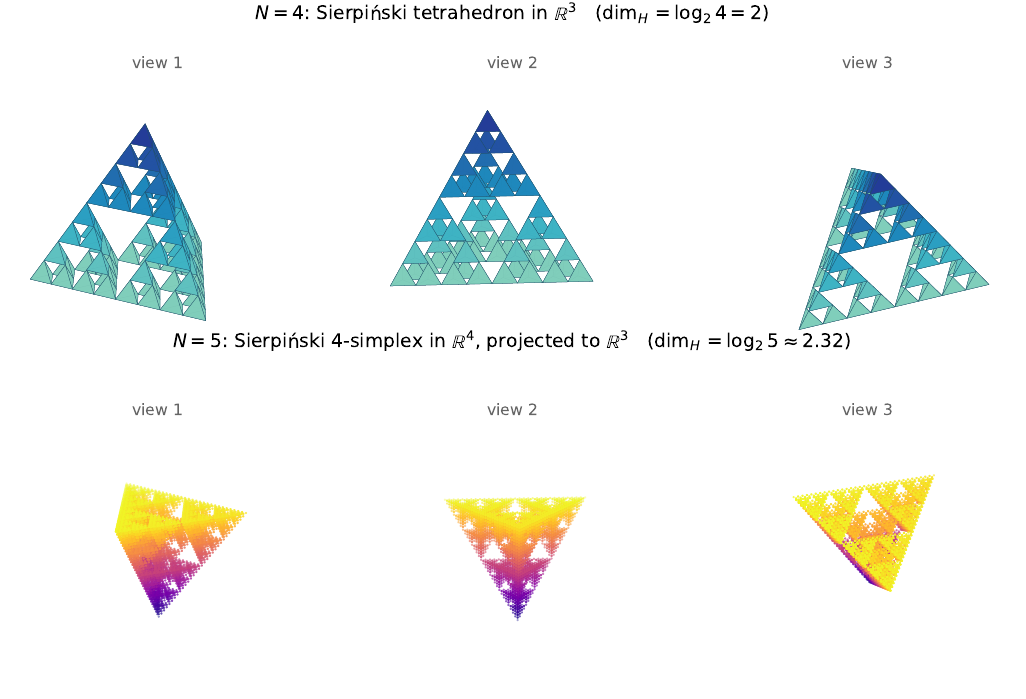}
\caption{The limit set $\mathcal{S}\Delta_N$ from three viewpoints, $N = 4$ in $\mathbb{R}^{3}$ (top) and $N = 5$ in $\mathbb{R}^{4}$ orthogonally projected to $\mathbb{R}^{3}$ (bottom). Orthogonal projection commutes with corner similitudes.}
\label{fig:sierpinski-ndim}
\end{figure*}

\begin{figure}[t]
\centering
\includegraphics[width=0.82\linewidth]{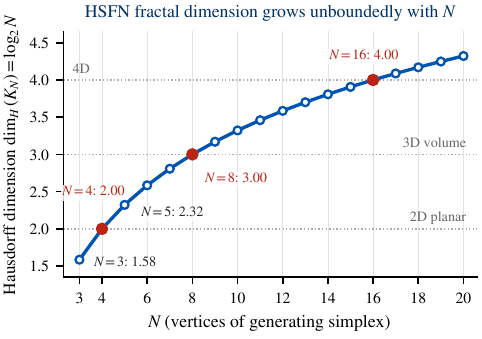}
\caption{Hausdorff dimension $\dim_H(K_N) = \log_2 N$ of the HSFN limit set against simplex size $N$, highlighting the exact integer crossings at $N = 4$ ($2\text{D}$), $N = 8$ ($3\text{D}$), and $N = 16$ ($4\text{D}$).}
\label{fig:hsfn-dimension}
\end{figure}

Fig.~\ref{fig:barycentric-ifs} illustrates the barycentric open set condition of Theorem~\ref{thm:dim-II}, Fig.~\ref{fig:sierpinski-ndim} shows the self-similar limit set $\mathcal{S}\Delta_N$ for $N = 4$ in $\mathbb{R}^{3}$ and $N = 5$ in $\mathbb{R}^{4}$ (orthogonally projected to $\mathbb{R}^{3}$), and Fig.~\ref{fig:hsfn-dimension} plots $\dim_{\mathrm{H}}(\mathcal{S}\Delta_N) = \log_2 N$ across $N = 3, \ldots, 20$.

\subsection{Midpoint Realization and Family Monotonicity}
\label{sec:realizations}
\label{subsec:realization-maps}
\label{subsec:family-containments}

Every deployed node belongs to one \emph{birth cell}, and its \emph{peers} are the other $N{-}1$ members of that cell, ordered $y_1 \prec \cdots \prec y_{N-1}$ by increasing final digit.

\begin{definition}[Midpoint realization]
\label{def:embedding-maps}
Fix $N \geq 3$, $m \geq 1$, and a seed simplex $\Delta_N = \operatorname{conv}\{p_1,\ldots,p_N\} \subset \mathbb{R}^{N-1}$ with $\iota(d) = p_{d+1}$ for $d \in \Sigma_N$. The \emph{midpoint realization} $\iota \colon \mathcal{W}_{N,m} \to \mathbb{R}^{N-1}$ extends recursively by
\begin{equation}
\label{eq:iota-dual}
    \iota(a0) = \iota(a), \qquad
    \iota(ad) = \tfrac{1}{2}\bigl(\iota(a) + \iota(y_d)\bigr)
\end{equation}
for $1 \leq d \leq N{-}1$, where $y_1 \prec \cdots \prec y_{N-1}$ are the peers of $a$. Each tier-$t$ cell realizes as a simplex scaled by $2^{-(t-1)}$.
\end{definition}

\begin{figure}[t]
\centering
\includegraphics[width=0.85\linewidth]{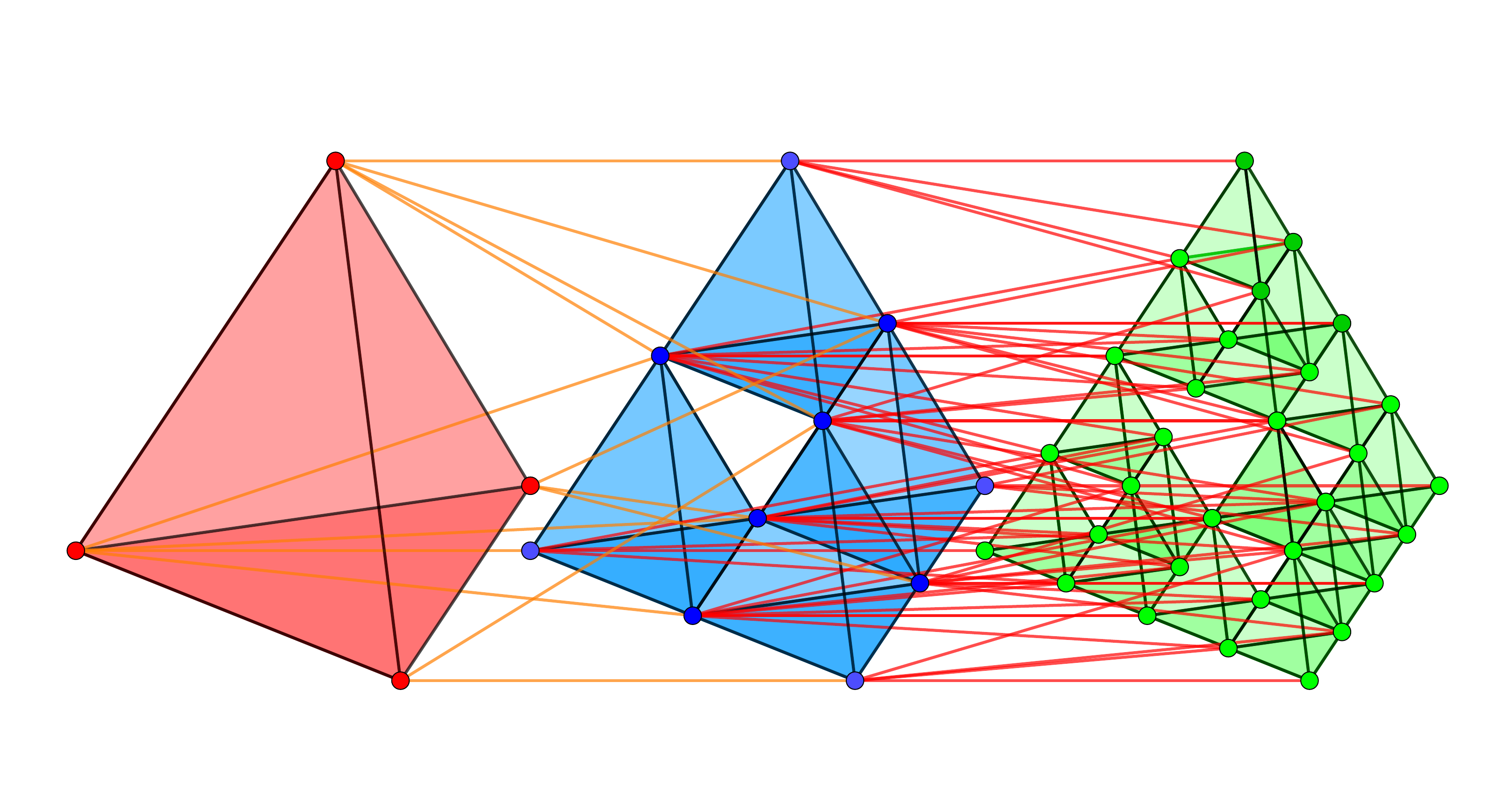}
\caption{Exploded midpoint realization for $N = 4, m = 3$, separating co-located duplicate slots vertically.}
\label{fig:dual-exploded}
\end{figure}

\begin{figure}[t]
\centering
\subfloat[Inter-tier uplinks (green). Each spawned cell links vertically only to its anchor.]{\includegraphics[width=0.47\linewidth]{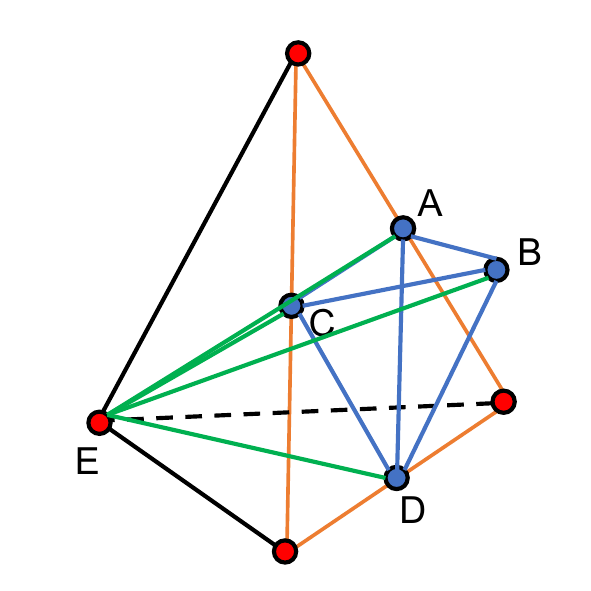}\label{fig:interlayer-links}}
\hfill
\subfloat[Horizontal coincidence. Adjacent peers place distinct, non-adjacent offspring at the same edge midpoint.]{\includegraphics[width=0.47\linewidth]{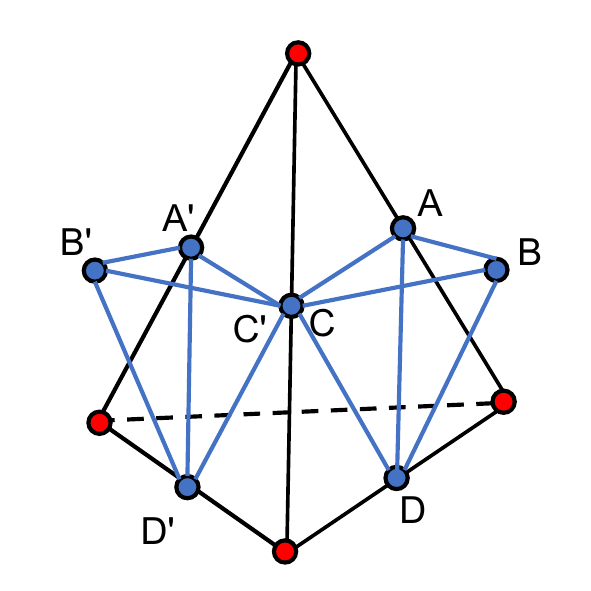}\label{fig:coincident-nodes}}
\caption{Local geometry of the midpoint realization.}
\label{fig:realization-detail}
\end{figure}

Fig.~\ref{fig:dual-exploded} draws the exploded midpoint realization for $N = 4, m = 3$, while Fig.~\ref{fig:realization-detail} isolates vertical uplinks and horizontal midpoint coincidences.

\begin{remark}[Coincident nodes remain logically distinct]
\label{rem:coincident}
The map $\iota$ is non-injective in two structured ways. \emph{Vertically}, $\iota(a0) = \iota(a)$ collapses every marked duplicate chain $a, a0, a00, \ldots$ onto a single point. \emph{Horizontally}, every peer pair $\{x, y\}$ at tier $t < m$ places one offspring of $x$ and one offspring of $y$ at the same midpoint $\tfrac{1}{2}(\iota(x) + \iota(y))$ (Fig.~\ref{fig:coincident-nodes}). All coincident nodes carry distinct addresses, and geometric coincidence never adds an edge. All $N$ members $a0, \ldots, a(N{-}1)$ link to anchor $a$ via uplinks, while only the marked duplicate $a0$ has a zero-length Euclidean uplink.
\end{remark}

\emph{Running example ($N = 5$, $m = 3$).} At node $a = 42$ with peers $40 \prec 41 \prec 43 \prec 44$, $\iota(420) = \iota(42)$ and $\iota(423) = \tfrac{1}{2}(\iota(42) + \iota(43))$. At peer $43$ (peers $40 \prec 41 \prec 42 \prec 44$), $\iota(433)$ lands at the same midpoint, creating one horizontal coincidence between distinct non-adjacent nodes $423$ and $433$.

\begin{IEEEproof}[Proof of Lemma~\ref{lem:family-mono}]
Every word of length at most $m$ over $\{0,\ldots,N{-}1\}$ lies in $\mathcal{W}_{N',m'}$. The predicates defining sibling edges, uplink edges, and marked duplicates (final digit $0$ at length $\geq 2$) depend only on lengths, prefixes, and digits, all preserved and reflected by the inclusion. Thus $F(N,m)$ is an induced, mark-preserving subnetwork of $F(N',m')$. By Theorem~\ref{thm:distance-II}, distances depend only on tiers and longest common prefix length, proving isometry. Composition of inclusions is immediate.
\end{IEEEproof}

\emph{Classical counterpart.} This generalizes hypercube nesting $Q_d \subset Q_{d+1}$ simultaneously in alphabet size $N$ and depth $m$ while transporting the marking $M$.

\section{Protocol Cost and Reliability Proofs}
\label{app:protocol}

This appendix records the normalized message-proxy observable used by the operations calculus, collects the explicit combinatorial recursions, and supplies the proofs of Sections~\ref{sec:time} and~\ref{sec:fault}.

\subsection{Normalized Message Proxy}

\begin{proposition}[Normalized message proxy]
\label{thm:complexity}
Under a normalized accounting model charging $N^2$ intra-cell PBFT-scale messages per $N$-member cell and one leader message per node, a fully populated $G(N,m)$ of $V = V(N,m)$ nodes incurs
\begin{equation*}
    C_{\mathrm{full}} = (N + 1) V
\end{equation*}
proxy messages per round, scaling as $\Theta(NV)$.
\end{proposition}

\begin{IEEEproof}
There are $V/N$ cells, contributing $(V/N)N^2 = NV$ intra-cell proxy messages plus $V$ inter-layer messages, totaling $(N{+}1)V$.
\end{IEEEproof}

Compared with the exact executed count $C_{\mathrm{exec}} = 2(N{-}1)V + 2s(V - N)$ of Proposition~\ref{prop:exec-count}, at $s = 1$ we have $C_{\mathrm{exec}} = 2NV - 2N < 2(N{+}1)V$, and at $s = f{+}1$ the ratio stays below $2(N{+}f)/(N{+}1)$. Table~\ref{tab:hierbft} compares message and per-replica reception counts across flat and hierarchical BFT protocols.

\begin{table*}[!t]
\centering
\caption{Message Complexity Across Flat and Hierarchical BFT Protocols. Columns 3 and 4 apply a uniform per-decision message count where a source specifies exact phases (steady-state pipelining for HotStuff and Kauri), leaving aggregate-only formulations open.}
\label{tab:hierbft}
\footnotesize
\setlength{\tabcolsep}{3pt}
\renewcommand{\arraystretch}{1.2}
\begin{tabular}{@{}>{\raggedright}p{0.12\textwidth}>{\raggedright}p{0.30\textwidth}>{\raggedright}p{0.17\textwidth}>{\raggedright}p{0.14\textwidth}>{\raggedright\arraybackslash}p{0.20\textwidth}@{}}
\toprule
Protocol & Cost as stated by source & Messages per decision & Max received by one replica & Fault model and depth \\
\midrule
Flat PBFT~\cite{castro1999practical} & $O(n^{2})$ authenticators normal, $O(n^{3})$ view change~\cite{yin2019hotstuff} & $2n(n{-}1)$ & $2(n{-}1)$ & $n = 3f{+}1$, flat \\
HotStuff~\cite{yin2019hotstuff} & $O(n)$ authenticators per view & $2(n{-}1)$ & $n{-}1$ (at leader) & $n = 3f{+}1$, flat \\
Kauri~\cite{neiheiser2021kauri} & pipelined tree of fanout $b$ & $2(n{-}1)$ & $b{+}1$ & $n = 3f{+}1$, depth $\log_b n$ \\
Multi-layer PBFT~\cite{li2020scalable} & $(m{+}1)^{2} + m(n{+}1)^{2}$ for $m$ groups of $n$ & open & open & $3f{+}1$ per group, $2$ layers \\
Steward~\cite{amir2010steward} & $O(S^{2})$ wide-area across $S$ sites & open & open & $3f{+}1$ per site, $2$ levels \\
HSFN (this paper) & $2(N{-}1)V + 2(f{+}1)(V{-}N)$ messages, root commit in $4m{-}1$ overlay delays ($N \geq 4$, fault-free, timer-free) & $2(N{-}1)V + 2(f{+}1)(V{-}N)$ & $2(N{-}1) + (N{+}1)(f{+}1)$ & $f$ per cell of $N$, depth $m$ \\
\bottomrule
\end{tabular}
\end{table*}

\subsection{Cobweb Dynamics and Registration Proof}
\label{app:aph}

Fig.~\ref{fig:rg-cobweb} in Section~\ref{subsec:fpd} illustrates the one-dimensional dynamical system $x \mapsto f_b(x)$ of Theorem~\ref{prop:threshold-b} on the HSFN at $b = N = 8$. Below $q_0(b)$, iterates remain trapped in the certified invariant interval $[q, 2q]$. Beyond the saddle-node fold $q_c(b)$, the survivable fixed point disappears and deep networks collapse toward $1$.

\emph{Classical counterpart.} Iterated recursive majority mirrors von Neumann's multiplexed majority organs~\cite{vonneumann1956probabilistic,moore1956reliable} and tree percolation~\cite{lyons1990random}, with explicit factor-$2$ certificate below $q_0(b)$ and a saddle-node transition at $q_c(b)$.

\begin{figure}[t]
\centering
\includegraphics[width=0.92\linewidth]{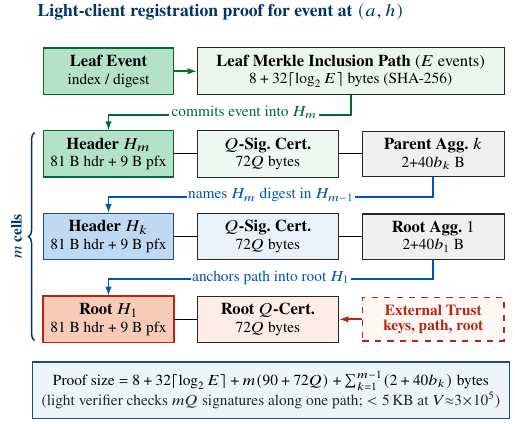}
\caption{Wire anatomy of a light-client registration proof matching~\eqref{eq:proof-size}. Given external trust inputs (committee keys, path, and root anchor), a verifier checks one SHA-256 Merkle inclusion path inside the leaf block plus $m$ signed cell headers ($81$ B header + $9$ B certificate prefix each) and $m{-}1$ aggregate links up to the root.}
\label{fig:reg-proof-packet}
\end{figure}

Fig.~\ref{fig:reg-proof-packet} details the concrete wire layout of the light-client registration proof sized in~\eqref{eq:proof-size}.

\subsection{Explicit Combinatorial Recursions}
\label{app:explicit-recursions}

For convenient reference, we record the explicit combinatorial forms of the anchor-coupled recursions and path registration probabilities summarized in Section~\ref{sec:fault}. Set $r = \lceil N/2 \rceil - 1$. Under single-anchor crash faults on $G(N,m)$ (Proposition~\ref{prop:phys-recursion}),
\begin{equation}
\label{eq:gN-def}
\begin{split}
    g_N(x) = 1 - \sum_{j=0}^{\min(f, r)} &\tbinom{N}{j} P_f^{\,j} (1-P_f)^{N-j} \\
    &\cdot \Pr\bigl[\mathrm{Bin}(N{-}j, x) \leq r - j\bigr].
\end{split}
\end{equation}
Under dual homing (Proposition~\ref{prop:dual}), let $a(J)$ count digits $d \in \Sigma_N$ with both $d \in J$ and $(d{+}1) \bmod N \in J$, and set $\mathcal{J} = \{J \subseteq \Sigma_N : |J| \leq f,\; a(J) \leq r\}$. Then
\begin{equation}
\label{eq:hN-def}
\begin{split}
    h_N(x) = 1 - \sum_{J \in \mathcal{J}} &P_f^{\,|J|} (1-P_f)^{N-|J|} \\
    &\cdot \Pr\bigl[\mathrm{Bin}(N{-}a(J), x) \leq r - a(J)\bigr].
\end{split}
\end{equation}
For certified leaf-to-root event registration with site outage probability $p_s$ and $q_s = p_s + (1{-}p_s)q$ (Proposition~\ref{prop:registration}), the exact overlay probability is
\begin{equation}
\label{eq:reg-overlay}
\begin{split}
    A = (1 - q_s) \prod_{k=1}^{m-1} \Bigl( &(1 - q_s) \\
    &\cdot \Pr\bigl[\mathrm{Bin}(N{-}1, 1 - P_{k+1}) \geq \lfloor N/2 \rfloor\bigr] \Bigr),
\end{split}
\end{equation}
and the exact single-anchor physical probability is
\begin{equation}
\label{eq:reg-phys}
\begin{split}
    A^{\mathrm{ph}} = (1{-}q_s) \prod_{k=1}^{m-1} \Biggl( &(1{-}p_s)\! \sum_{j=0}^{\min(f,r)}\! \tbinom{N-1}{j} P_f^{\,j} (1{-}P_f)^{N-j} \\
    &\cdot \Pr\bigl[\mathrm{Bin}(N{-}1{-}j, P^{\mathrm{phys}}_{k+1}) \leq r - j\bigr] \Biggr),
\end{split}
\end{equation}
where $P_k^{\mathrm{phys}}$ iterates~\eqref{eq:gN-def} with survival factor $1{-}p_s$ per cell.

\subsection{Proofs}
\label{app:protocol-proofs}

\begin{IEEEproof}[Proof of Proposition~\ref{prop:safety}]
Recall $Q = \lceil (N{+}f{+}1)/2 \rceil$, so $2Q \geq N{+}f{+}1$, $2Q - f - N \geq 1$, and $Q - f \geq f{+}1$. (i)~Two prepared certificates in the same view each aggregate $Q$ distinct cell signers (including the primary). Their signer sets intersect in at least $2Q - N \geq f{+}1$ members, including at least one correct member that prepares at most one digest per view. If correct members commit $d$ in view $v$, let $S$ be the set of correct members that sent commit for $d$ in view $v$, so $|S| \geq Q{-}f$. By the view-change lock of step~1, every member of $S$ sent that commit before moving above view $v$, so every subsequent view change it sends carries a prepared certificate of view at least $v$. Any $Q$ view changes for a higher view intersect $S$ in at least $2Q - f - N \geq 1$ correct member. Induction over views forces every higher prepared certificate and new-view proposal to carry $d$~\cite{castro1999practical}. (ii)~A commit certificate contains $Q$ signatures, at least $Q{-}f \geq f{+}1$ from correct members. A correct member signs a commit only for a digest it prepared in that view, which by~(i) is the child's unique certified block in that round (and matches the block committed by every correct member of that child that finishes the round). Validity requires a commit certificate per included child, so whenever every included child cell along the path stays within budget $f$, induction along the path establishes the final statement.
\end{IEEEproof}

\begin{IEEEproof}[Proof of Proposition~\ref{prop:exec-count}]
Each of the $V/N$ cells sends $N{-}1$ pre-prepares, $(N{-}1)^2$ prepares, and $N(N{-}1)$ commits, totaling $2N(N{-}1)$ intra-cell messages, or $2(N{-}1)V$ network-wide. Each of the $V/N - 1$ non-root cells sends $sN$ reports upward and receives $sN$ disseminations downward, contributing $2s(V - N)$ messages. On $G(N,m)$, every intra-cell message takes $1$ sibling link. Each sender's $N$ upward reports cross $1$ uplink to the anchor plus $2$ links (uplink and sibling link) to each of the $N{-}1$ other parent members, traversing $1 + 2(N{-}1) = 2N{-}1$ physical hops in total. For downward dissemination from a parent cell to its $N$ child cells, each sender anchors exactly one child cell ($N$ direct uplinks) and reaches the other $N{-}1$ child cells via $2N$ physical hops each (one sibling hop to that child's anchor plus $N$ uplinks); summed over all $N$ child cells, each parent sender contributes $N + (N{-}1)\cdot 2N = N(2N{-}1)$ hops, averaging $2N{-}1$ physical hops per child cell per sender and yielding the stated total.
\end{IEEEproof}

\begin{IEEEproof}[Proof of Proposition~\ref{thm:committee-tree}]
The message inventory of Proposition~\ref{prop:exec-count} is strictly local to each committee $c$ and each parent--child edge $\{c, p(c)\}$. For committees with $N_c \geq 4$ ($Q_c - 1 \geq 2$), every backup requires at least one peer prepare in addition to its own before sending commit, so local PBFT takes exactly $3\delta$ (pre-prepare, prepare, commit). Every interior committee starts $1\delta$ after its slowest child commits and finishes $3\delta$ later, advancing $4\delta$ per tier to reach the root at $3\delta + 4(L_{\max}{-}1)\delta = (4L_{\max}{-}1)\delta$. Dissemination descends one tier per $\delta$, adding $(L_{\max}{-}1)\delta$. (At $N_c = 3$, $f_c = 0$ and $Q_c = 2$, so a backup's own prepare completes $Q_c - 1 = 1$ immediately upon receiving pre-prepare, committing locally in $2\delta$.)
\end{IEEEproof}

\begin{IEEEproof}[Proof of Proposition~\ref{prop:latency}]
For $N \geq 4$, leaf cells commit after $3\delta$ (pre-prepare, prepare, commit). On the overlay, reports reach parent cells in $1\delta$ and dissemination descends at $1\delta$ per tier, matching Proposition~\ref{thm:committee-tree} at $L_{\max} = m$. On $G(N,m)$, a report reaches the parent primary (digit $0$) in $2\delta$ for every child cell except the one anchored at that primary, so each interior primary begins proposing $2\delta$ after its children commit and finishes $3\delta$ later. Reaching the root takes $3\delta + 5(m{-}1)\delta = (5m{-}2)\delta$. Dissemination takes $2\delta$ per tier for non-anchor destinations, adding $2(m{-}1)\delta$ to reach $t_{\mathrm{conf}} = (7m{-}4)\delta$. (At $N = 3$, replacing local $3\delta$ by $2\delta$ yields $(3m{-}1)\delta$ and $(4m{-}2)\delta$ on the overlay, and $(4m{-}2)\delta$ and $(6m{-}4)\delta$ on $G(3,m)$.)
\end{IEEEproof}

\begin{IEEEproof}[Proof of Theorem~\ref{thm:security}]
A level-$k$ cell satisfies $S$ iff it stays within budget $f$ (probability $1 - q$) and at most $\lceil b_k/2 \rceil - 1$ of its $b_k$ children fail. Because member sets of distinct cells and subtrees below distinct children are pairwise disjoint under independent node faults, internal survival and child indicators are mutually independent. By induction from the leaves, every level-$(k{+}1)$ child fails with probability $P_{k+1}$, giving
\[
\begin{aligned}
    1 - P_k &= (1 - q)\,\Pr\bigl[\mathrm{Bin}(b_k, P_{k+1}) \leq \lceil b_k/2 \rceil - 1\bigr] \\
            &= (1 - q)\, S_{b_k}(P_{k+1}),
\end{aligned}
\]
which is $P_k = f_{b_k}(P_{k+1})$. On the uniform $N$-ary cell tree of $G(N,m)$, $b_k = N$ for all $1 \leq k < m$, yielding $P_1(N,m) = f_N^{\,m-1}(q)$. The heterogeneous formula follows identically by replacing the binomial sum with the sum of independent indicators $X_k$ of mean $P_k$.
\end{IEEEproof}

\begin{IEEEproof}[Proof of Theorem~\ref{prop:threshold-b}]
\emph{Monotone convergence.} Since $S_b(x)$ is a binomial lower tail, it is nonincreasing in $x$, making $f_b(x) = q + (1-q)(1 - S_b(x))$ nondecreasing. Because $f_b(q) \geq q$, the iterates $f_b^{\,h}(q)$ are nondecreasing and bounded above by $1$, hence converge by continuity to a fixed point $x_b^*(q) \in [q,1]$. Induction against any fixed point $y \geq q$ shows $x_b^*(q) \leq y$.

\emph{Depth-uniform certificate.} By the union bound over $c$-subsets ($c = \lceil b/2 \rceil$),
\[
    1 - S_b(x) = \Pr\bigl[\mathrm{Bin}(b, x) \geq c\bigr] \leq \binom{b}{c} x^{c}.
\]
For $x \in [q, 2q]$, monotonicity of $f_b$ yields
\[
    f_b(x) \leq q + \binom{b}{c}(2q)^{c} = q + \left[2^{c}\binom{b}{c} q^{c-1}\right] q.
\]
Since $b \geq 3$ gives $c - 1 \geq 1$, the bracketed factor is strictly less than $1$ exactly when $q < q_0(b)$. Together with $f_b(x) \geq q$, this proves $f_b([q,2q]) \subseteq [q,2q]$, so every iterate stays at most $2q$. Note $2q < 2 q_0(b) \leq 2 \cdot 2^{-c/(c-1)} < 1$.

\emph{Critical threshold.} Pointwise monotonicity of $f_b$ in $q$ makes $x_b^*(q)$ nondecreasing in $q$, so $\{q : x_b^*(q) < 1\}$ is down-closed and contains $[0, q_0(b))$, making its supremum $q_c(b) \geq q_0(b)$ well defined. Counting surviving children ($\geq \lfloor b/2 \rfloor + 1$) gives
\[
\begin{aligned}
    S_b(x) &= \Pr\bigl[\mathrm{Bin}(b, 1{-}x) \geq \lfloor b/2 \rfloor + 1\bigr] \\
           &\leq \binom{b}{\lfloor b/2 \rfloor + 1} (1 - x)^{\lfloor b/2 \rfloor + 1}.
\end{aligned}
\]
Whenever $q > 1 - \binom{b}{\lfloor b/2 \rfloor + 1}^{-1}$, we obtain $(1-q) S_b(x) < 1 - x$ for all $x \in [0,1)$, hence $f_b(x) > x$ on $[0,1)$, forcing $x_b^*(q) = 1$.
\end{IEEEproof}

\begin{IEEEproof}[Proof of Corollary~\ref{cor:threshold-II}]
By Theorem~\ref{thm:security}, $P_1(N,m) = f_N^{\,m-1}(q)$. Applying Theorem~\ref{prop:threshold-b} at $b = N$ traps all iterates inside $[q, 2q]$ whenever $q < q_0(N)$.
\end{IEEEproof}

\begin{IEEEproof}[Proof of Proposition~\ref{prop:phys-recursion}]
Under crash faults, a cell succeeds iff at most $N{-}Q$ members fail and at most $r$ children are lost. When $3 \nmid N$, $N{-}Q = f$. When $3 \mid N$, $N{-}Q = f{+}1$, so restricting to $j \leq f$ omits surviving outcomes and yields an upper bound. Member $i$ anchors child $i$, so $j$ faulty members automatically sever $j$ distinct children. The remaining $N{-}j$ children head disjoint subtrees failing independently with probability $P_{k+1}^{\mathrm{phys}}$, requiring at most $r - j$ failures among them. Summing over $j$ gives~\eqref{eq:gN-def}. Under natural coupling, $\mathrm{Bin}(N, x) \leq \mathrm{Bin}(N{-}j, x) + j$, so each summand is bounded above by $\binom{N}{j} P_f^j (1{-}P_f)^{N-j} S_N(x)$, whose sum over $j \leq f$ equals $(1-q)S_N(x) = 1 - f_N(x)$.
\end{IEEEproof}

\begin{IEEEproof}[Proof of Proposition~\ref{prop:dual}]
A child cell is disconnected by parent faults only when both of its anchors belong to $J$, severing exactly $a(J)$ children. Conditioning on $J$ in place of $|J|$ yields~\eqref{eq:hN-def}. When $f = 1$, $|J| \leq 1$ forces $a(J) = 0$, collapsing $h_N$ identically to $f_N$.
\end{IEEEproof}

\begin{IEEEproof}[Proof of Proposition~\ref{prop:registration}]
Under independent node faults and cell site outages, disjoint off-path subtrees and distinct cells along the leaf-to-root path fail independently. The leaf cell succeeds with probability $1 - q_s$. Given that the child on the path delivers its report to its level-$k$ ancestor, that ancestor must survive its own faults and receive reports from at least $\lfloor N/2 \rfloor$ of its $N{-}1$ off-path children (so that total received reports reach $\lfloor N/2 \rfloor + 1$). On the overlay (and on dual-homed $G(N,m)$ with $f = 1$, where $h_N = f_N$), internal survival ($1 - q_s$) is independent of off-path children, giving~\eqref{eq:reg-overlay}. On single-anchor $G(N,m)$, the ancestor must avoid site outage ($1 - p_s$), its member anchoring the on-path child must stay alive ($1 - P_f$), and $j \in \{0,\ldots,\min(f,r)\}$ of the remaining $N{-}1$ members may fail, automatically severing their $j$ off-path children while leaving $N{-}1{-}j$ independent off-path subtrees of failure probability $P_{k+1}^{\mathrm{phys}}$ of which at most $r - j$ may fail. Multiplying across $k = 1, \ldots, m{-}1$ yields~\eqref{eq:reg-phys}.
\end{IEEEproof}

\section{Hosting Service Topologies: Proofs and Constructions}
\label{app:hosting}
\label{app:universality}

\subsection{Universality of the HSFN}
\label{sec:universality}

Sections~\ref{sec:model}--\ref{sec:space} built the HSFN from the top down, one generating rule unfolding into a specific, highly structured graph. This appendix develops the full proof of Theorem~\ref{thm:blockcut} and Lemma~\ref{lem:tw-nec} from Section~\ref{sec:reconfig}, showing which finite topologies $F(N,m)$ can host under disciplined reductions. The cell size consumed is governed by two classical width invariants of the target, its largest biconnected block and its treewidth, while the depth is set by the descent weight of its block--cut tree. This provides the structural foundation for the lifting theorem of Appendix~\ref{app:calculus}, where edits on a hosted graph transfer to calculus words on the HSFN.

Throughout, $F(N,m)$ denotes the marked HSFN network of Definition~\ref{def:address-II}, while $H$ is a finite simple graph. We write $\beta(H)$ for the \emph{block order} of $H$, the maximum number of vertices in a biconnected component (block) of $H$, and $\mathrm{tw}(H)$ for its treewidth. As established below (Lemma~\ref{lem:maxclique}), $F(N,m)$ is a block graph whose blocks have orders $N$ and $N{+}1$.

The main hosting result is Theorem~\ref{thm:blockcut} in Section~\ref{subsec:hosting-main}. Below we record the formal realization semantics, the supporting structural lemmas, the two-full boundary obstruction, and the full proofs and constructions.

\subsubsection{Disciplined Realizations}
\label{subsec:realization}

Universality is vacuous unless the reduction is constrained. Under arbitrary edge selections and quotients, every finite graph is a quotient of a subgraph of a sufficiently large clique, which collapses all resource bounds. The correct notion contracts the co-located duplicate chains that the construction itself produces and nothing else.

\begin{definition}[Realization]
\label{def:realization}
A \emph{duplicate chain} in $F(N,m)$ is a path $x, x0, x00, \ldots$ in which each node is the digit-$0$ child (the co-located duplicate of Lemma~\ref{lem:dup-digit}) of its predecessor. Let $S$ be a subgraph of $F(N,m)$, a set of nodes together with a set of selected links among them. Let $\mathcal{C}$ be a set of vertex-disjoint duplicate chains whose nodes and edges lie in $S$. The \emph{realization} $\mathrm{Real}(S,\mathcal{C})$ is the simple graph obtained from $S$ by contracting each chain in $\mathcal{C}$ to a single vertex and passing to the simple quotient, in which self-loops created by a contracted chain edge are discarded and parallel edges are merged. A finite graph $H$ is \emph{realized} by $F(N,m)$ if $H \cong \mathrm{Real}(S,\mathcal{C})$ for some such $S$ and $\mathcal{C}$. Selection deletes links and never adds them, while contraction is restricted to the designated chains. The subgraph freedom is necessary and harmless. It is necessary because $F(N,m)$ is claw-free (every neighborhood is covered by the birth clique and the spawn cell). Hence already the star $K_{1,3}$ has no induced copy in any family member. Neither has any graph carrying a non-clique block, since an induced subgraph of the block graph $F(N,m)$ is again a block graph. It is harmless because an overlay is free to leave a provisioned link unused (Remark~\ref{rem:static-substitution}).
\end{definition}

Two points make the definition well posed. First, only \emph{designated} chains are contractible. The horizontal geometric coincidences of the realization map $\iota$ (Section~\ref{sec:realizations}), distinct sibling midpoints sharing a position, are never contracted. Contracting one would fuse independent logical nodes and inject a $5$-cycle into $F(3,m)$, destroying every width bound below. Second, contraction of a designated pair produces a simple graph. The chain edge becomes a discarded loop, while the only parallel edges are the two copies of an edge to a shared neighbor, which merge (verified on the worked case $\{0, 00\}$ in $F(3,2)$, where $V'=11$ and $E'=18$).

The clique structure of $F(N,m)$ controls what a realization can contain. Recall from Section~\ref{sec:construction} that a cell is the set of $N$ children of a common spawner. The \emph{anchored clique} of a non-leaf node $v$ is $\{v\} \cup \operatorname{cell}(v)$, of order $N+1$.

\begin{lemma}[Blocks and maximal cliques]
\label{lem:maxclique}
For $N \geq 3$ and $m \geq 2$, the blocks of $F(N,m)$ are exactly the tier-$1$ cell (order $N$) and the anchored cliques $\{v\}\cup\operatorname{cell}(v)$ (order $N+1$). These are precisely its maximal cliques. Hence the clique number is $\omega(F(N,m)) = N+1$, attained only by anchored cliques.
\end{lemma}

\emph{Classical counterpart.} The lemma places $F(N,m)$ in Harary's block graphs, in which every block is a clique~\cite{harary1963block} and the block--cut tree is the clique tree~\cite{gavril1974intersection}. Chordality and treewidth one below the clique number carry over. What differs is that this member is self-similar with blocks of bounded order named by addresses.

The tier-$1$ exception is load-bearing. It is the mechanism behind the first free block at each cut vertex below. It is also why the sibling clause of Proposition~\ref{thm:adjacency-II} must be stated with the virtual root $\varepsilon$.

\subsubsection{Trees and the Fan-out Bound}
\label{subsec:trees}

The acyclic case is exact and cheap. It fixes the depth cost of a realization.

\begin{lemma}[Tree embedding]
\label{lem:tree-embed}
Every rooted tree of fan-out at most $N$ and depth $d$ is a subgraph of $F(N,d+1)$, realized with no contractions.
\end{lemma}

\emph{Classical counterpart.} This is the dilation-one case of embedding a guest tree into a host tree~\cite{bhatt1986optimal}. That fan-out and depth are the only obstructions carries over. What differs is that the host is the prescribed member $F(N,d{+}1)$, with the depth cost returning as the descent weight of Theorem~\ref{thm:blockcut}.

\subsubsection{The Hosting Characterization}
\label{subsec:blockcut}

A graph with cycles forces cliques, which cost dimension. This subsection determines which topologies the family hosts, exactly so for the topologies in which no cut vertex lies in two blocks of order $N{+}1$. The answer is governed by the block order, with the boundary case of order-$(N{+}1)$ blocks controlled by a placement law at cut vertices.

For a connected $H$, root its block--cut tree $\mathcal{T}(H)$ at a block. Each cut vertex $v$ then has a unique parent block and some number $c_v \geq 1$ of child blocks. Define the \emph{descent weight} of a root-to-leaf path in $\mathcal{T}(H)$ as the sum of $c_v$ over the cut vertices it passes. Let $\operatorname{dw}(H)$ be the maximum descent weight over all such paths. The $r$-th child block at a cut vertex is hosted one tier below the $(r{-}1)$-st slot of its duplicate chain. A cut vertex therefore charges up to $c_v$ tiers of descent, which is exactly what the sum counts. Call a block of order exactly $N+1$ a \emph{full} block.

The necessity direction rests on a structure lemma for how contraction interacts with blocks.

\begin{lemma}[Block images]
\label{lem:block-image}
Let $H = \mathrm{Real}(S, \mathcal{C})$ be realized by $F(N,m)$. Then every block of $H$ is the simple quotient of a subgraph of a single block of $F(N,m)$ in which at most one pair of vertices, the anchor--duplicate pair of that block, is identified. Consequently every block of $H$ has order at most $N+1$. A block of order $N+1$ occupies all $N+1$ vertices of one full anchored clique of $F(N,m)$, untouched, in that it is the isomorphic image of a spanning subgraph of that clique and no chain of $\mathcal{C}$ contains two of its host vertices.
\end{lemma}

\begin{IEEEproof}[Proof sketch]
A duplicate chain meets a block of the host in at most the anchor--duplicate pair. Contracting an edge whose endpoints lie in one block never merges two blocks. Iterating over the chain edges confines every block of $H$ to the image of a single host clique. Full proof in Appendix~\ref{app:hosting-proofs}.
\end{IEEEproof}

\emph{Classical counterpart.} The mechanism is classical, contracting an edge inside a block never merging two blocks of the block--cut tree~\cite{harary1963block}. That confinement carries over. What differs is that only designated duplicate chains contract, which turns a soft bound into the exact ceiling $N{+}1$.

The one-unit gap between $\beta \leq N$ and $\beta \leq N+1$ is real. The boundary is governed by a two-slot law at cut vertices. A non-leaf vertex of $F(N,m)$ lies in exactly two maximal cliques, its birth clique and its anchored clique, while a leaf lies in its birth clique alone (Lemma~\ref{lem:maxclique}). This budget is what a cut vertex of $H$ spends.

\begin{proposition}[Placement of full blocks]
\label{prop:blockcut-caveat}
Let $H$ be realized by $F(N,m)$ and let $v$ be a cut vertex of $H$.
\begin{enumerate}
\item[(i)] $v$ lies in at most \emph{two} full blocks, one realized in the birth clique of the head of $v$'s hosting chain and one realized as the anchored clique of the final vertex of that chain. Three full blocks at one cut vertex are not realizable in any $F(N,m)$.
\item[(ii)] Both slots are attainable simultaneously, since two full blocks sharing a cut vertex are realizable, e.g.\ as $\{a\} \cup \operatorname{cell}(a)$ and $\{ad\} \cup \operatorname{cell}(ad)$ glued at $ad$ with no contractions.
\item[(iii)] If $v$ lies in a full block realized as the anchored clique $\{x\}\cup\operatorname{cell}(x)$ of the head $x$ of its hosting chain (rather than of the chain's final vertex), then the chain is trivial and every block of $H$ at $v$ other than that full block is hosted inside the birth clique of $x$.
\end{enumerate}
\end{proposition}

\emph{Classical counterpart.} In a chordal graph the maximal cliques at a vertex form a subtree of the clique tree~\cite{gavril1974intersection}, whereas in a block graph a cut vertex may lie in arbitrarily many blocks~\cite{harary1963block}. The subtree picture carries over. What differs is that this subtree is a path on two vertices. Further child blocks therefore cost depth.

\emph{Classical counterpart.} The question is virtual network embedding, the mapping of a guest topology onto a substrate under capacity constraints. That literature treats it as an optimization attacked by heuristics~\cite{fischer2013virtual}. The framing carries over. What differs is the answer's form, an if and only if in one named invariant for the class in which no cut vertex lies in two full blocks, with necessary obstructions outside it.

\begin{remark}[The two-full boundary]
\label{rem:two-full}
At exactly two full blocks per cut vertex, realizability turns from a per-vertex count into a condition on the orientation of the block--cut tree. The two full slots at a cut vertex are one parent-side slot and one descendant-side slot (Proposition~\ref{prop:blockcut-caveat}). A realization therefore orients $\mathcal{T}(H)$ globally so that every doubly-full cut vertex has one of its full blocks on its parent side. The pair of full cliques of Proposition~\ref{prop:blockcut-caveat}(ii) is realizable. What the family refuses is the same budget spread over a shared block. That refusal is a theorem, Proposition~\ref{prop:two-full-sep} in Appendix~\ref{app:hosting-proofs}. Let $A$ be a single non-full block containing two cut vertices $v \neq w$, with two further full blocks attached at $v$ and two at $w$, in which case each of $v, w$ lies in exactly two full blocks. The configuration is then realized by no member of the family. The reason is a slot count. By Lemma~\ref{lem:block-image} a full block consumes all $N{+}1$ vertices of one anchored clique. The two full blocks at $v$ therefore consume the birth clique of $v$'s chain head and the terminal anchored clique of that chain outright (Proposition~\ref{prop:blockcut-caveat}(i)), which leaves $A$ only an interior clique of that chain. An interior clique carries the two consecutive chain slots $x0^{i}, x0^{i+1}$ of $v$. The same argument at $w$ puts a second, disjoint chain's consecutive pair inside the same anchored clique, whereas an anchored clique holds only one such pair. The orientation condition above is necessary at the two-full boundary, of which Proposition~\ref{prop:two-full-sep} is the local obstruction. Whether it is also sufficient remains open. We leave its closed form aside because part~(iii) of Theorem~\ref{thm:blockcut} already covers the regime in which Appendix~\ref{app:calculus} edits hosted topologies.
\end{remark}

\subsubsection{Treewidth Necessity}
\label{subsec:treewidth}

The necessary direction is cleanest through treewidth (Lemma~\ref{lem:tw-nec} in Section~\ref{subsec:hosting-main}), which rules out large grids and $K_{N+2}$ in one stroke.

\emph{Classical counterpart.} Treewidth is the invariant of Robertson and Seymour~\cite{robertson1986graph}, the graphs of treewidth at most $k$ being the partial $k$-trees~\cite{bodlaender1998partial}. Its use as an obstruction carries over. What differs is that one bound refuses large grids and $K_{N+2}$ at once.

\subsubsection{Two Reframed Separations}
\label{subsec:examples}

The two examples below restate informal hosting claims in the realization semantics of Definition~\ref{def:realization}. Each becomes a precise statement that differs from its informal reading.

\emph{Cycle length does not govern $N$.} Let $H$ be the $8$-vertex graph formed by two hexagons sharing two opposite edges. It is a single block of order $8$. Its longest cycle has length $6$, yet by Theorem~\ref{thm:blockcut}(ii) it cannot be realized in $F(6,m)$ for any $m$ ($\beta(H)=8 > 7$). It does embed as a plain subgraph, with no duplicates, of the anchored clique $\{0\}\cup\operatorname{cell}(0) = K_8$ in $F(7,2)$. It is the block order, not the girth or the circumference, that sets the required dimension.

\emph{Depth pays what dimension cannot.} Let $H$ be $n$ triangles sharing one common cut vertex (an $n$-petal flower). Every block has order $3$. Hence $N=3$ suffices for all $n$. Yet the shared cut vertex carries $c_v = n-1$ child blocks after rooting, forcing a duplicate chain of length $n-2$ and $m = \Theta(n)$. Cell size $3$ is the smallest the family admits, so no block-order argument is at work and depth alone grows linearly (Fig.~\ref{fig:blockcut} draws the case $n = 3$). Dimension and depth are distinct resources.

\begin{remark}[Scope of static substitution]
\label{rem:static-substitution}
A realized $H$ inherits the full addressing infrastructure of the host, in that node names, the prefix hierarchy and the length-prefixed locator (Proposition~\ref{prop:locator-II}) transfer verbatim. Lowest-common-ancestor forwarding does not, since its hops are host links that $H$ need not select. The metric transfers as a quotient, in that distances in $H$ are bounded below by the distances of the host \emph{after contracting the designated chains}. For contraction-free realizations this quotient metric is $d_F$ itself (Theorem~\ref{thm:distance-II}). The raw $d_F$ is not a lower bound once a chain is contracted. In the three-petal flower the hub rides the chain $0 \sim 00$. Its distance to a third-petal node is $1$ in $H$ but $2$ in the host. What never transfers is metric equality. Because $H$ selects a subset of links, its own distances may exceed the quotient bound. Routing must therefore stay within the selected links. The calculus of Section~\ref{sec:calculus} exploits exactly the part that transfers, the address algebra, while remaining silent on the part that does not.
\end{remark}

\subsection{Proofs}
\label{app:hosting-proofs}

\begin{IEEEproof}[Proof of Lemma~\ref{lem:maxclique}]
Two nodes are adjacent iff they are siblings in one cell or joined by an anchor uplink (Proposition~\ref{thm:adjacency-II}). Let $K$ be a clique meeting two distinct cells, with $x \in K \cap \operatorname{cell}(u)$ and $y \in K \cap \operatorname{cell}(v)$, $u \neq v$. Since $x$ and $y$ are not siblings, they are joined by an uplink, say $y = xd$ (the other case is symmetric). Hence $\operatorname{cell}(v) = \operatorname{cell}(x)$ is the cell spawned by $x$. A further member $z \in K \cap \operatorname{cell}(u)$ with $z \neq x$ would satisfy $|z| = |y| - 1$ and $z \neq \operatorname{pre}(y)$. Neither clause of Proposition~\ref{thm:adjacency-II} then joins $z$ to $y$. Hence $K \cap \operatorname{cell}(u) = \{x\}$. If $K$ met a third cell, a member $t$ of it would be adjacent to $x$ through an uplink, forcing either $t \in \operatorname{cell}(x)$ or $t = \operatorname{pre}(x)$. The first contradicts the choice of a third cell. In the second $|y| - |t| = 2$, which leaves $t$ and $y$ not adjacent. Therefore $K \subseteq \{x\} \cup \operatorname{cell}(x)$. A clique inside a single cell lies in that cell's anchored clique whenever the cell has an anchor. Thus every maximal clique is an anchored clique $\{v\}\cup\operatorname{cell}(v)$, of order $N+1$, \emph{with one exception}. The initial cell has no anchor (its virtual root $\varepsilon$ of Definition~\ref{def:address-II} is not a node), which makes the tier-$1$ cell $K_N$ itself maximal and contained in no anchored clique. As for biconnectivity, each anchored clique and the tier-$1$ clique is $2$-connected, while any two share at most one node. They are therefore exactly the blocks.
\end{IEEEproof}

\begin{IEEEproof}[Proof of Lemma~\ref{lem:tree-embed}]
Map the root to a tier-$1$ node $u_0$. A node placed at a tier-$t$ address $w$ has $N$ available children in its spawn cell $\operatorname{cell}(w)$, each adjacent to $w$ by an uplink. Assign the $\le N$ tree-children of $w$ to distinct members of $\operatorname{cell}(w)$ and select exactly the uplinks $\{w, wd\}$, leaving the sibling links inside $\operatorname{cell}(w)$ unselected. Distinct spawn cells are node-disjoint and their members carry no cross edges (Proposition~\ref{thm:adjacency-II}). The selected links therefore form a subgraph copy of the tree. Depth $d$ reaches tier $d+1$. The digit-$0$ duplicate is used here as an ordinary, uncontracted child, which attains fan-out $N$ rather than $N-1$. Selection is essential, because two tree-children in one spawn cell are siblings in the host and hence adjacent there. For fan-out $3$ or more no induced copy therefore exists (the claw-freeness noted after Definition~\ref{def:realization}).
\end{IEEEproof}

\begin{IEEEproof}[Proof of Lemma~\ref{lem:block-image}]
\emph{Blocks of $S$ sit in blocks of the host.} A $2$-connected subgraph of $S$ is $2$-connected in $F(N,m)$, hence contained in one block of $F(N,m)$. By Lemma~\ref{lem:maxclique} the blocks of $F(N,m)$ are the tier-$1$ cell and the anchored cliques, all cliques of order at most $N+1$.

\emph{One chain meets one block in at most a consecutive pair.} A block $B = \{v\} \cup \operatorname{cell}(v)$ spans exactly two tiers, while a duplicate chain contains at most one vertex per tier. A chain therefore meets $B$ in at most two vertices. If in two, they are consecutive on the chain, of the form $(w, w0)$ with both in $B$, which forces $w = v$ and makes the pair the anchor--duplicate pair $(v, v0)$. The tier-$1$ cell spans one tier and meets each chain at most once. Distinct chains are vertex-disjoint and contract to distinct vertices of $H$. Within one block the only identification that a contraction can perform is therefore $v \sim v0$.

\emph{Contraction does not merge blocks.} Pass from $S$ to $H$ by contracting the chain edges one at a time. Each chain edge $e = \{w, w0\}$ lies inside the single block $B_w = \{w\} \cup \operatorname{cell}(w)$ of the host, hence inside one block $B$ of the current graph. Contracting $e$ replaces $B$ by $B/e$ (which may split into smaller blocks, as when contracting the central chord of a $4$-vertex diamond yields a $3$-vertex path of two bridges) and never merges distinct blocks across a cut vertex, since two blocks meeting at a cut vertex still meet in at most one vertex after the contraction. No $2$-connected subgraph can therefore cross between images of distinct blocks, and the terminal simple quotient (discarding loops, merging parallel edges) preserves block containment. By induction, every block of $H$ is contained in the contracted image of a single block of $S$, hence is the image of a subgraph of one host block under the identifications available inside it, namely at most the pair $v \sim v0$. An image of a subgraph of a clique of order at most $N+1$ with at most one pair identified has order at most $N+1$. Order exactly $N+1$ forces the block to occupy every vertex of a full anchored clique with no identification performed.
\end{IEEEproof}

\begin{IEEEproof}[Proof of Proposition~\ref{prop:blockcut-caveat}]
Let the hosting of $v$ be the chain $x, x0, \ldots, x0^{k}$ ($k \geq 0$, a single vertex when $k = 0$). By Lemma~\ref{lem:block-image}, a full block at $v$ occupies all vertices of an untouched anchored clique containing exactly one chain vertex. The chain meets three kinds of clique. The birth clique of $x$ contains the chain vertex $x$ only, since $x0 \notin \operatorname{cell}(\operatorname{pre}(x))$. The anchored cliques $\{x0^{i}\} \cup \operatorname{cell}(x0^{i})$ for $i < k$ each contain the \emph{two} chain vertices $x0^{i}, x0^{i+1}$, which the contraction identifies, whence their images have order at most $N$. The terminal anchored clique $\{x0^{k}\} \cup \operatorname{cell}(x0^{k})$ contains the chain vertex $x0^{k}$ only. Untouched, vertex-saturated anchored cliques therefore occur only in the two stated slots, proving (i). The count is independent of $k$. For (ii), take $S = \{a\} \cup \operatorname{cell}(a) \cup \operatorname{cell}(ad)$ with $d \neq 0$, all links among these nodes selected and $\mathcal{C} = \emptyset$. The only vertex common to the two cliques is $ad$, while no edges join $\operatorname{cell}(a) \setminus \{ad\}$ to $\operatorname{cell}(ad)$ by the two-clause law (Proposition~\ref{thm:adjacency-II}). Hence $H = S$ is exactly two full blocks glued at the cut vertex $ad$. For (iii), $k = 0$ and the block $\{x\}\cup\operatorname{cell}(x)$ contains $x0$, which means that the chain cannot be extended. Then $x$ lies in two maximal cliques only, while by the argument of Lemma~\ref{lem:block-image} every block of $H$ at $v$ lives in one of them. Every block at $v$ other than the full block, which fills $\{x\}\cup\operatorname{cell}(x)$, therefore lives in the birth clique of $x$ (several may, since selection can drop the links between them).
\end{IEEEproof}

\begin{IEEEproof}[Proof of Theorem~\ref{thm:blockcut}]
\emph{Sufficiency (i).} Process $\mathcal{T}(H)$ from the root. A block $B$ with $\le N$ vertices is a subgraph of a clique $K_N$, hence of a cell of $F(N,m)$ by Lemma~\ref{lem:maxclique}. Place $B$ inside a cell whose anchor is the image of $B$'s parent cut vertex $v$, which makes $v$'s copy adjacent to all of $B$. At a cut vertex $v$ hosted at tier $t$, the parent block already realizes $v$. The $r$-th of its $c_v$ child blocks ($r = 1, \ldots, c_v$) is hosted in the spawn cell of the $(r{-}1)$-st slot of $v$'s duplicate chain, hence on tier $t+r$. The first child block uses $v$'s own spawn cell, while each further child extends the chain by one co-located digit-$0$ slot, which spawns a fresh cell and after contraction (Definition~\ref{def:realization}) hosts its block as though attached at $v$. A root-to-leaf path therefore descends $\sum_v r_v \leq \sum_v c_v$ tiers below tier~$1$, at most its descent weight. Hence $m \geq 1 + \operatorname{dw}(H)$ suffices. The bowtie and the three-petal flower (Fig.~\ref{fig:blockcut}) are the boundary cases. The flower admits no realization at $m = 2$, which makes the bound tight there (Appendices~\ref{app:universality} and~\ref{app:verification}).

\emph{Necessity (ii).} Immediate from Lemma~\ref{lem:block-image} and Proposition~\ref{prop:blockcut-caveat}(i).

\emph{Characterization (iii).} Necessity is (ii). For sufficiency, assume every block has order at most $N+1$ and no cut vertex lies in two full blocks. If $H$ is a single block, realize it inside one anchored clique. Otherwise root $\mathcal{T}(H)$ at a non-full block if one exists. If all blocks are full, $H$ has no cut vertex, since a cut vertex would lie in two full blocks, and $H$ is again a single block. Place the root block $B_0$ (order $\leq N$) directly inside the tier-$1$ seed cell $K_N$, leaving every placed vertex's own spawn cell and duplicate chain untouched. Recursively, at a cut vertex $v$ hosted at a vertex (or chain point) with a free duplicate chain, attach the child blocks of $v$ along the chain. Each non-full child block occupies an interior chain clique, using the chain point and at most $N-1$ further members while reserving the next chain slot. The unique full child block, if present, is scheduled last and occupies the terminal anchored clique in full, with the parent cut vertex arriving at its anchor. Every non-anchor vertex of every placed block sits at a member whose own spawn cell and duplicate chain are untouched, which lets the recursion continue. In a non-root full block the one constrained vertex, the anchor, is exactly the parent cut vertex, whose remaining children were placed on earlier chain links. The result is a realization of $H$ at some finite depth $m$.
\end{IEEEproof}

\begin{proposition}[Two-full separation]
\label{prop:two-full-sep}
Let $H$ be realized by $F(N,m)$ and let $v \neq w$ be cut vertices of $H$ that lie in a common block $A$ with $|A| \leq N$. Then at least one of $v, w$ lies in at most one full block. In particular the configuration of Remark~\ref{rem:two-full}, a non-full block carrying two cut vertices with two full blocks attached at each, is realized by no member of the family.
\end{proposition}

\begin{IEEEproof}[Proof of Proposition~\ref{prop:two-full-sep}]
Suppose both $v$ and $w$ lie in two full blocks. Write $\operatorname{chain}(v) = x, x0, \ldots, x0^{k}$ and $\operatorname{chain}(w) = y, y0, \ldots, y0^{l}$ for the hosting chains, which are vertex-disjoint by Definition~\ref{def:realization}. By Lemma~\ref{lem:block-image} every full block is the isomorphic image of a spanning subgraph of one untouched anchored clique. It therefore consumes all $N{+}1$ slots of that clique, one distinct vertex per slot. By Proposition~\ref{prop:blockcut-caveat}(i) the two full blocks at $v$ are hosted in the birth clique of $x$ and in the terminal anchored clique $\{x0^{k}\}\cup\operatorname{cell}(x0^{k})$. The same holds at $w$.

The block $A$ is distinct from all four, since $|A| \leq N$. By Lemma~\ref{lem:block-image} it is hosted inside a single block $K$ of $F(N,m)$, which therefore contains a slot of $\operatorname{chain}(v)$ and a slot of $\operatorname{chain}(w)$. The blocks of $F(N,m)$ met by $\operatorname{chain}(v)$ are the birth clique of $x$, the interior anchored cliques $\{x0^{i}\}\cup\operatorname{cell}(x0^{i})$ with $i < k$ and the terminal anchored clique (proof of Proposition~\ref{prop:blockcut-caveat}(i)). Now $K$ is neither of the two consumed cliques. Indeed, distinct blocks of $H$ meet in at most one vertex, whence $A$ meets each full block at $v$ in $v$ alone, while $|A| \geq 2$ forces $A$ to occupy a slot of $K$ with a vertex other than $v$. Every slot of a consumed clique already carries a vertex of its full block. No slot carries two vertices, since the chains are pairwise disjoint. Hence $K = \{x0^{i}\}\cup\operatorname{cell}(x0^{i})$ for some $i < k$. Then $k \geq 1$ and $K$ contains the two consecutive chain slots $x0^{i}$ and $x0^{i+1}$ of $v$. Reading the same argument at $w$ gives $y0^{j}, y0^{j+1} \in K$ for some $j < l$.

Let $K = \{z\}\cup\operatorname{cell}(z)$. A pair of slots of $K$ consecutive on a duplicate chain has the form $u, u0$ with both in $K$, which forces $u = z$, because for a member $zd$ the slot $zd0$ lies outside $K$. The tier-$1$ cell contains no such pair at all. So $\{x0^{i}, x0^{i+1}\} = \{z, z0\} = \{y0^{j}, y0^{j+1}\}$. The chains of $v$ and $w$ then share a slot, contradicting their disjointness. Therefore one of $v, w$ lies in at most one full block. The configuration of Remark~\ref{rem:two-full} has $|A| \leq N$ and two full blocks at each of $v, w$. It is therefore excluded.
\end{IEEEproof}

\begin{IEEEproof}[Proof of Lemma~\ref{lem:tw-nec}]
The anchored cliques and the tier-$1$ cell form a clique tree. Take one bag per cell, containing the cell's $N$ members together with its anchor (the tier-$1$ bag has $N$ members and no anchor). Adjacent cells share exactly their common anchor node. The bags therefore satisfy the running-intersection property along the cell tree, giving a tree decomposition of width $N$ (bag size $N+1$). For $m \geq 2$ no smaller width is possible because $F(N,m)$ contains $K_{N+1}$ (Lemma~\ref{lem:maxclique}), whence $\mathrm{tw}(F(N,m)) = N$. At $m = 1$ the graph is $K_N$, of treewidth $N-1$. A realized $H$ is obtained from a subgraph by edge contraction and simple quotient, i.e.\ $H$ is a minor of $F(N,m)$. Since treewidth is minor-monotone, $\mathrm{tw}(H) \leq N$.
\end{IEEEproof}

\subsection{Constructions}

This appendix records the placement recursion behind Theorem~\ref{thm:blockcut} and three worked realizations, among them the two boundary cases named in its proof. All are reproduced by the enumeration script of Appendix~\ref{app:verification}.

\emph{Placement recursion.} Given a connected $H$ with $\beta(H)\le N$, root the block--cut tree $\mathcal{T}(H)$ at any block $B_0$. Embed $B_0$ into the tier-$1$ cell of $F(N,m)$ (Lemma~\ref{lem:maxclique}), which is a $K_N$ and hosts any block of order $\le N$. Recurse on $\mathcal{T}(H)$. For a cut vertex $v$ already placed at address $w$, its first child block is embedded in the anchored clique $\{w\}\cup\operatorname{cell}(w)$. Each further child block is embedded after extending the duplicate chain $w \to w0 \to w00 \to \cdots$ one step and using the fresh spawn cell of the new chain node, whose contraction (Definition~\ref{def:realization}) identifies it with $v$. Under the hypotheses of Theorem~\ref{thm:blockcut}(iii), a block of order $N+1$ is placed only as a terminal step consuming the full anchored clique (Proposition~\ref{prop:blockcut-caveat}). The $r$-th child block below a cut vertex sits one tier deeper than the $(r{-}1)$-st chain slot. Hence the tier reached along any root-to-leaf path is at most $1 + \sum c_v$, one more than its descent weight, which shows that every $m \ge 1+\operatorname{dw}(H)$ suffices.

\emph{Worked cases.} (a) \emph{Bowtie} (two triangles at one cut vertex) in $F(3,2)$. Triangle one is the tier-$1$ clique $\{0,1,2\}$. Triangle two is $\{0,01,02\}$ in the spawn cell of $0$. The duplicate slot $00$ is untouched, hence no contraction is needed. Every host link among the selected nodes is selected, an induced copy. (b) \emph{$K_4$ plus a pendant} in $F(4,2)$. The $K_4$ is the tier-$1$ clique. The pendant hangs from a non-duplicate child of one vertex. No duplicate slot is used, hence again no contraction is needed. (c) \emph{Three-petal flower} in $F(3,3)$. The chain $0 \to 00$ supplies one fresh cell. The three triangles therefore occupy $\{0,1,2\}$, $\{0,01,02\}$ and $\{00,001,002\}$, while the selected links are the nine triangle edges and the chain edge. Contracting $0 \sim 00$ merges the third triangle onto the cut vertex, one chain loop discarded in the simple quotient. Exhaustive search over $F(3,2)$ confirms that no realization of the flower exists at $m = 2$ (Appendix~\ref{app:verification}). The descent-weight bound $m \geq 3$ is therefore tight here.

\section{Operations Calculus, the Group $V_N$ and Lifting}
\label{app:calculus}

This appendix develops the reconfiguration operations of Section~\ref{sec:reconfig} as a calculus on addresses, identifies its invertible part with the Higman--Thompson group $V_N$, proves the coordinate persistence of Section~\ref{sec:spacetime} and states the lifting theorem for hosted topologies. The configurations and the eight operations are those of Definition~\ref{def:calc-state} and Table~\ref{tab:typing}.

\subsection{Network Dynamics: An Operations Calculus}
\label{sec:calculus}

Everything so far is statics. The generating rule of Definition~\ref{def:event} constructs fully populated networks. A deployed system is instead some partially grown, partially occupied member of the family, changing as validators churn, as capacity is added and as load is rebalanced. This section develops the dynamics as an operations calculus over the address grammar of the HSFN. The development proceeds in two movements. \emph{Kinematics} treats configurations as finite prefix-closed sets of atomically deployed cells with an occupancy marking, on which eight typed generators act (Definition~\ref{def:calc-state}, Table~\ref{tab:typing}). \emph{Dynamics} covers the graft laws, the observables, commutation, canonical words and the representation on the address tree. Section~\ref{sec:algebra} then proves Theorem~\ref{thm:vn-identification} identifying the invertible core as the Higman--Thompson group $V_N$, proves coordinate persistence (Proposition~\ref{prop:coordinate-persistence}), and proves the lifting dichotomy outside one open subcase and a boundary regime whose closed-form condition remains open. Fig.~\ref{fig:ops-filmstrip} in Section~\ref{subsec:calculus-main} shows a four-frame film strip of the calculus in action.

\subsubsection{Kinematics: Configurations and Generators}

Throughout this section $\Sigma_N = \{0, 1, \ldots, N{-}1\}$ is the digit alphabet, addresses are digit words as in Definition~\ref{def:address-II}, $\operatorname{pre}(x)$ denotes the parent word of a nonempty word $x$, $C(w) = \{wd : d \in \Sigma_N\}$ is the cell at $w$ and $\operatorname{cone}(a) = \{aw : w \in \Sigma_N^{*}\}$ is the address cone below $a$ (including $a$). We write $F(N,m)$ for the fully populated HSFN (Section~\ref{subsec:family-containments}), as in Section~\ref{sec:universality}.

\emph{Classical counterpart.} The table is a guarded operation set of the kind structured overlays publish, Chord and Pastry giving join and leave procedures whose preconditions preserve the ring or the prefix invariant~\cite{stoica2003chord,rowstron2001pastry}. Declared preconditions and restored invariants carry over. What differs is that these eight generators are cell-atomic and closed under inversion.

Three reading notes on Table~\ref{tab:typing}. First, $\mathsf{spawn}$ deploys a whole cell in one step. No generator adds a single slot. Cell-atomicity is therefore an invariant, as are prefix closure of $D$ and the disjointness $\zeta \cap (\mu \cup D) = \emptyset$ (routine induction over the rows). Second, in $\mathsf{graft}(a \to b)$ the condition $a \not\preceq b$ means that $a$ is not a prefix of $b$, so the target lies outside the moved cone. The reverse inclusion $b \preceq a$ cannot occur, since $D$ is prefix-closed and contains $a$ while $b \notin D$. The quantity $h_x(a) = \max\{|s| : s \in D \cap \operatorname{cone}(a)\} - |a| + 1$ is the height of the deployed cell subtree below $a$. The depth-fit guard therefore keeps every image cell inside $\Sigma_N^{\leq m-1}$. The effect replaces every deployed cell $aw \in D$ ($w \in \Sigma_N^{*}$) by $bw$ and transports the occupancy and decommission bits of all slots strictly below $a$. The slot $a$ itself, a member of the static cell $C(\operatorname{pre}(a))$, stays behind with its participant, while the arriving cells are re-anchored at $b$. Both anchors must be occupied, since $a$ hands the subtree off while $b$ takes custody and afterwards carries all child--parent traffic of the arriving cells (Remark~\ref{rem:comm-abstraction}). The guard $\operatorname{pre}(b) \in D$ names the parent of the target site and is what makes $b$ a slot at all. Third, the footprint column already includes the guard reads. Proposition~\ref{prop:commute} shows that this is exactly what a sound commutation criterion needs. The $\operatorname{Stab}(0)$ restriction on non-root $\mathsf{permute}$ letters is Lemma~\ref{lem:dup-digit}(ii), precisely the relabeling freedom that preserves the marked set $M(x)$. Under $\mathsf{graft}$ the marking transports by itself, since a prefix substitution changes no trailing digit (Proposition~\ref{thm:graft-iso}).

\paragraph*{Derived operations}
The eight generators form a basis, and the operations a protocol designer wants are short words in them. \emph{Migrate} is a synonym for a single graft. \emph{Swap}$(a,b)$ exchanges the anchored subnetworks and the custodians at two incomparable occupied slots. Let $c$ be a staging slot, meaning occupied, cell-free and deep enough (if none exists, $\mathsf{spawn}$ and $\mathsf{join}$ create one at the frontier). The structural part is fixed by how many of the two slots anchor a deployed cell. When both do, it is the three-graft word $\mathsf{swap}(a,b;c) = \mathsf{graft}(c \to b) \circ \mathsf{graft}(b \to a) \circ \mathsf{graft}(a \to c)$, whose middle letter reads $b \in D$ as a guard and is therefore available in this case alone. When exactly one of them does, say $a$, that middle graft has no source cell. The structural part is then the two-graft word $\mathsf{graft}(c \to b) \circ \mathsf{graft}(a \to c)$, which parks $\operatorname{cone}(a)$ at the staging slot before re-anchoring it at $b$. When neither does, the structural part is empty. In all three cases the two slots remain occupied and both anchors are fixed pointwise, the grafts moving only what hangs strictly below them. The occupancy marking alone therefore cannot say which of two occupied slots holds which participant. We accordingly declare participant identity as separate protocol metadata, an injective labeling of the occupancy set written $\mu(s)$ for the participant occupying $s \in \mu$, with the prescribed swap update $\mu' = \mu \circ (a\,b)$ transposing the custodians of $a$ and $b$ while leaving every other label alone. That update is the exchange of custodianship. The four-letter occupancy word $\mathsf{join}\,\mathsf{join}\,\mathsf{leave}\,\mathsf{leave}$, composed right to left as two leaves followed by two joins, commits it at the level of memberships while leaving the occupancy set itself unchanged. \emph{Merge} has two variants. The \emph{re-rooting} variant moves the deployed child cells of $a$ one by one, by grafts, to vacant slots of $C(b)$'s subtree (subject to slot availability), after which $a$'s emptied cell is retracted. The \emph{co-located-quotient} variant instead extends the designated duplicate chain at $b$ (Section~\ref{sec:universality}) so that the two hosted vertices reduce to one vertex of the simple quotient. This second variant is the one consumed by Theorem~\ref{thm:lifting}. \emph{Split} is its inverse, in which the child cells of one slot are partitioned, by grafts, between the slot and a staging slot. \emph{Clone}$(a \to b)$ re-spawns the deployed address pattern below $a$ underneath $b$ (structure only, participants joining separately). \emph{Isolate}$(s)$ is $\mathsf{disable}(s) \circ \mathsf{leave}(s)$, applied to an anchor only after its subtree has been grafted away or coarsened. This is the calculus-level echo of the anchor exposure of Proposition~\ref{prop:conn-price}. \emph{Heal} is $\mathsf{join} \circ \mathsf{enable}$. \emph{Refine}$(a)$ deepens locally, being $\mathsf{spawn}(a)$ followed by joins into $C(a)$. \emph{Coarsen}$(a)$ is the local inverse of refinement, namely leaves over the subtree below $a$ followed by retracts, bottom-up, collapsing the cell subtree below $a$ back into $a$ (Fig.~\ref{fig:ops-filmstrip}(d)). \emph{Embed}$(H)$, the word that deploys a realization of a target graph $H$ from scratch, is supplied by Proposition~\ref{prop:canon-word} applied to the constructions of Appendix~\ref{app:hosting}.

This calculus is specific to the node-anchored HSFN grammar. In every HSFN configuration the anchor correspondence is a bijection, in that each deployed non-seed cell has exactly one anchor node and each node anchors at most one cell ($w \mapsto C(w)$ on $D$). Because no cell attaches across multiple host vertices, prefixes substitute freely without grammar vetoes, yielding the unrestricted suffix covariance of Proposition~\ref{thm:graft-iso}.

\begin{table}[!t]
\centering
\caption{Operational Dictionary. Each row names a maintenance operation of a deployed system, the calculus word or derived operation that performs it here and the result that guarantees the outcome. The table carries no numerical claim.}
\label{tab:opdict}
\footnotesize
\setlength{\tabcolsep}{2pt}
\renewcommand{\arraystretch}{1.25}
\begin{tabular}{@{}>{\raggedright}p{0.27\columnwidth}>{\raggedright}p{0.27\columnwidth}>{\raggedright\arraybackslash}p{0.40\columnwidth}@{}}
\toprule
Operation elsewhere & Calculus word & Guarantee \\
\midrule
Chord node join~\cite{stoica2003chord} & $\mathsf{spawn}$ then $\mathsf{join}$ & cell-atomic deployment under declared guards, the fresh slots adjacent by Proposition~\ref{thm:adjacency-II} (Table~\ref{tab:typing}) \\
Chord node departure~\cite{stoica2003chord} & $\mathsf{leave}$, then $\mathsf{retract}$ once the cell is empty & exact inverses of the arrival generators (Theorem~\ref{thm:rep-tree}(i)) \\
Chord stabilize and fix-fingers~\cite{stoica2003chord} & none for adjacency & adjacency and distance are computed from the two addresses, while occupancy is committee state (Proposition~\ref{thm:adjacency-II} and Theorem~\ref{thm:distance-II}) \\
Pastry node arrival~\cite{rowstron2001pastry} & $\mathsf{stage}$ then $\mathsf{join}$ & the staging word supplies an anchored clique of order $N{+}1$ adjacent to the host chain (Lemma~\ref{lem:staging}) \\
Kademlia bucket refresh~\cite{maymounkov2002kademlia} & none for adjacency & no adjacency state is cached, hence none can go stale, while occupancy is committee state (Proposition~\ref{thm:adjacency-II}) \\
Migration of a virtual node~\cite{yu2008rethinking,fischer2013virtual} & $\mathsf{graft}$ & prefix substitution, the moved set arriving isomorphic and marked (Proposition~\ref{thm:graft-iso}), boundary exactly $N$ uplinks (Lemma~\ref{lem:edge-cut}) \\
Re-embedding of a virtual link~\cite{yu2008rethinking} & mask update, with grafts when the block must be re-hosted & edit transfer with the lifting dichotomy (Theorem~\ref{thm:lifting}) \\
Load-balancing swap~\cite{karger2004simple} & $\mathsf{swap}(a,b)$, three grafts when both slots anchor cells & a staged transposition of incomparable cones. These transpositions generate the invertible dynamics (Lemma~\ref{lem:transp-gen}, Theorem~\ref{thm:vn-identification}) \\
\bottomrule
\end{tabular}
\end{table}

\subsubsection{The Graft Laws}

The graft generator carries the structural load of the calculus. Its laws are exact. Two graphs are in play. The \emph{state graph} $G(x)$ is the physical graph on $\operatorname{Sl}(x)$ with the sibling cliques and anchor uplinks of Section~\ref{sec:graphs}. Its adjacency is the two-clause law of Proposition~\ref{thm:adjacency-II} restricted to the deployed slots. The \emph{deployed-cell graph} $\Gamma(x)$ has vertex set $D$ (the seed cell $\varepsilon$ is a vertex, the base case that keeps the boundary law uniform) and an edge between $w \neq \varepsilon$ and $\operatorname{pre}(w)$ (\emph{parent edges}) and between deployed sibling cells $wd, wd' \in D$ (\emph{sibling edges}). Away from $\varepsilon$, $\Gamma(x)$ is nothing new. Two deployed cells are $\Gamma$-adjacent iff their anchor nodes are adjacent in $G(x)$. Hence $\Gamma(x) - \varepsilon$ is the subgraph of $G(x)$ induced on the anchors, the \emph{anchor overlay} (this identification is re-proved inside Proposition~\ref{prop:kpi-affine}).

\begin{lemma}[Boundary edge-cut law]
\label{lem:edge-cut}
Let $\mathsf{graft}(a \to b)$ be applicable at $x$ with $|a| = |b|$, let $x'$ be the result and let the \emph{moved set} be the deployed cells in $\operatorname{cone}(a)$ (in $x$) and their images in $\operatorname{cone}(b)$ (in $x'$). Write $\sigma_x(a) = |\{d \in \Sigma_N : \operatorname{pre}(a)d \in D,\ \operatorname{pre}(a)d \neq a\}|$ for the number of deployed sibling cells of $a$.
\begin{enumerate}
    \item \emph{Node level.} In $G(x)$, exactly $N$ edges join the moved slots to the rest, namely the $N$ uplinks from $C(a)$ into the anchor $a$. In $G(x')$ the created boundary is likewise the $N$ uplinks into $b$.
    \item \emph{Cell level.} In $\Gamma(x)$, the graft cuts exactly $1 + \sigma_x(a)$ edges, all incident to the moved root cell $a$, namely the parent edge $\{a, \operatorname{pre}(a)\}$ and the $\sigma_x(a)$ sibling edges at $a$. In $\Gamma(x')$ it creates exactly $1 + \sigma_{x'}(b)$ edges, all incident to $b$, namely the parent edge $\{b, \operatorname{pre}(b)\}$ and the sibling edges at $b$. For $|a| = 1$ the parent edge is the edge to the seed cell $\varepsilon$, which is why $\varepsilon$ is a vertex of $\Gamma$. All other edges of $\Gamma$ are untouched. Edges with both endpoints outside the moved set are equal in $\Gamma(x)$ and $\Gamma(x')$, while the substitution maps the internal edges bijectively.
\end{enumerate}
\end{lemma}

\emph{Classical counterpart.} Moving a subtree by relinking its root is the rope operation, a split or a concatenation touching $O(\log n)$ pointers and never reading the moved interior~\cite{boehm1995ropes}. Locality carries over, with the cost given by the boundary rather than the size. What differs is that the boundary is an identity, exactly $N$ uplinks and $1 + \sigma_x(a)$ cell edges.

\begin{IEEEproof}[Proof of Proposition~\ref{thm:graft-iso}]
(i) is the definition of the effect in Table~\ref{tab:typing} read off. The substitution acts on the prefix only, while $M(x)$ is the set of deployed addresses with trailing digit $0$ below a deployed parent, a property of the suffix.

(ii) Two moved slots have equal length iff their images do. They agree on their first $t{-}1$ digits iff their images do. Write $u = aw_1$ and $v = aw_2$. If $|w_1| = |w_2|$ the compared prefixes are $a\operatorname{pre}(w_1)$ and $a\operatorname{pre}(w_2)$, which agree iff $\operatorname{pre}(w_1) = \operatorname{pre}(w_2)$, a condition invariant under $a \mapsto b$. Clause~(P) likewise reads $w_2 = w_1 d$, invariant under $a \mapsto b$. By Proposition~\ref{thm:adjacency-II} adjacency among deployed slots is decided by these clauses. Hence edges and non-edges correspond.

(iii) With $|a| = |b|$ the substitution preserves address length, which leaves per-tier cell counts and per-tier occupancy histograms unchanged (moved cells carry their memberships, static cells keep theirs). The four affine indicators of Proposition~\ref{prop:kpi-affine} that factor through them are therefore constant. For the anchoring histograms, a cell outside the moved set has the same deployed children after the graft except for the two boundary parents, where $\operatorname{pre}(a)$ loses the child cell $a$ and $\operatorname{pre}(b)$ gains the child cell $b$. If $\operatorname{pre}(a) = \operatorname{pre}(b)$ the two changes cancel inside one cell. Moved cells keep their child counts by (i). The anchor-overlay statement is Lemma~\ref{lem:edge-cut}(ii) combined with the identification of $\Gamma(x) - \varepsilon$ with the anchor overlay.
\end{IEEEproof}

\emph{Classical counterpart.} The generator acts as the prefix substitution $s_b s_a^{*}$ of the polycyclic inverse monoid~\cite{lawson1998inverse}, its operational picture the rope relink~\cite{boehm1995ropes}. An untouched interior carries over. What differs is that a rope keeps its subtree's names while a graft substitutes the prefix, which is why the arrival is an isomorphism.

\emph{Running example ($N = 5$, $m = 3$).} In $F(5,3)$, take the configuration with $D = \{\varepsilon, 1, 4, 42, 44\}$, whose $5$ deployed cells provide $25$ slots. Let all anchors and the members of $C(42)$ be occupied. The graft $42 \to 13$ is applicable once $13$ is occupied, since $|42| = |13| = 2$, cell $13$ is undeployed and its parent cell $C(1)$ is deployed. The moved set is the single cell $42$ with its five members $420, \ldots, 424$, which the substitution renames $130, \ldots, 134$. The slot $420$, the designated duplicate of $42$, becomes $130$, the designated duplicate of $13$. At the node level exactly $N = 5$ uplinks are cut (into $42$) and five created (into $13$). In the deployed-cell graph $\sigma_x(42) = 1$ (the deployed sibling cell $44$), which cuts $1 + 1 = 2$ edges, $\{42, 4\}$ and $\{42, 44\}$. At the target $\sigma_{x'}(13) = 0$, which creates exactly one edge, $\{13, 1\}$. The tier-$2$ anchoring histogram shows the two-bin transfer $\kappa(4)\colon 2 \to 1$, $\kappa(1)\colon 0 \to 1$. Every other indicator of Proposition~\ref{prop:kpi-affine} is unchanged.

\begin{remark}[Grafts splice the update cycle]
\label{rem:graft-splice}
The canonical update order of Proposition~\ref{thm:update-main} is a depth-first traversal of the cell tree. On a partially deployed configuration the digit-ordered depth-first list of the deployed cells (skipping absent cells) plays the same role. Because a depth-first list places the members of a cell subtree contiguously, the moved set of a graft occupies one contiguous block of the cycle. By Proposition~\ref{thm:graft-iso}(i)--(ii) the successor relation \emph{inside} the block is untouched, in that addresses change while the order does not. A graft therefore repairs the cycle with at most three successor updates, one pointer closing the gap at the source and two pointers splicing the block in at the target position determined by the digit of $b$. When the block re-enters at the position it vacated, no pointer changes at all.
\end{remark}

\subsubsection{Observables: What Is Affine and What Is Not}

A control plane steering the calculus needs the network's indicators after every operation. The right bookkeeping objects are two families of per-tier histograms. For a configuration $x$ and tier $t \in \{1, \ldots, m\}$, the tier-$t$ cells are the $w \in D$ with $|w| = t{-}1$. Write $M0_t$ for their number and $o_t(k)$ for the number of them with exactly $k$ occupied members ($k = |C(w) \cap \mu|$, the \emph{occupancy histograms}). Write $h_t(k)$ for the number of them with exactly $k$ deployed child cells ($k = \kappa(w) = |\{d : wd \in D\}|$, the \emph{anchoring histograms}). The \emph{tier profile} of $x$ is the family $\{M0_t, o_t, h_t\}_t$. We write $M1, M2$ for first and second moments, e.g., $M2(h_t) = \sum_k k^2 h_t(k)$.

\begin{proposition}[Affine indicators and $O(1)$ maintenance]
\label{prop:kpi-affine}
For every configuration $x$ of $F(N,m)$:
\begin{enumerate}
    \item \emph{Slot and participant counts.}
    $|\operatorname{Sl}(x)| = N \sum_t M0_t$ and $|\mu| = \sum_t M1(o_t)$.
    \item \emph{Provisioned links.} The state graph $G(x)$ has
    \begin{equation*}
        L(x) \;=\; \Bigl(\tbinom{N}{2} + N\Bigr) \sum_t M0_t \;-\; N
    \end{equation*}
    edges.
    \item \emph{Message proxies.} The full-occupancy hierarchical-PBFT proxy of Proposition~\ref{thm:complexity} evaluates to $(N{+}1)\, N \sum_t M0_t$. The occupancy-refined proxy, each cell contributing the PBFT-scale square of its live membership plus one aggregation message per participant, is
    \begin{equation*}
        C_{\mathrm{load}}(x) \;=\; \sum_t \sum_k \bigl(k^2 + k\bigr)\, o_t(k).
    \end{equation*}
    Under full occupancy ($o_t(N) = M0_t$), $C_{\mathrm{load}}$ reduces directly to $(N{+}1)V$. The executed count of Proposition~\ref{prop:exec-count} is a distinct protocol quantity given for arbitrary occupancy by Proposition~\ref{thm:committee-tree}.
    \item \emph{Anchor overlay.} $\Gamma(x) - \varepsilon$ is the subgraph of $G(x)$ induced on the anchor nodes. Its link count is
    \begin{equation*}
        L_{\mathrm{anch}}(x) \;=\; \sum_t \sum_k \tbinom{k}{2}\, h_t(k) \;+\; \sum_{t \geq 3} M0_t .
    \end{equation*}
\end{enumerate}
All four displays are affine functionals of the tier profile. Every generator updates the profile in $O(1)$ arithmetic operations for fixed $N$. A $\mathsf{join}$ or $\mathsf{leave}$ makes one two-bin transfer in one $o_t$. A $\mathsf{spawn}$ or $\mathsf{retract}$ makes one two-bin transfer in $h_{|a|}$ plus one bin at $(o_{|a|+1}(0), h_{|a|+1}(0))$. An equal-tier $\mathsf{graft}$ makes the two-bin transfer of Proposition~\ref{thm:graft-iso}(iii). $\mathsf{permute}$, $\mathsf{disable}$ and $\mathsf{enable}$ make none. The one exception is a cross-tier graft, which translates the moved subtree's rows across $|b| - |a|$ tiers in $O(m)$ operations.
\end{proposition}

\emph{Classical counterpart.} The statement is incremental view maintenance, refreshing an aggregate from the update rather than from the data~\cite{gupta1993maintaining}. That criterion carries over. What differs is that the tier profile is identified as the sufficient statistic, in which four named indicators are affine.

\emph{The affine indicators need only the moments $(M0, M1, M2)$.} For the four indicators of Proposition~\ref{prop:kpi-affine} the histograms may be compressed to their first three moments. Slot count, provisioned links and the full-occupancy proxy factor through $M0$ alone. The participant count is $\sum_t M1(o_t)$. The second moments enter in exactly two places, $M2(o_t)$ through the quadratic intra-cell load term of $C_{\mathrm{load}}$ and $M2(h_t)$ through the pair count $\sum_k \binom{k}{2} h_t(k) = \tfrac{1}{2}(M2 - M1)(h_t)$ of the anchor overlay. Decommissioned slots are vacant and anchor nothing by invariant. Hence $\zeta$ enters no primary indicator.

Reliability is the deliberate exception. It is an observable of the calculus, but not an affine one. Indeed it is not a function of the tier profile at all.

\begin{proposition}[Reliability, exact recursion without a profile formula]
\label{prop:rel-exact}
Equip a configuration $x$ with the FPD reliability model of Section~\ref{sec:security} restricted to deployed cells (Remark~\ref{rem:comm-abstraction}), in which each deployed cell fails internally with probability $q$, independently, while a cell with $c$ deployed child cells is faulty iff either its internal consensus fails or at least $\lceil N/2 \rceil$ of its deployed child cells are faulty, a fixed quorum that coincides with the half rule of Section~\ref{sec:security} when all $N$ children are deployed (childless cells fail internally only). Here $R(x)$ is the probability that the seed cell is not faulty. Then:
\begin{enumerate}
    \item \emph{No profile formula.} $R$ is not a function of the tier profile, a fortiori not affine in it. A witness pair (Appendix~\ref{app:verification}) is the following. For $N = 3$ and $q = 0.20$, deploy all cells through tier~$3$ ($1 + 3 + 9$ cells) and fully expand three tier-$3$ cells ($9$ tier-$4$ cells), all slots occupied. Placing the three heavy cells one per tier-$2$ branch gives $R = 0.6226$, whereas placing all three under one branch gives $R = 0.6202$. The two configurations nonetheless have identical per-tier cell counts $(1,3,9,9)$ and identical per-tier anchoring and occupancy histograms. A companion pair for $N = 4$, $q = 0.125$ ($1 + 4 + 16 + 32$ cells, eight heavy tier-$3$ cells, spread $(2,2,2,2)$ versus $(4,4,0,0)$) yields $R = 0.6737$ versus $0.6691$. In both pairs the balanced allocation is the more reliable, a design rule the profile cannot see.
    \item \emph{Exact evaluation.} The bottom-up recursion
    \begin{equation*}
        p_w \;=\; q + (1 - q) \Pr\Bigl[\textstyle\sum_{d :\, wd \in D} X_{wd} \geq \lceil N/2 \rceil\Bigr],
    \end{equation*}
    with independent indicators $\Pr[X_{wd} = 1] = p_{wd}$ evaluated by Poisson--binomial convolution, computes $R(x) = 1 - p_\varepsilon$ exactly in $O(N |D|) = O(V)$ arithmetic operations. With the fixed quorum $\lceil N/2 \rceil$ used here, the recursion agrees with the level iterations $f_{b_k}$ of Theorems~\ref{thm:security} and~\ref{prop:threshold-b} at cells whose $N$ child cells are all deployed and is a separate model elsewhere. Every generator invalidates only the $p_w$ on the anchor paths of its footprint (one path of length at most $m$, or two paths for a graft). Recomputing Poisson--binomial convolutions directly along each affected path takes $O(N^2 m)$ arithmetic operations in the worst case (or $O(N m)$ when per-cell child-polynomial coefficients are cached and updated by removing/inserting at most two Bernoulli factors with separate zero/one factor counts).
    \item \emph{Monotone interval bounds from the profile.} Define from the tier profile alone $\overline{P}_T = \underline{P}_T = q$ at the deepest deployed tier $T$ and, downward in $t$,
    \begin{equation*}
    \begin{split}
        \overline{P}_t &= \max\{F_k(\overline{P}_{t+1}) : h_t(k) > 0\}, \\
        \underline{P}_t &= \min\{F_k(\underline{P}_{t+1}) : h_t(k) > 0\},
    \end{split}
    \end{equation*}
    where $F_k(p) = q + (1-q) \Pr[\mathrm{Bin}(k, p) \geq \lceil N/2 \rceil]$. Then every tier-$t$ deployed cell has failure probability in $[\underline{P}_t, \overline{P}_t]$. In particular $1 - \overline{P}_1 \leq R(x) \leq 1 - \underline{P}_1$, computable in $O(mN)$ from the histograms.
\end{enumerate}
\end{proposition}

\emph{Classical counterpart.} Network reliability is classically hard, the probability that a graph stays connected being $\#\mathrm{P}$-complete~\cite{provan1983complexity}, while series-parallel structures yield to one bottom-up pass~\cite{moore1956reliable}. Independence across subtrees carries over and gives item~2. What differs is item~1, the witness pairs making the answer depend on the shape of the deployment.

\subsubsection{Commutation, the Address Monoid and Canonical Words}

\begin{proposition}[Footprint-disjoint operations commute]
\label{prop:commute}
Let $\omega_1, \omega_2$ be generators whose footprints (Table~\ref{tab:typing}, guard reads included) are disjoint. Then $\omega_1 \omega_2 = \omega_2 \omega_1$ as partial maps, in that the two composites have the same domain and agree on it.
\end{proposition}

\emph{Classical counterpart.} Making concurrent operations commute is the aim of operational transformation~\cite{ellis1989concurrency} and of conflict-free replicated data types~\cite{shapiro2011conflict}, the same test appearing in databases as read--write conflict serializability. Disjoint footprints commuting carries over. What differs is which accesses count, since the pair exhibited below has disjoint write sets and still fails to commute.

\emph{Guard reads are load-bearing.} The criterion would be unsound with write-sets alone. One pair shows why. Let $b$ be an occupied, cell-free slot and $a \in D$ a movable cell root elsewhere. Compare $\mathsf{graft}(a \to b)$ with $\mathsf{leave}(b)$, the departure of the anchor of the target site, which is the parent of the arriving cell in the cell tree. Their \emph{written} attributes are disjoint, since the graft rewrites cell structure and transported bits inside the two cones and leaves the occupancy bit of $b$ untouched. The leave rewrites exactly that bit. Yet the two do not commute as partial maps, since $\mathsf{leave}(b)$ followed by $\mathsf{graft}(a \to b)$ is undefined (the guard $b \in \mu$ fails), while $\mathsf{graft}(a \to b)$ followed by $\mathsf{leave}(b)$ is defined and strands the arriving subtree on a vacated anchor. The conflict is invisible to any write--write test and is caught precisely because the footprint of the graft \emph{reads} $b$'s occupancy. That is why Table~\ref{tab:typing} charges guard reads to the footprint and why the stranded-anchor state is flagged as a protocol-layer concern in the execution model below.

The generators are partial maps of a special kind. Each is injective with a two-sided partial inverse in the calculus, which places the whole apparatus inside inverse semigroup theory. Through the address tree it lands in a classical object, the polycyclic inverse monoid $P_N$ of Nivat and Perrot presented by $s_i^{*} s_j = \delta_{ij}$ on generators $s_0, \ldots, s_{N-1}$. Its nonzero elements have the normal form $s_u s_v^{*}$ and act on the $N$-ary address tree by prefix substitution $v w \mapsto u w$~\cite{lawson1998inverse}. Call a word \emph{rigid} if it contains no $\mathsf{permute}$ letter. Let $B_N$ denote the Boolean inverse monoid of partial injections of $\Sigma_N^{*}$ expressible as finite disjoint unions of prefix substitutions $s_u s_v^{*}$ restricted to finite Boolean combinations of cones, a realization of the Booleanization of $P_N$~\cite{lawson1998inverse}. By Lemma~\ref{lem:family-mono} a word of $F(N,m)$ is a word of every $F(N, m')$ with $m' \geq m$. Hence its action on addresses is meaningfully a partial map of the full tree $\Sigma_N^{*}$.

\begin{theorem}[Representation on the address tree]
\label{thm:rep-tree}
Fix $N \geq 3$, $m \geq 2$.
\begin{enumerate}
    \item \emph{Inverse structure.} Every generator is an injective partial map of the configuration space with a two-sided partial inverse in the calculus:
    $\mathsf{retract}(a) = \mathsf{spawn}(a)^{-1}$,
    $\mathsf{leave}(s) = \mathsf{join}(s)^{-1}$,
    $\mathsf{enable}(s) = \mathsf{disable}(s)^{-1}$,
    $\mathsf{permute}(w, \tau)^{-1} = \mathsf{permute}(w, \tau^{-1})$ and
    $\mathsf{graft}(a \to b)^{-1} = \mathsf{graft}(b \to a)$.
    Consequently the partial maps defined by words form an inverse monoid $\mathcal{M}(N,m)$ of partial bijections of the configuration space.
    \item \emph{Prefix-substitution cocycles.} Each rigid word representative $w$ induces an element $\Phi(w) \in B_N$ such that for every $x \in \operatorname{dom}(w)$ and every slot $s$ of $x$ that survives $w$ (is not retracted), the address of $s$ in $w(x)$ is $\Phi(w)(s)$. On generators, $\Phi(\mathsf{graft}(a \to b))$ restricts on the strict open cone $a\Sigma_N^{+}$ to the polycyclic element $s_b s_a^{*}$ and is the identity off $a\Sigma_N^{+} \cup b\Sigma_N^{+}$ (fixing the finite anchors $a$ and $b$). The six bookkeeping generators have $\Phi = \mathrm{id}$. Because the shadow tracks the word representative rather than only its structural partial map on configurations, two rigid words that agree on configurations can carry distinct partial shadows where guards exclude witnesses; the assignment is multiplicative up to domain restriction, in that $\Phi(w_2 w_1)$ agrees with $\Phi(w_2) \Phi(w_1)$ at every address deployed and surviving through the composite.
    \item \emph{Faithfulness on relocation.} $\Phi$ is sound for relocation, in that two rigid words whose $\Phi$-elements agree on the addresses deployable in $F(N,m)$ relocate every surviving slot identically on every state they share. The converse holds only where the shared states supply a witness. If rigid words $w$ and $w'$ share a state with a slot $p\,d$ that survives both of them, then the address of that slot after execution is $\Phi(w)(p\,d)$ on one side and $\Phi(w')(p\,d)$ on the other. Relocating it identically therefore forces the two elements to agree at $p\,d$. Sharing a state does not by itself supply such a witness for every deployable address, because the vacancy guards of a word can exclude every configuration with a slot below a prescribed address. The claim is therefore agreement exactly at the addresses that some shared state witnesses. The kernel of the representation is pure bookkeeping, since rigid words with $\Phi(w) \leq \mathrm{id}$ act by deployment, occupancy and decommission changes only, never by renaming.
\end{enumerate}
\end{theorem}

\begin{IEEEproof}[Proof sketch]
(i) Each listed pair is checked guard-against-effect. The invariant $\zeta \cap (\mu \cup D) = \emptyset$ and the vacancy guards make each pair exact. After $\mathsf{spawn}(a)$, for instance, the new cell is vacant, enabled and childless, which is precisely the guard of $\mathsf{retract}(a)$. After $\mathsf{graft}(a \to b)$ the guards of $\mathsf{graft}(b \to a)$ hold with the roles of the two occupied anchors exchanged. A set of partial bijections closed under composition and inversion generates an inverse monoid inside the symmetric inverse monoid of the configuration space. (ii) Induction on the word length. The generator cases are the two displayed clauses. The composite of two elements of $B_N$ is again a finite union of restricted prefix substitutions, because cones pull back under $s_u s_v^{*}$ to cones or points and Boolean combinations are preserved. The complement of finitely many cones is again a finite union of cones and points, which keeps the off-cone identity inside $B_N$. (iii) Soundness is (ii) read slot by slot. For the partial converse, let $x$ be a state shared by $w$ and $w'$ with a slot $p\,d$ surviving both. By (ii) the address of that slot in $w(x)$ is $\Phi(w)(p\,d)$ while its address in $w'(x)$ is $\Phi(w')(p\,d)$, which pins the two elements together at $p\,d$ whenever the two relocations agree. When the guards leave no shared state with such a slot, the shared states say nothing about $\Phi$ at $p\,d$. Full bookkeeping for (i)--(iii) is in Appendix~\ref{app:calculus}.
\end{IEEEproof}

\emph{Classical counterpart.} The target is the polycyclic inverse monoid $P_N$ of Nivat and Perrot, whose elements $s_u s_v^{*}$ act on the $N$-ary tree by prefix substitution~\cite{lawson1998inverse}. The normal form carries over. What differs is that the six bookkeeping generators have $\Phi = \mathrm{id}$, which gives the representation a kernel.

\emph{Permutations, the wreath layer.} The excluded letters are exactly the automorphism layer. A $\mathsf{permute}(w, \tau)$ letter acts on addresses as a cell-wise relabeling. The group generated by all of them on the depth-$m$ tree is the iterated permutational wreath product of copies of $S_N$, the automorphism group of the rooted $N$-ary tree in Nekrashevych's self-similar-group formalism~\cite{nekrashevych2005self}. The duplicate designation restricts it to $\operatorname{Stab}_{S_N}(0) \cong S_{N-1}$ in every non-root coordinate (Lemma~\ref{lem:dup-digit}(ii)). General words therefore represent into the inverse monoid generated by $B_N$ together with this marked wreath group. We keep the two layers separate because they answer different questions, $B_N$ moving structure and the wreath layer renaming it.

The realization map $\iota$ of Section~\ref{sec:realizations} carries each prefix substitution to a partial similarity of $\mathbb{R}^{N-1}$, branch by branch on the labeled cells rather than on the glued point set. On the labeled cells a one-digit extension acts by the corner similitude of Theorem~\ref{thm:dim-II} attached to that cell's local digit order, which makes $\Phi(w)$ a family of partial similarities indexed by branches. This family descends to a single-valued map of the self-similar limit set $\mathcal{S}\Delta_N$ only when it preserves the fibers of $\iota$, which a graft need not do. The map $\iota$ collapses every designated duplicate onto its anchor and identifies the horizontally coincident vertices contributed by neighboring cells, while a graft may move one member of such a pair and fix the other (Lemma~\ref{lem:dup-digit}). Distinct calculus states and distinct relocations can therefore share one Euclidean picture. The address tree, not the embedding, is the faithful carrier of the dynamics.

\begin{proposition}[Canonical words and fixed-origin decidability]
\label{prop:canon-word}
Order addresses length-lexicographically.
\begin{enumerate}
    \item Every configuration $x = (D, \mu, \zeta)$ of $F(N,m)$ is reachable from the seed configuration $x_0$ by the \emph{canonical word}
    \begin{equation*}
        w_{\mathrm{can}}(x):\;
        \text{spawn } D \setminus \{\varepsilon\};\;
        \text{join } \mu;\;
        \text{disable } \zeta,
    \end{equation*}
    each block in length-lexicographic order. Conversely every state reachable from $x_0$ is a configuration. The state space of the calculus is therefore exactly the configuration space, with $w_{\mathrm{can}}$ a normal form for states.
    \item Equivalence relative to a fixed origin is decidable by evaluation. Given words $w, w'$ and a configuration $x$, executing both decides whether they are applicable at $x$ with $w(x) = w'(x)$, in time linear in the word lengths times the footprint sizes, because configurations are finite objects.
\end{enumerate}
\end{proposition}

\emph{Classical counterpart.} Reaching a unique reduced word by a fixed order is the aim of Knuth--Bendix completion~\cite{knuth1970simple}. In the address shadow the polycyclic normal form $s_u s_v^{*}$ plays that part~\cite{lawson1998inverse}. A length-lexicographic canonical representative carries over. What differs is that $w_{\mathrm{can}}$ normalizes states rather than words.

Agreement of two words as partial maps on \emph{every} configuration is a rewriting question we leave open. The address shadow suggests the route. Rigid words project by Theorem~\ref{thm:rep-tree} onto $B_N$, where the polycyclic normal form $s_u s_v^{*}$ and its Boolean joins are complete invariants~\cite{lawson1998inverse}. What remains is a confluent treatment of the bookkeeping letters against Proposition~\ref{prop:commute}'s footprint criterion, which is a sufficient but deliberately incomplete commutation test. A confluence theorem remains open.

\subsubsection{Execution Model}

Every calculus operation is proposed and committed by local PBFT of the affected cells. A $\mathsf{spawn}$, $\mathsf{join}$, $\mathsf{leave}$, $\mathsf{disable}$ or $\mathsf{enable}$ affects the single cell whose state it touches. A $\mathsf{graft}(a \to b)$ affects the two committees $C(\operatorname{pre}(a))$ and $C(\operatorname{pre}(b))$ hosting the anchors. A swap is therefore gated by the committees of both endpoints, so that a repositioning commits only with the agreement of both. The calculus specifies atomic before/after states. The transient interleavings of a distributed commit, the reliability of a subtree \emph{during} a graft and the stranded-anchor states flagged after Proposition~\ref{prop:commute} are protocol-layer concerns, deliberately outside the static-epoch theory (Remark~\ref{rem:comm-abstraction}) and deferred to the follow-up work anticipated in Section~\ref{sec:discussion}. The motivating scenarios are the usual three, namely validator churn, elastic scaling of capacity and shard rebalancing. Each is a sentence at the protocol layer and a short word in the calculus (Fig.~\ref{fig:ops-filmstrip}).

\subsection{The Algebra of the Dynamics and the Lifting Theorem}
\label{sec:algebra}

The representation of Theorem~\ref{thm:rep-tree} places the calculus inside a classical algebra. This section identifies that algebra exactly and shows that the spacetime coordinates of Section~\ref{subsec:spacetime} are equivariant under it. It then returns to the hosting theory of Section~\ref{sec:universality} to ask how a hosted topology evolves and proves the lifting dichotomy outside one open subcase and a boundary regime whose closed-form condition remains open.

\subsubsection{The Invertible Dynamics Is Thompson's Group \texorpdfstring{$V_N$}{VN}}
\label{subsec:vn}

Let $\partial T = \Sigma_N^{\omega}$ be the boundary of the address tree, the Cantor space of infinite digit strings with the cone topology. Every element of $B_N$, being a finite union of prefix substitutions restricted to Boolean combinations of cones, acts canonically on $\partial T$. We call the induced partial injection of $\partial T$ the \emph{boundary shadow} of a rigid word. On the boundary, $B_N$ is precisely the \emph{Cuntz inverse monoid} $C_N$ of Lawson, the tight (orthogonal) completion of the polycyclic inverse monoid $P_N$, whose nonzero elements are the finite compatible joins of restricted prefix substitutions and whose group of units is the Higman--Thompson group $G_{N,1} = V_N$~\cite{lawson2007polycyclic,lawson2020revisited}. Elements of $V_N$ are the table bijections of $\partial T$, given by two complete prefix codes $\{u_1, \ldots, u_k\}$, $\{w_1, \ldots, w_k\}$ and the map $u_i \xi \mapsto w_i \xi$. For incomparable words $a, b$ (neither a prefix of the other), the \emph{transposition} $\tau_{a,b} \in V_N$ exchanges the two cones by prefix substitution, $a\xi \mapsto b\xi$ and $b\xi \mapsto a\xi$, fixing everything else~\cite{bleak2017infinite}.

\begin{figure}[t]
\centering
\includegraphics[width=0.95\linewidth]{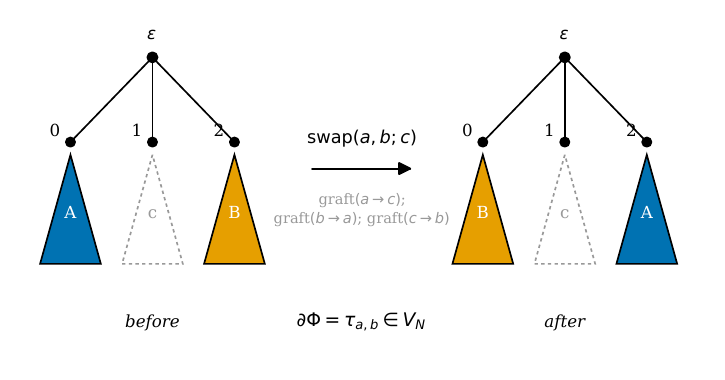}
\caption{The derived swap as a partial cone transposition (Lemma~\ref{lem:bridge}). Three grafts through a staging slot $c$ exchange $\operatorname{cone}(a)$ and $\operatorname{cone}(b)$ off $\operatorname{cone}(c)$; joining two differently staged swaps yields the total transposition $\tau_{a,b} \in V_N$, and these transpositions generate $V_N$ (Theorem~\ref{thm:vn-identification}).}
\label{fig:swap}
\end{figure}

\begin{lemma}[Bridge, swaps are transpositions]
\label{lem:bridge}
Let $a, b$ be incomparable addresses (Fig.~\ref{fig:swap}).
\begin{enumerate}
    \item[(i)] The boundary shadow of every rigid word lies in $C_N$.
    \item[(ii)] For any staging address $c$ incomparable to both $a$ and $b$, the derived word $\mathsf{swap}(a,b;c) = \mathsf{graft}(c \to b) \circ \mathsf{graft}(b \to a) \circ \mathsf{graft}(a \to c)$ has boundary shadow $\tau_{a,b}$ restricted to $\partial T \setminus \operatorname{cone}(c)$.
    \item[(iii)] For two staging addresses $c_1 \neq c_2$, incomparable to each other and to $a, b$, the shadows of $\mathsf{swap}(a,b;c_1)$ and $\mathsf{swap}(a,b;c_2)$ are compatible and their join in $C_N$ is exactly $\tau_{a,b}$.
\end{enumerate}
\end{lemma}

\begin{IEEEproof}
(i) is Theorem~\ref{thm:rep-tree}(ii) read on the boundary, since a finite union of prefix substitutions restricted to Boolean combinations of cones is a compatible join of restricted polycyclic elements, which is an element of $C_N$~\cite{lawson2020revisited}.

(ii) Track the three grafts on the boundary. The shadow of $\mathsf{graft}(x \to y)$ is $s_y s_x^{*}$ on $\operatorname{cone}(x)$, the identity off $\operatorname{cone}(x) \cup \operatorname{cone}(y)$ and undefined on $\operatorname{cone}(y)$ (the guard $y \notin D$ keeps the target cone empty, which prevents any slot from being relocated into the domain there). Composing right to left, a point $a\xi$ goes to $c\xi$ under the first graft, is untouched by the second and goes to $b\xi$ under the third. A point $b\xi$ is untouched by the first, goes to $a\xi$ under the second and is untouched by the third. A point of $\operatorname{cone}(c)$ is outside the domain of the first graft, while every other point is fixed by all three. The composite is therefore $\tau_{a,b}$ restricted to $\partial T \setminus \operatorname{cone}(c)$.

(iii) The two shadows agree with $\tau_{a,b}$ on the intersection of their domains. Hence they are compatible. Their domains $\partial T \setminus \operatorname{cone}(c_1)$ and $\partial T \setminus \operatorname{cone}(c_2)$ cover $\partial T$ because $c_1, c_2$ are incomparable. The join is therefore total and equals $\tau_{a,b}$.
\end{IEEEproof}

\emph{Classical counterpart.} The transpositions $\tau_{a,b}$ of incomparable cones are the generators by which Bleak and Quick present Thompson's group $V$~\cite{bleak2017infinite}. The germ carries over, with the derived swap performing such a cone exchange. What differs is that one swap is available only off its staging cone, since a deployment has to park a subtree somewhere. The shadow of a single staged swap is therefore a proper partial injection. It is invertible on its range without being a unit of $C_N$. The transposition $\tau_{a,b}$ itself is reached only through the compatible join of part (iii), which joins two differently staged shadows rather than reading off the shadow of any one word (Fig.~\ref{fig:swap}).

\begin{lemma}[Transpositions generate]
\label{lem:transp-gen}
For every $N \geq 2$, the transpositions $\{\tau_{a,b} : a, b \text{ incomparable}\}$ generate the Higman--Thompson group $G_{N,1} = V_N$.
\end{lemma}

\begin{IEEEproof}[Proof sketch, with details in Appendix~\ref{app:calculus}]
Let $T$ be the set of transpositions and $\langle T \rangle$ the subgroup they generate. Conjugation preserves $T$ up to products. For $g \in V_N$, refine the table of $g$ so that $g$ acts by a single prefix substitution on each of finitely many cones partitioning $\operatorname{cone}(a) \sqcup \operatorname{cone}(b)$. Then $g \tau_{a,b} g^{-1}$ is the product of the disjoint transpositions of the image pieces. Hence $\langle T \rangle$ is a nontrivial normal subgroup. For $N$ even, $G_{N,1}$ is simple~\cite{higman1974finitely} and $\langle T \rangle = V_N$. For $N$ odd, $G_{N,1}$ has a unique proper nontrivial normal subgroup, Higman's simple subgroup of index two~\cite{higman1974finitely}, the kernel of the parity character that assigns to a table bijection the sign of its row permutation. That character is well defined because expanding one row into $N$ rows inserts lexicographically consecutive blocks mapping in order, which changes the inversion count by a multiple of $N - 1$, an even number. A transposition has a single $2$-cycle as its row permutation and parity $-1$. Hence $\langle T \rangle$ is not contained in the kernel. A normal subgroup not inside the unique index-two kernel and of order greater than two equals the whole group. The binary case is also Theorem~1.1 of~\cite{bleak2017infinite}, which presents $V$ by these transpositions.
\end{IEEEproof}

\begin{IEEEproof}[Proof of Theorem~\ref{thm:vn-identification}]
(i) Shadows lie in $C_N$ by Lemma~\ref{lem:bridge}(i). Conversely, $C_N$ is generated under compatible joins by its restricted prefix substitutions and its idempotents~\cite{lawson2020revisited}, each of which is reachable from shadows. The round trip $\mathsf{graft}(b \to a) \circ \mathsf{graft}(a \to b)$ has shadow $\mathrm{id}$ restricted to $\partial T \setminus \operatorname{cone}(b)$. Products of such shadows give $\mathrm{id}$ restricted to the complement of any finite union of cones, while every clopen subset of $\partial T$ is such a complement. A prefix substitution $s_u s_v^{*}$ with $u, v$ incomparable is the restriction of $\tau_{u,v}$ (Lemma~\ref{lem:bridge}(ii), (iii)) to $\operatorname{cone}(v)$. Let $u \neq v$ be comparable. If both are nonempty, they share their first digit, and any $c$ with a different first digit is incomparable to both, whence $s_u s_v^{*} = s_u s_c^{*} \cdot s_c s_v^{*}$. If one of them is empty, $s_u s_v^{*}$ is the compatible join over $d \in \Sigma_N$ of the substitutions $s_{ud} s_{vd}^{*}$, whose prefixes are nonempty, and each term falls under one of the two cases above. For $u = v$ the substitution is an idempotent.

(ii) Compatible joins of shadows lie in $C_N$ by (i). A bijective element of $C_N$ is a unit, hence lies in $G_{N,1} = V_N$ by Lawson's theorem~\cite{lawson2007polycyclic,lawson2020revisited}. Conversely every element of $V_N$ is a product of transpositions (Lemma~\ref{lem:transp-gen}), each of which is a join of two swap shadows (Lemma~\ref{lem:bridge}(iii)). Products of compatible joins are compatible joins of products in a Boolean inverse monoid, whence every element of $V_N$ is a finite compatible join of shadows.

(iii) is Lemma~\ref{lem:transp-gen} together with Lemma~\ref{lem:bridge}(ii)--(iii). The reading for deployed networks is Definition~\ref{def:calc-state}, under which the net relocation of a finite operation sequence is the boundary shadow of its rigid part. That shadow is a partial injection of $\partial T$. It is a unit of $C_N$ precisely when its domain and its range are both all of $\partial T$, in which case it is a bijection and (ii) places it in $V_N$. Invertibility on its range does not make it a unit. A history with a proper partial shadow therefore acquires an element of $V_N$ only once a compatible total completion has been declared. Lemma~\ref{lem:bridge}(iii) supplies one for the staged swap.
\end{IEEEproof}

\emph{Classical counterpart.} The identification is Lawson's theorem that $G_{N,1} = V_N$~\cite{higman1974finitely} is the group of units of the Cuntz inverse monoid $C_N$~\cite{lawson2007polycyclic,lawson2020revisited}, whose $C^{*}$-completion is $\mathcal{O}_N$~\cite{cuntz1977simple}. That algebra carries over. What differs is the carrier, since the rearrangement groups of Belk and Forrest act on glued limit spaces and are typically proper subgroups~\cite{belk2019rearrangement}.

Theorem~\ref{thm:vn-identification} is the precise sense in which the dynamics of the HSFN is classical mathematics. On the address boundary, the invertible reconfiguration dynamics of a self-similar network is the Higman--Thompson group $V_N$~\cite{higman1974finitely,bleak2017infinite}, the symmetry group of exactly this kind of tree rearrangement. One remark positions the identification.

\begin{remark}[Positioning the identification]
\label{rem:vn-positioning}
Three boundaries locate the theorem. \emph{No gluing, full group.} Thompson-like groups of homeomorphisms of fractal \emph{limit spaces}, the rearrangement groups of Belk and Forrest~\cite{belk2019rearrangement}, are constrained by the gluing equivalence of the limit space and are typically proper subgroups of $V_N$. The HSFN dynamics acts on the logical address tree, where geometrically coincident nodes are distinct by decree (Lemma~\ref{lem:dup-digit}) and no gluing is imposed. That design decision, made in Section~\ref{sec:construction} for protocol reasons, is exactly what delivers the \emph{full} group. \emph{Marking does not shrink the group.} The duplicate marking restricts non-root $\mathsf{permute}$ letters to $\operatorname{Stab}_{S_N}(0)$ (Lemma~\ref{lem:dup-digit}(ii)). The generating swaps are nevertheless built from grafts alone. The marked calculus therefore still realizes all of $V_N$. The wreath layer itself sits inside $V_N$ as table bijections with equal-length rows. \emph{Cuntz completion.} The $C^{*}$-completion of the polycyclic relations $s_i^{*} s_j = \delta_{ij}$, $\sum_i s_i s_i^{*} = 1$ is the Cuntz algebra $\mathcal{O}_N$~\cite{cuntz1977simple}. A hosted topology selects the projection $P_H$ given by the join of the cone projections of its hosting slots. The dynamics that Theorem~\ref{thm:lifting} attaches to it is the compression $P_H\, w\, P_H$ of the ambient calculus word, a lens we record rather than a theorem.
\end{remark}

\subsubsection{Coordinate Persistence}
\label{subsec:persistence}

The identification closes the loop with the spacetime coordinates of Section~\ref{subsec:spacetime}, since the group that moves the network preserves the coordinate system.

\begin{IEEEproof}[Proof of Proposition~\ref{prop:anchored-time}]
(i)~Uniqueness per round follows from Proposition~\ref{prop:safety}(i) for both normal commits and closing view changes. Any block $B$ committed in round $r$ was prepared by at least $Q{-}f$ correct members, intersecting any $Q$ view changes in at least one correct replica ($2Q - f - N \geq 1$), so closing round $r$ always preserves $B$. In-order locking and predecessor hashes force all correct chains to extend one unique history. (ii)~Follows from Proposition~\ref{prop:safety} and round-bound certificate validity. (iii)~Uniqueness of the root decision and collision-resistant hashing of aggregate bodies fix the nested snapshot uniquely.
\end{IEEEproof}

\begin{IEEEproof}[Proof of Proposition~\ref{thm:update-main}]
Induction on $m - |a|$ shows that $\mathcal{D}(a)$ lists every word with prefix $a$ and length in $[|a|{+}1, m]$ once, since words of length at least $|a|{+}2$ appear uniquely inside $\mathcal{D}(ad)$ for their $(|a|{+}1)$-st digit $d$. The wrap-around successor on a repetition-free finite list is a single cycle.
\end{IEEEproof}

\begin{IEEEproof}[Proof of Proposition~\ref{prop:coordinate-persistence}]
(i) The only address-changing generators are $\mathsf{graft}$ and $\mathsf{permute}$. The former acts by prefix substitution on the strict moved cone $a\Sigma_N^{+}$ while fixing both finite anchors $a$ and $b$, leaving suffixes intact (Theorem~\ref{thm:rep-tree}(ii), Proposition~\ref{thm:graft-iso}(i)--(ii)), while the latter is the cell-wise relabeling of Lemma~\ref{lem:dup-digit}(ii). Composing the per-letter maps $\widehat{\Phi}(\omega_j)$ in the order in which the letters act therefore relocates every surviving finite address, which is the definition of the finite-slot address transport $\widehat{\Phi}(w)$. The letters are composed in place rather than sorted into a rigid block followed by relabelings, because a $\mathsf{graft}$ letter and a $\mathsf{permute}$ letter need not commute. (ii) The effects column of Table~\ref{tab:typing} transports the membership sets of moved cells verbatim. The chain is held by the participants of the cell, which the transport keeps, and the cell identifier and the certificates move with the cell by the convention of Definition~\ref{def:calc-state}. No generator writes chain state, and historical verification takes the round's committee keys and configuration record as authenticated inputs. (iii) A graft moves one contiguous block of the deployed cell-tree traversal (and its block-expanded node cycle) and preserves the successor relation inside it (Remark~\ref{rem:graft-splice}). At most three pointer repairs therefore re-close the cycle, which is again the digit-ordered depth-first traversal of the resulting deployment by Proposition~\ref{thm:update-main} applied to $w(x)$. (iv) The first component of the finite-slot transport form $\chi(w(s)) = (\widehat{\Phi}(w)(a(s)),\, h(s))$ is (i) and the second is (ii), settling every applicable word. For any rigid word representative $w$, $\widehat{\Phi}(w)$ coincides on surviving finite slots with the finite-slot cocycle $\Phi(w) \in B_N$ of Theorem~\ref{thm:rep-tree}(ii). Passing to infinite digit strings in $\partial T = \Sigma_N^{\omega}$, the word representative $w$ induces a partial boundary shadow in $C_N$, and any total bijective compatible join completion defines a unit $g \in V_N$ (Theorem~\ref{thm:vn-identification}(ii)) agreeing with that partial boundary shadow on its domain (and, as a sufficient condition, matching uniform prefix replacement by $\widehat{\Phi}(w)(s)$ on $\operatorname{cone}(s)$ whenever $\operatorname{cone}(s)$ remains inside a single substitution piece of $w$ and no letter of $w$ fixes an intermediate finite image of $s$ as a graft anchor).
\end{IEEEproof}

\emph{Classical counterpart.} Process migration requires a relocated process to keep its logical clock~\cite{lamport1978time,fidge1991logical}. Virtual network embedding requires a re-embedding to leave the guest intact~\cite{fischer2013virtual}. Persistence of identity and history under a change of location carries over. What differs is that the relocation has a closed form, a prefix substitution computed from the old address.

\subsubsection{The Lifting Theorem: Operation Transfer}

Section~\ref{sec:universality} tells us \emph{which} topologies a family member can host. The calculus should also tell us how a hosted topology \emph{evolves}. Fix the realization semantics of Definition~\ref{def:realization}. In calculus terms a realization of a finite graph $H$ in $F(N,m)$ is a configuration $x_\rho$ together with two further data. The first is a \emph{selected-link set} $E_\rho$, a subset of the provisioned links among the occupied slots that contains the edges of the designated duplicate chains. The second is the reduction $\rho: (x_\rho, E_\rho) \rightsquigarrow H$ that contracts the chains and takes the simple quotient of the selected subgraph, inducing an isomorphism onto $H$. Like the chains and the scores of the repositioning protocol, $E_\rho$ is protocol metadata riding on the memberships rather than part of the configuration. A graft relocates by one prefix substitution both endpoints of every selected link interior to the moved set, which therefore transports verbatim. A selected link between the moved set and the rest is re-anchored from $a$ to $b$, while a selected link disjoint from the moved set is untouched (Proposition~\ref{thm:graft-iso}(i)--(ii), Lemma~\ref{lem:edge-cut}(i)). An \emph{elementary edit} of $H$ is one of vertex deletion, edge deletion, vertex addition (with a prescribed set of neighbors), edge addition, vertex split and vertex merge. The transfer principle says that every such edit is implemented by a calculus word together with a mask update and states when the implementation is local, with one subcase and the closed form of one boundary condition left open.

\begin{theorem}[Operation transfer, the lifting dichotomy]
\label{thm:lifting}
Let $\rho$ be a \emph{tidy} realization of $H$ in $F(N,m)$ with configuration $x_\rho$, tidiness being the invariant of Appendix~\ref{app:calculus} that Theorem~\ref{thm:blockcut} delivers and that the lifts of additions, merges, and block-preserving splits preserve. Realizations outside that class are not covered by the constructions of that appendix. Let $\omega$ be an elementary edit of $H$.
\begin{enumerate}
    \item \emph{Existence.} There are parameters $N' \geq N$, $m' \geq m$, a calculus word $w$ of $F(N', m')$ (in which $x_\rho$ is a configuration by family monotonicity, Lemma~\ref{lem:family-mono}) and a realization $\rho'$ of $\omega(H)$ with configuration $x_{\rho'} = w(x_\rho)$, such that the square commutes with vertex tracking, meaning $\rho' \circ w = \omega \circ \rho$ on the deployment. That is, for every vertex $v$ of $H$ that $\omega$ does not delete, the hosting chain of $v$ under $\rho'$ is the $w$-relocation of its hosting chain under $\rho$, up to extension or truncation of designated duplicate chains at either end (a chain extended at its head by a freshly joined participant hosts the same vertex by construction). No surviving vertex is destroyed and re-created.
    \item \emph{Locality.} If $\omega$ is either a vertex deletion or an edge deletion, then $w$ can be chosen \emph{local}, unconditionally. The reason is that deletions never enlarge a block. An edge deletion is a mask update ($E_\rho$ drops every selected link whose two reduced endpoints are the deleted edge) with an empty structural word, while a vertex deletion empties one chain. In the local case $N' = N$. For the four remaining edits, assume $\rho$ is also \emph{block-separated and slot-clean}, meaning that distinct blocks of $H$ occupy distinct host cliques and no extraneous occupied, disabled, or child-carrying vacant slots occupy unused members of the affected host cliques (both delivered automatically by the recursive placement of Theorem~\ref{thm:blockcut} and preserved by vertex/edge additions, merges, and block-preserving splits under $\mathsf{clean}$, whereas deletions and block-splitting splits can leave multiple smaller blocks inside one clique without a separate consolidation pass). When the configuration keeps an enabled cell-free staging slot at tier at most $|s|$ outside the edited cones for each invoked descent, each chain extension or single re-rooting adds at most two tiers (and $k$ successive descents add at most $k$ tiers to the active height, giving $m' \leq m + 2$ whenever no nested chain of descents occurs). Without such spare capacity each re-rooting obeys the current-state bound of Lemma~\ref{lem:descent}, and a uniform global depth bound for composed re-rootings is not claimed. Tidiness is what makes the local construction available. It holds for every realization produced by Theorem~\ref{thm:blockcut} and for the output of every lift of an addition, a merge, or a block-preserving split below, whereas a deletion or a block-fracturing split can leave a realization that is no longer tidy without a separate consolidation pass (Appendix~\ref{app:calculus}). The word touches only the cells hosting the chains of the edited vertices and their $H$-neighbors, at most a bounded number of freshly spawned cells adjacent to them ($\mathsf{spawn}$ plus the duplicate chains of Lemma~\ref{lem:dup-digit} manufacture fresh slots adjacent to any occupied slot) and finitely many grafts of hanging sub-realizations, while leaving $\rho$ unchanged outside the touched cells. The word length is bounded by a function of $N$, the degrees involved in the edit and the lengths of the hosting chains of every vertex that the word moves, uniformly in $|H|$, $m$ and the depth of the deployment. If $\omega$ is a vertex addition, an edge addition, a split or a merge, the edit can enlarge or fuse blocks. An added vertex with neighbors in two or more blocks fuses the block--cut path between them, exactly as an added edge does, while a split can grow a block by one vertex whether or not it keeps its two copies adjacent (Lemma~\ref{lem:gate}). For these four edits the largest block $B'$ of $\omega(H)$ meeting the edit decides through the hosting conditions at cell size $N$. Those conditions are that every block of $\omega(H)$ has order at most $N{+}1$ (Lemma~\ref{lem:block-image}), that no cut vertex lies in three full blocks (Proposition~\ref{prop:blockcut-caveat}(i)) and that no two cut vertices of a common non-full block each lie in two full blocks (Proposition~\ref{prop:two-full-sep}). \emph{(Stay.)} Suppose either that $|B'| \leq N$ or that $|B'| = N{+}1$ and no cut vertex of $\omega(H)$ lies in two full blocks, which is the regime of Theorem~\ref{thm:blockcut}(iii). For additions and merges (and for splits that either keep every incident block wholly on one copy or satisfy $|V_\pi| \leq N$, or $|V_\pi| = N{+}1$ at a free terminal slot), a local lift at $N' = N$ exists as above whenever $|B'| \leq N$ (on either an anchored host clique or the tier-$1$ seed cell $C(\varepsilon)$). It exists as well when $|B'| = N{+}1$ and either the target block $B_1$ already occupies an anchored clique whose anchor is the end of its apex's chain, or $B_1 \subseteq C(\varepsilon)$ contains a tier-$1$ vertex $u_1$ whose spawn cell is not yet deployed (re-rooting $B_1$ into $\{z_0\} \cup C(z_0)$ at tier~$2$ with $z_0 = \operatorname{head}(u_1)$). For a split that fractures a block into smaller blocks inside one clique, the output realizes $\omega(H)$ locally at $N' = N$ without unconditionally preserving tidiness (T1) or block separation unless followed by consolidation, while splits with $|V_\pi| > N$ that remain hostable via separate cliques (such as splitting a vertex of a hosted $5$-cycle at $N = 4$ into a $6$-vertex path) are handled by climb or re-embedding rather than the single-clique Stay word. When instead $|B'| = N{+}1$ and the target anchor chain continues below the clique (or every vertex of $B_1 \subseteq C(\varepsilon)$ already carries child blocks), the construction needs the crossing transport of Conjecture~\ref{conj:crossing-transport}, which leaves the local lift open in that subcase. \emph{(Climb.)} If one of the hosting conditions fails for $\omega(H)$, then no realization of $\omega(H)$ exists in $F(N, m')$ for any $m'$. No lift, local or global, is therefore available at cell size $N$. Every lift climbs the family to some $N' > N$ through item~1, which always succeeds by Lemma~\ref{lem:family-mono}. \emph{(Boundary.)} In the remaining regime, $|B'| = N{+}1$ with some cut vertex of $\omega(H)$ in two full blocks and all three hosting conditions satisfied, the orientation of Remark~\ref{rem:two-full} is necessary for any realization of $\omega(H)$ at cell size $N$. Whether it suffices, and hence whether a local lift exists there, is open. Within the regimes proved here, what a hosting condition forbids at cell size $N$ cannot be bought at any depth, only by a larger cell size.
\end{enumerate}
\end{theorem}

\begin{IEEEproof}[Proof sketch, with the case bookkeeping in Appendix~\ref{app:calculus}]
Two ingredients drive existence. (\emph{Reach}) Any configuration can be carried to any other, since by Proposition~\ref{prop:canon-word} and the exact inverses of Theorem~\ref{thm:rep-tree}(i), $w_{\mathrm{can}}(y) \circ w_{\mathrm{can}}(x)^{-1}$ maps $x$ to $y$. (\emph{Room}) $\beta(\omega(H)) \leq |V(\omega(H))| \leq |V(H)| + 1$. Choosing $N' = N + |V(H)| + 1$ adds $|V(H)| + 1$ guaranteed vacant, enabled, cell-free digits $\{N, \ldots, N + |V(H)|\}$ to every deployed cell of $x_\rho$ (including the tier-$1$ seed cell $C(\varepsilon)$) and puts $\omega(H)$ inside the sufficiency regime of Theorem~\ref{thm:blockcut}. Raw reach would destroy participants. The constructive proof instead edits in place. The block--cut tree of $\omega(H)$ differs from that of $H$ only at the blocks meeting the edit. Deletions split or shrink blocks and never enlarge one. A vertex addition with at most one neighbor attaches a pendant. A vertex addition with neighbors in two or more blocks, an added edge, a merge or a split can fuse or enlarge the blocks along one path of the block--cut tree. The canonical realization of Section~\ref{sec:universality} is recursive along the block--cut tree. The lift therefore re-runs the construction on the affected blocks only. Fresh cells are spawned next to the hosting chains, the re-embedded blocks are populated by joins and chains are extended or truncated to re-route the quotient. The untouched sub-realizations hanging at cut vertices are carried across whole by grafts (Proposition~\ref{thm:graft-iso} guarantees they arrive isomorphic, marked and correctly re-anchored), after which the vacated cells are emptied and retracted. Vertex tracking holds because every surviving vertex's chain is only grafted, extended or truncated, never vacated. Locality is unconditional for the two deletions because no block grows. For the other four edits the fused or enlarged blocks decide, through the same three-way gate as the hosting characterization. In the Stay regime all surgery happens in the touched cells plus fresh tiers, one tier of staging slots and one of chain extensions on a tidy realization, a duplicate slot occupied by a foreign participant being pushed one tier deeper by one descent inside the same budget (Lemma~\ref{lem:descent}). For the four non-deletion edits, the largest affected block $B$ decides. A block with $|B| \leq N$ re-embeds into a single anchored clique next to any of its chains. Order $N{+}1$ is governed by the two-slot placement law of Proposition~\ref{prop:blockcut-caveat}, the full block occupying a terminal anchored clique scheduled last on the affected chain (or the parent-side birth clique when that slot is free). A block with $|B| > N{+}1$ contradicts the necessity direction of Theorem~\ref{thm:blockcut} at cell size $N$, leaving only the climb of item~1, with the whole affected path re-embedded at $N'$. The four ingredients just used are isolated as Lemmas~\ref{lem:staging}--\ref{lem:gate} in Appendix~\ref{app:calculus}, namely a staging lemma manufacturing fresh slots adjacent to a hosting chain, a transport lemma carrying a hanging sub-realization across a graft with its mask and its chains, a re-hosting lemma moving one vertex with its whole sub-realization into a free member slot of a target clique and a gate lemma computing the blocks of $\omega(H)$ and deciding Stay against Climb. The six per-edit constructions, with their guard checks and word-length counts, follow there.
\end{IEEEproof}

\emph{Classical counterpart.} The theorem is the reconfiguration side of virtual network embedding, serving a change in the guest by re-embedding part of it~\cite{fischer2013virtual}, with the width theory of Section~\ref{sec:universality} on the impossibility side~\cite{bodlaender1998partial,robertson1986graph}. That a local edit should cost local surgery carries over. What differs is that the Climb criterion is exact, being the hosting conditions at cell size $N$, which is why depth never substitutes for cell size.

Three caveats bound the theorem. First, lifts are not canonical, since the word of Theorem~\ref{thm:lifting} depends on staging-slot choices, on the order in which hanging sub-realizations are grafted and on where fresh cells are spawned. A deployment obtains a deterministic lift only by fixing a policy such as always taking the length-lexicographically least admissible slot. Different policies yield equally correct but different words. Second, the theorem quantifies only over tidy realizations, a subclass of the disciplined reduction class of Definition~\ref{def:realization}. For arbitrary quotient maps the statement trivializes, because unconstrained identifications can encode any graph anywhere. Third, no optimality is claimed, since neither the depth $m'$ nor the vertex budget nor the word length of a lift is asserted minimal. Section~\ref{sec:discussion} records the remaining structural open problems.

\subsection{Proofs of the Calculus Statements}
\label{app:calculus-proofs}

\begin{IEEEproof}[Proof of Lemma~\ref{lem:edge-cut}]
(i) Sibling edges of $G(x)$ join two members of one cell, while each cell lies entirely inside or entirely outside the moved slot set $\operatorname{Sl}(x) \cap \operatorname{cone}(a) \setminus \{a\}$. No sibling edge therefore crosses the boundary. An uplink $\{wd, w\}$ crosses iff exactly one endpoint is a moved slot. If $wd$ is moved then $a \preceq wd$ properly. If moreover $w \neq a$ then $a \preceq w$ properly and $w$ is moved too. The only crossing uplinks are therefore those with $w = a$, namely the $N$ edges $\{ad, a\}$, which all exist since $a \in D$. Conversely if $w$ is moved then so is $wd$. The same computation in $x'$ gives the $N$ uplinks into $b$.

(ii) Let $D_a = D \cap \operatorname{cone}(a)$. A parent edge $\{s, \operatorname{pre}(s)\}$ with $s \in D_a$, $s \neq a$ has $a \preceq \operatorname{pre}(s)$, hence both endpoints in $D_a$. With $s \notin D_a$, if $\operatorname{pre}(s) \in D_a$ then $a \preceq \operatorname{pre}(s) \preceq s$ puts $s \in D_a$, a contradiction. So the unique crossing parent edge is $\{a, \operatorname{pre}(a)\}$. A sibling edge $\{wd, wd'\}$ with $wd \in D_a$, $wd \neq a$ has $a \preceq w$, hence $wd' \in D_a$. The crossing sibling edges are therefore exactly those at $wd = a$, one per deployed sibling of $a$, giving $1 + \sigma_x(a)$ cut edges, all incident to $a$. The created count is the same computation in $x'$. Edges outside $D_a$ mention no address with prefix $a$ or, in $x'$, prefix $b$. Since the substitution changes only such addresses and $D \setminus D_a$ is unchanged, those edges persist. On $D_a$ the substitution is a bijection preserving both the parent relation and siblinghood, which maps internal edges bijectively.
\end{IEEEproof}

\begin{IEEEproof}[Proof of Proposition~\ref{prop:kpi-affine}]
(i) is cell-atomicity. (ii) Each of the $\sum_t M0_t$ deployed cells contributes $\binom{N}{2}$ sibling links, while each non-seed cell contributes $N$ uplinks. There are $\sum_t M0_t - 1$ non-seed cells, giving $\binom{N}{2}\sum M0_t + N(\sum M0_t - 1)$. (iii) The first display is $(N{+}1) |\operatorname{Sl}(x)|$, while the second is definitional and evaluates to $(N{+}1)V$ at full occupancy since $\sum_t M1(o_t) = V$ and each full cell contributes $N^2$. (iv) Two deployed cells $u, v \in D \setminus \{\varepsilon\}$ are $\Gamma$-adjacent iff $\{u = \operatorname{pre}(v)$ or $v = \operatorname{pre}(u)\}$ or $\{u, v$ deployed siblings$\}$, which by Proposition~\ref{thm:adjacency-II} is exactly adjacency of the anchor nodes $u, v$ in $G(x)$, since clause~(P) is the parent relation and clause~(S) is siblinghood. For the count, each cell $w$ with $\kappa(w) = k$ hosts $\binom{k}{2}$ sibling pairs of child cells, summing to $\sum_t \sum_k \binom{k}{2} h_t(k)$. The parent edges avoiding $\varepsilon$ are those of cells at tiers $t \geq 3$ (a tier-$t$ cell's parent edge lands on an address of length $t{-}2$, which is a node iff $t \geq 3$), summing to $\sum_{t \geq 3} M0_t$. The update costs are read off the effects in Table~\ref{tab:typing}, since each generator changes the membership of at most two cells' occupancy bins and at most two cells' anchoring bins at fixed tiers, except a cross-tier graft, whose moved rows shift tier index by the constant $|b| - |a|$.
\end{IEEEproof}

\begin{IEEEproof}[Proof of Proposition~\ref{prop:rel-exact}]
(ii) Cells fail internally independently, while the fault status of a cell depends on its own coin and its children's statuses. The tree recursion is therefore exact by induction on the subtree. A cell with $c$ children costs $O(c^2) \leq O(Nc)$ convolution steps. Since $\sum_w c_w < |D|$, the total is $O(N|D|)$. An operation changes $D$ and hence the recursion inputs only inside its footprint, whose affected cells lie on at most two anchor paths of length at most $m$ to $\varepsilon$, giving $O(N^2 m)$ direct recomputation along the path (or $O(Nm)$ with cached per-cell convolution polynomials).

(iii) Induction upward from $t = T$, where every cell is childless and fails with probability exactly $q$. For the step, let $w$ be a tier-$t$ cell with $c = \kappa(w)$ children whose failure probabilities $p_i$ lie in $[\underline{P}_{t+1}, \overline{P}_{t+1}]$ by hypothesis. Coupling each child's coin through a uniform variable shows $\Pr[\sum X_i \geq r]$ is nondecreasing in each $p_i$. Hence $p_w \leq F_c(\overline{P}_{t+1}) \leq \overline{P}_t$ since $h_t(c) > 0$. Symmetrically $p_w \geq \underline{P}_t$. The root case $t = 1$ is the seed cell.

(i) The two stated pairs share their tier profiles by inspection, since heavy cells are full, all others are childless and the cell counts per tier match. Every profile functional, including the bounds of (iii) and all indicators of Proposition~\ref{prop:kpi-affine}, therefore agrees on each pair, while the exact recursion of (ii) separates them. Structurally, the recursion composes $F_k$ along the tree. Composition order matters, since the profile records which $F_k$ occur per tier but not which child feeds which parent.
\end{IEEEproof}

\begin{IEEEproof}[Proof of Proposition~\ref{prop:commute}]
Two locality facts hold row by row in Table~\ref{tab:typing}. (\emph{Guard locality}) Applicability of $\omega$ at $x$ depends only on the restriction of $x$ to the footprint $\mathrm{fp}(\omega)$, since each guard is a membership or emptiness test at addresses of $\mathrm{fp}(\omega)$, while the depth-fit guard of a graft reads only $D \cap \operatorname{cone}(a)$. (\emph{Effect locality}) $\omega$ changes the membership data of $D, \mu, \zeta$ at addresses of $\mathrm{fp}(\omega)$ only, while the new values depend only on the old restriction to $\mathrm{fp}(\omega)$. Footprints are fixed address sets (points and cones), independent of the state. Now let $\mathrm{fp}(\omega_1) \cap \mathrm{fp}(\omega_2) = \emptyset$ and let $x \in \operatorname{dom}(\omega_2)$ with $\omega_2 x \in \operatorname{dom}(\omega_1)$. Since $\omega_2$ does not alter the restriction of the state to $\mathrm{fp}(\omega_1)$, guard locality gives $x \in \operatorname{dom}(\omega_1)$. Since $\omega_1$ does not alter the restriction to $\mathrm{fp}(\omega_2)$, also $\omega_1 x \in \operatorname{dom}(\omega_2)$. The two composite states agree address by address, since on $\mathrm{fp}(\omega_1)$ both equal the $\omega_1$-rewrite of $x$'s restriction, on $\mathrm{fp}(\omega_2)$ both equal the $\omega_2$-rewrite and elsewhere both equal $x$. Exchanging the roles of $\omega_1, \omega_2$ gives the equality of domains, hence of the partial maps.
\end{IEEEproof}

\begin{IEEEproof}[Proof of Proposition~\ref{prop:canon-word}]
(i) Applicability of the canonical word is checked block by block. When $\mathsf{spawn}(a)$ fires, $\operatorname{pre}(a) \in D$ has already been spawned (it precedes $a$ length-lexicographically). Hence $a$ is a slot, is not yet a cell root, is enabled ($\zeta$ is still empty) and satisfies $|a| \leq m{-}1$ since $a \in D$. The joins then find existing vacant enabled slots, while the disables find existing vacant childless slots by the invariant $\zeta \cap (\mu \cup D) = \emptyset$ of the target. The result is $x$ by construction. Conversely every generator preserves well-formedness, which makes reachable states configurations. (ii) is immediate from Table~\ref{tab:typing}, since each letter is executable in time proportional to its footprint intersected with the current deployment.
\end{IEEEproof}

This appendix completes the proofs of Theorems~\ref{thm:rep-tree} and~\ref{thm:lifting} and of Lemma~\ref{lem:transp-gen}.

\subsubsection{Inverse Pairs (Theorem~\ref{thm:rep-tree}(i))}
Each generator is injective on its domain, since its effect is a reversible set operation determined by its parameters. The listed partners are exact partial inverses. We verify the two nontrivial pairs, the rest being identical in pattern. \emph{Spawn/retract.} If $x \in \operatorname{dom}(\mathsf{spawn}(a))$, then in $x' = \mathsf{spawn}(a)(x)$ the cell $C(a)$ is deployed, vacant, enabled and childless. Hence $\mathsf{retract}(a)$ is applicable and returns $x$. Conversely, if $x \in \operatorname{dom}(\mathsf{retract}(a))$ then after the retract $a$ is a slot ($a \neq \varepsilon$ and $\operatorname{pre}(a) \in D$ by prefix closure), is not a cell root, is not decommissioned ($a \in D$ implied $a \notin \zeta$) and $|a| < m$ since $a \in D \subseteq \Sigma_N^{\leq m-1}$. Therefore $\mathsf{spawn}(a)$ is applicable and restores $x$, the erased attributes having been trivial by the retract guards. \emph{Graft/graft.} Let $x \in \operatorname{dom}(\mathsf{graft}(a \to b))$ and $x'$ the image. In $x'$ we have $b \in D$, while $a$ is still a slot ($C(\operatorname{pre}(a))$ is static) with $a \notin D$. Also $\operatorname{pre}(a) \in D$ (static) and $a, b \in \mu$ (occupancies of the anchors are untouched). Finally $b \not\preceq a$, which already held in $x$, where prefix closure of $D$ would otherwise force $b \in D$. The depth fit reverses because $h_{x'}(b) = h_x(a)$, while $|a| + h_x(a) \leq m$ holds automatically from $a \in D$ and $D \subseteq \Sigma_N^{\leq m-1}$ (the deepest deployed cell below $a$ has length $|a| + h_x(a) - 1 \leq m - 1$). Applying the substitution $b \mapsto a$ to $x'$ restores $x$ address by address. For closure, composites of partial bijections are partial bijections, while every word's inverse is the reversed word of partner letters. A set of partial bijections closed under composition and inversion is an inverse monoid in the symmetric inverse monoid of the state space.

\subsubsection{Cocycles and the Boolean Normal Form (Theorem~\ref{thm:rep-tree}(ii)--(iii))}
For a rigid letter $\omega$ define $\Phi(\omega) \in B_N$ as in the theorem. For a rigid word $w = \omega_k \cdots \omega_1$ define the \emph{survivor set} of $x \in \operatorname{dom}(w)$ as the slots not erased by any $\mathsf{retract}$ along the way. The cocycle is the induced address map. We prove by induction on $k$ that (a) the cocycle at every $x$ is the restriction of $\Phi(\omega_k) \cdots \Phi(\omega_1)$ to the survivors and (b) this product lies in $B_N$ and admits a normal form $\bigvee_{i=1}^{r} \bigl(s_{u_i} s_{v_i}^{*}\bigr)\big|_{E_i}$ with $r$ finite and each $E_i$ a finite Boolean combination of cones and points, pairwise disjoint. For (a), Table~\ref{tab:typing} shows that the letter $\omega_k$ relocates exactly per $\Phi(\omega_k)$ on all slots it preserves. For (b), it suffices that $B_N$ is closed under composition. The composite $(s_u s_v^{*})|_E \circ (s_p s_q^{*})|_F$ has domain $F \cap (s_p s_q^{*})^{-1}(E)$. The preimage of a cone $\operatorname{cone}(c)$ under $s_p s_q^{*}$ is $\operatorname{cone}(q c')$ if $c = p c'$, is $\operatorname{cone}(q)$ intersected with a point set or all of $\operatorname{cone}(q)$ if $c \preceq p$ and is empty otherwise, in every case a Boolean combination of cones and points. Complements of cones decompose as $\Sigma_N^{*} \setminus \operatorname{cone}(c) = \bigcup_{j < |c|} \bigl(\{\operatorname{pre}_j(c)\} \cup \bigcup_{d \neq c_{j+1}} \operatorname{cone}(\operatorname{pre}_j(c) d)\bigr)$, a finite union of the same shape. Disjointifying the finitely many pieces yields the normal form. For (iii), soundness is immediate from (a), since two elements of $B_N$ that agree on the deployable addresses restrict equally to the survivors of any shared state. For the partial converse, fix rigid words $w, w'$ and a state $x \in \operatorname{dom}(w) \cap \operatorname{dom}(w')$. Let $p\,d$ be a slot of $x$ that survives both words. By (a) the address of that slot in $w(x)$ is $\Phi(w)(p\,d)$, while its address in $w'(x)$ is $\Phi(w')(p\,d)$. If the two words relocate the survivors of $x$ identically then the two normal forms agree at $p\,d$. Whether a shared state witnesses a given address is a question about the guards rather than about $B_N$. The guards can rule every witness out. A word whose first executed letter is $\mathsf{spawn}(a)$ requires $a \notin D$. By prefix closure no state of its domain has a slot strictly below $a$, hence no state of its domain witnesses $\Phi$ anywhere in the open cone $a\Sigma_N^{+}$. For a concrete pair, take $N = 3$, $m = 2$ and the state $D = \{\varepsilon\}$, $\mu = \{0,1\}$, $\zeta = \emptyset$, on which both $\mathsf{spawn}(0)$ alone and $\mathsf{spawn}(0)$ followed by $\mathsf{graft}(0 \to 1)$ are applicable. Both words fix every surviving slot of that state, yet $\Phi$ of the first sends $00$ to $00$ and $\Phi$ of the second sends $00$ to $10$, both images being addresses deployable in $F(3,2)$. Agreement on shared states therefore forces agreement of the normal forms at the surviving slots those states supply and at no further address. In the other direction (a) shows the normal form determines the relocation of every survivor. Words with $\Phi(w) \leq \mathrm{id}$ relocate nothing and therefore act by the bookkeeping sets alone.

\subsubsection{Generation by Transpositions (Lemma~\ref{lem:transp-gen})}
\label{app:transp-gen}
Let $T = \{\tau_{a,b} : a, b \in \Sigma_N^{*} \text{ incomparable}\}$ and $\langle T\rangle \leq G_{N,1}$ the subgroup it generates.

\emph{Normality.} Let $g \in G_{N,1}$ and $\tau_{a,b} \in T$. Refine the table of $g$ until its domain code refines the partition $\{\operatorname{cone}(a), \operatorname{cone}(b), \text{rest}\}$. Then $\operatorname{cone}(a) = \bigsqcup_{i=1}^{r} \operatorname{cone}(a p_i)$ and $\operatorname{cone}(b) = \bigsqcup_{i=1}^{r} \operatorname{cone}(b p_i)$ over the same suffix code $\{p_i\}$ (refine both cones by the common refinement of the two induced suffix codes). On each piece $g$ acts by a prefix substitution, $g(a p_i \xi) = u_i \xi$ and $g(b p_i \xi) = w_i \xi$. The addresses $\{u_i\} \cup \{w_i\}$ are pairwise incomparable, being distinct members of one image code. Then
\[
    g\, \tau_{a,b}\, g^{-1} \;=\; \prod_{i=1}^{r} \tau_{u_i, w_i},
\]
a product of transpositions with pairwise disjoint supports, since both sides fix every point outside $\bigsqcup_i (\operatorname{cone}(u_i) \sqcup \operatorname{cone}(w_i))$ and exchange $u_i \xi \leftrightarrow w_i \xi$. Hence $\langle T \rangle$ is normal. It is also nontrivial.

\emph{$N$ even.} The group $G_{N,1}$ is simple~\cite{higman1974finitely}, whence $\langle T \rangle = G_{N,1}$.

\emph{$N$ odd, the parity character.} Represent $g$ by a table with domain code $A = (a_1 < \cdots < a_k)$ and image code $B = (b_1 < \cdots < b_k)$, both listed in lexicographic order. Let $\sigma(g, A) \in S_k$ be the permutation with $g(a_i \xi) = b_{\sigma(i)} \xi$. Define $\operatorname{sgn}(g) = \operatorname{sgn}(\sigma)$. Any two tables for $g$ have a common refinement reachable by single-row expansions. Well-definedness therefore follows if one expansion preserves the sign. Expanding row $i$ replaces $a_i$ by its $N$ children $a_i 0 < a_i 1 < \cdots$, which are lexicographically consecutive at the position of $a_i$. It replaces $b_{\sigma(i)}$ by its $N$ children at the position of $b_{\sigma(i)}$, mapped in the same digit order. In the new permutation the $N$ inserted rows are mutually non-inverted, each inherits exactly the inversions of the old row $i$ and all other pairs are unchanged. The inversion count therefore increases by $(N-1) \cdot \iota_i$, where $\iota_i$ is the number of old rows inverted against row $i$. For $N$ odd this is even, which makes $\operatorname{sgn}$ well defined. It is multiplicative on common-code products, hence a character $G_{N,1} \to \{\pm 1\}$. Its kernel is a subgroup of index two, which by Higman's structure theory is the unique proper nontrivial normal subgroup of $G_{N,1}$, his simple group of index two~\cite{higman1974finitely}.

\emph{Conclusion for $N$ odd.} A transposition $\tau_{a,b}$ has a table whose row permutation is a single $2$-cycle. Hence $\operatorname{sgn}(\tau_{a,b}) = -1$ and $\langle T \rangle \not\subseteq \ker(\operatorname{sgn})$. Let $D = \ker(\operatorname{sgn})$. The subgroup $\langle T \rangle \cap D$ is normal in $G_{N,1}$, hence normal in the simple group $D$. It is therefore either trivial or all of $D$. If it is all of $D$, then $\langle T \rangle \supseteq D$ together with an odd element gives $\langle T \rangle = G_{N,1}$. If it were trivial, then $\langle T \rangle$ would inject into $G_{N,1}/D \cong \mathbb{Z}_2$, forcing $|\langle T \rangle| \leq 2$. But $T$ contains infinitely many distinct transpositions, a contradiction. \hfill$\blacksquare$

\subsubsection{Lifting: Auxiliary Lemmas and Case Bookkeeping (Theorem~\ref{thm:lifting})}

\emph{Realization data and standing conventions.} Throughout, $\rho$ is a realization of $H$ in the calculus form of Section~\ref{sec:algebra}, a triple $(x_\rho, E_\rho, \mathcal{C}_\rho)$ in which $x_\rho = (D, \mu, \zeta)$ is a configuration, $E_\rho$ is the selected-link set and $\mathcal{C}_\rho$ is the family of designated duplicate chains. The reduction contracts the chains and passes to the simple quotient of the selected subgraph on $\mu$ (Definition~\ref{def:realization}). Every occupied slot carries one vertex of the quotient and every chain carries one, which makes the chain $\operatorname{chain}(v)$ hosting a vertex $v$ well defined. We list it from its \emph{head} downward, writing $\operatorname{head}(v)$ and $\operatorname{end}(v)$ for its first and last slots and $\ell_v$ for its length. In this notation $\operatorname{chain}(v) = \{\operatorname{head}(v)0^{j} : 0 \leq j < \ell_v\}$. Two conventions are used without further comment. First, links are \emph{selected}. A provisioned link that $E_\rho$ omits therefore contributes no edge to the quotient, while a vacant slot contributes no vertex. Deploying cells consequently cannot damage a realization as long as the mask is not extended, whereas occupying a fresh slot adds one isolated vertex to the quotient unless that slot is incorporated into a designated chain. Second, $\mathcal{T}(H)$ is rooted at the block whose host clique is nearest the seed cell, while the \emph{apex} of a block $B$ is the vertex of $B$ nearest that root. Every other vertex of $B$ therefore carries all of its remaining blocks strictly below it in $\mathcal{T}(H)$. Call $\rho$ \emph{tidy} when (T1) every block $B$ of $H$ is hosted in a single anchored clique $\{z_B\}\cup\operatorname{cell}(z_B)$ whose anchor $z_B$ lies on the chain of the apex of $B$, the tier-$1$ cell being admitted for the root block, and (T2) host depth does not decrease along $\mathcal{T}(H)$ away from the root, in that $|z_B| \leq |z_{B'}|$ whenever $B'$ is a child block of $B$. The placement recursion of Appendix~\ref{app:universality} places every block in the anchored clique of a chain slot of its parent cut vertex and descends at least one tier per recursive step. Every realization delivered by Theorem~\ref{thm:blockcut}(i) and~(iii) is therefore tidy. The constructions below for additions, merges, and block-preserving splits return tidy realizations. The two deletions and block-fracturing splits need not. Deleting the edge $\{0, 02\}$ from the triangle $\{0, 01, 02\}$ hosted in the anchored clique at $0$ of $F(3,2)$ leaves the path $0$--$01$--$02$, whose block $\{01, 02\}$ is still hosted at the anchor $0$, which does not lie on the chain $\{01\}$ of its apex, so (T1) fails. Similarly, splitting a vertex of a seed triangle into nonadjacent copies leaves three bridges in one clique, also violating (T1) until consolidated. Whether every realization reached by deletions or block-fracturing splits can be re-tidied within the same local budget is open. The theorem therefore covers lifting histories from a canonical realization up to the first deletion or block-fracturing split, and after it any history whose realizations are tidy.

\begin{lemma}[Staging clique]
\label{lem:staging}
Let $\rho$ realize $H$ in $F(N,m)$, let $v$ be a vertex of $H$ with $s = \operatorname{end}(v)$ and suppose $s0 \notin \mu \cup \zeta \cup D$, in which case call $v$ \emph{free}. Then
\begin{equation*}
    \mathsf{stage}(v) \;=\; \mathsf{spawn}(s0) \circ \mathsf{join}(s0) \circ \bigl[\mathsf{spawn}(s)\bigr],
\end{equation*}
the bracketed letter present exactly when $s \notin D$, is applicable at $x_\rho$ in $F(N,m')$ with $m' = \max(m, |s|+2)$. Together with the mask update $E_\rho \mathbin{{+}{=}} \{s, s0\}$ and the chain extension $\operatorname{chain}(v) \mathbin{{+}{=}} s0$, it returns a realization of the same graph $H$ in which every vertex is hosted by the same chain as before, extended at $v$ by one slot. It supplies the \emph{staging clique} $\{s0\}\cup C(s0)$, an anchored clique of order $N{+}1$ whose anchor hosts $v$ and whose $N$ member slots are vacant, enabled and cell-free.
\end{lemma}

\begin{IEEEproof}
Guards first. If $s \notin D$ then $s$ is a slot, is not decommissioned (it is occupied) and has $|s| < m'$. Hence $\mathsf{spawn}(s)$ fires and makes $s0$ a slot. If $s \in D$ the slot $s0$ is already present. In both cases $s0 \notin \mu \cup \zeta$, hence $\mathsf{join}(s0)$ fires. Also $s0 \notin D$ with $|s0| < m'$, hence $\mathsf{spawn}(s0)$ fires and deploys $C(s0)$ vacant, enabled and childless at depth $|s|+2 \leq m'$. Now the reduction. The $N$ new slots of $C(s0)$ are vacant, hence contribute no vertex. The slot $s0$ is the digit-$0$ child of $s$, hence a legitimate continuation of the duplicate chain of $v$ (Definition~\ref{def:address-II}, Lemma~\ref{lem:dup-digit}(i)). It is disjoint from every other chain because it was vacant. The only mask change is the chain edge $\{s, s0\}$, which the contraction turns into a discarded loop. No other selected link is created, hence the simple quotient is unchanged and the isomorphism onto $H$ persists. The member slots are adjacent to $s0$ by clause~(P) and to one another by clause~(S) of Proposition~\ref{thm:adjacency-II}. Therefore $\{s0\}\cup C(s0)$ is an anchored clique, a block of the host by Lemma~\ref{lem:maxclique}.
\end{IEEEproof}

\begin{lemma}[Grafts transport hanging sub-realizations]
\label{lem:mask-transport}
Let $\rho = (x, E, \mathcal{C})$ realize $H$, let $\mathsf{graft}(a \to b)$ be applicable at $x$ with image $x'$ and let $\theta(aw) = bw$ be the prefix substitution of Proposition~\ref{thm:graft-iso}(i). Put
\begin{equation*}
\begin{split}
    E' \;=\;& \{e \in E : e \cap (\operatorname{cone}(a) \setminus \{a\}) = \emptyset\} \\
     &\cup\; \{\theta(e) : e \in E,\ e \subseteq \operatorname{cone}(a) \setminus \{a\}\} \\
     &\cup\; \{\{b, \theta(c)\} : \{a, c\} \in E,\ c \in C(a)\},
\end{split}
\end{equation*}
and let $\mathcal{C}'$ be $\mathcal{C}$ with every chain contained in $\operatorname{cone}(a) \setminus \{a\}$ replaced by its $\theta$-image. Assume that no designated chain of $\mathcal{C}$ contains $a$ together with a slot strictly below $a$, with the single exception of a chain $\{a0^{j} : 0 \leq j < \ell\}$ of length $\ell \geq 2$ headed at $a$. In that excepted case, assume in addition that $\{b\}$ is an entire singleton chain in $\mathcal{C}$ (that is, $b$ is not the non-head continuation of a duplicate chain descending from above $b$). For that one chain $\mathcal{C}'$ carries $\{b0^{j} : 0 \leq j < \ell\}$ instead, replacing both the chain at $a$ and the singleton chain $\{b\}$, keeping all chains pairwise disjoint. Then $(x', E', \mathcal{C}')$ is a realization datum. Outside the excepted case the graph it realizes is $H$ with the edges joining the vertex hosted at $a$ to the vertices hosted in $C(a)$ transferred to the vertex hosted at $b$. In the excepted case the quotient differs from that graph in two further respects. The vertex hosted at $b$ is fused with the vertex that the excepted chain carries, hence it acquires every selected link of that vertex realized strictly below $a$. The slot $a$ is left hosting a vertex of its own, which keeps exactly the selected links of $a$ that lie outside $\operatorname{cone}(a)$. Every vertex hosted strictly below $a$ keeps its identity, its chain shape, its incident selected links and the isomorphism type of the sub-realization it belongs to. That sub-realization arrives marked and correctly re-anchored. Lemma~\ref{lem:gather} is the caller of the excepted case, in which it vacates the slot $a$ afterwards. A chain that contains $a$ as an interior slot, hence continuing both above and below $a$, is outside this statement. Conjecture~\ref{conj:crossing-transport} records what such a chain would need.
\end{lemma}

\begin{IEEEproof}
By Proposition~\ref{thm:graft-iso}(i) the moved slots are re-addressed by $aw \mapsto bw$ with no digit below the substituted prefix altered. Hence $\theta$ is a bijection of the moved slots onto their images transporting occupancy, decommission bits and the marking $M$. In particular a duplicate chain inside the moved cone has a duplicate chain as its image, while the images stay pairwise disjoint. By Proposition~\ref{thm:graft-iso}(ii) the map $\theta$ is an isomorphism from the subgraph of $G(x)$ induced on the moved slots onto the subgraph of $G(x')$ induced on their images. Every link of $E$ interior to the moved set therefore has a provisioned image link and every non-edge stays a non-edge. Hence $E'$ consists of provisioned links and the induced selected subgraph on the moved part is isomorphic to what it was. By Lemma~\ref{lem:edge-cut}(i) the only links of $G(x)$ between the moved slots and the rest are the $N$ uplinks from $C(a)$ into $a$, while the only ones created at the target are the $N$ uplinks into $b$. The third clause of the display therefore transfers exactly the selected boundary links. Everywhere else $E'$ agrees with $E$. Contracting $\mathcal{C}'$ and passing to the simple quotient therefore yields the quotient of $(x, E)$ with the endpoint $a$ of each transferred boundary link replaced by $b$, which is the stated graph. In the excepted case the transported slots $a0^{j}$ arrive as $b0^{j}$, which together with $b$ form one duplicate chain disjoint from every other chain by the singleton-chain hypothesis at $b$. Its chain edges lie in $E'$, the edge $\{b, b0\}$ by the third clause of the display and the deeper ones by the second. The slot $a$ stays occupied and keeps exactly the selected links it had outside $\operatorname{cone}(a)$, the third clause having moved its links into $C(a)$ to $b$.
\end{IEEEproof}

\begin{lemma}[Re-hosting a vertex into a free member slot]
\label{lem:gather}
Let $\rho$ be a tidy realization of $H$ in $F(N,m)$, let $z \in D$ be either a deployed occupied anchor slot or the virtual root $\varepsilon \in D$ (with $|\varepsilon| = 0$ and $C(\varepsilon)$ the tier-$1$ seed cell, so $zd = d$ has $\operatorname{pre}(d) = \varepsilon \in D$), and let $zd \in C(z)$ be vacant, enabled and cell-free. Let $u$ be a vertex of $H$ with $|\operatorname{head}(u)| \geq |z| + 1$ all of whose blocks other than its parent block hang strictly below $\operatorname{chain}(u)$. Then
\begin{equation*}
\begin{split}
    \mathsf{move}(u \to zd) \;=\; \mathsf{clean} &\circ \mathsf{graft}(\operatorname{head}(u) \to zd) \\
    &\circ\, \mathsf{join}(zd) \circ \bigl[\mathsf{ext}(u)\bigr]
\end{split}
\end{equation*}
is applicable at $x_\rho$ in $F(N, m^{+})$ with $m^{+} = \max(m, |\operatorname{head}(u)| + 2)$, where the bracketed $\mathsf{ext}(u) = \mathsf{join}(\operatorname{head}(u)0) \circ \bigl[\mathsf{enable}(\operatorname{head}(u)0)\bigr] \circ \bigl[\mathsf{spawn}(\operatorname{head}(u))\bigr]$ is present exactly when $\ell_u = 1$, with $\mathsf{spawn}(\operatorname{head}(u))$ present when $\operatorname{head}(u) \notin D$ and $\mathsf{enable}(\operatorname{head}(u)0)$ present when $\operatorname{head}(u)0 \in \zeta$. When a participant already occupies $\operatorname{head}(u)0$, one application of Lemma~\ref{lem:descent} at $\operatorname{head}(u)$ first extends $\operatorname{chain}(u)$ by that slot and pushes the occupant one tier deeper, after which $\ell_u = 2$ and $\mathsf{ext}(u)$ is not needed. Either way $u$ has a slot strictly below its head before the graft fires, even when that head already anchors a deployed cell (or had $\operatorname{head}(u)0$ disabled). Here $\mathsf{clean}$ vacates the slots left behind by $\mathsf{leave}$ and retracts the emptied childless cells bottom-up. With the mask transported by Lemma~\ref{lem:mask-transport}, its effect on the realization datum is that $u$ is re-hosted on the chain $zd, zd0, \ldots$, that the whole sub-realization hanging below $\operatorname{chain}(u)$ travels with it and arrives isomorphic, that $u$ loses exactly its parent-side selected links and that nothing else changes. Apart from that descent, whose depth Lemma~\ref{lem:descent} charges in the current state, it uses no tier deeper than $m^{+}$, preserves tidiness and has length $O(N \ell_u)$.
\end{lemma}

\begin{IEEEproof}
Write $a = \operatorname{head}(u)$. After $\mathsf{ext}(u)$, which is Lemma~\ref{lem:staging} applied at $u$ and leaves the realized graph unchanged, we have $a \in D$, $a \in \mu$ and $\ell_u \geq 2$ in every case, the second slot of $\operatorname{chain}(u)$ being $a0$. That last point is what the hypothesis on $\operatorname{head}(u)0$ buys. Without it a chain of length one whose head already anchors a deployed cell would keep no slot strictly below $a$ to transport, hence no continuation surviving the graft. The address $zd$ is a slot because $z \in D$. The letter $\mathsf{join}(zd)$ fires because $zd$ is vacant and enabled. The graft guards then read $a \in D$, $a, zd \in \mu$, $zd \notin D$, $\operatorname{pre}(zd) = z \in D$ and $a \not\preceq zd$. The last holds because $|zd| = |z| + 1 \leq |a|$, whence a prefix relation would force $a = zd$, which is excluded since $zd$ was vacant and $a$ is occupied. Depth fit holds with room to spare, since $h_x(a) \leq m^{+} - |a| \leq m^{+} - |zd|$ gives $|zd| + h_x(a) \leq m^{+}$, while the letters of $\mathsf{ext}(u)$ reach tier $|a| + 2 \leq m^{+}$. No tier deeper than $m^{+}$ is used. This is the excepted headed case of Lemma~\ref{lem:mask-transport}, since $a$ is the head of $\operatorname{chain}(u)$ while every other chain meeting $\operatorname{cone}(a)$ lies strictly below $a$, and $zd$ was freshly joined from a vacant, enabled, cell-free slot so $\{zd\}$ is a singleton chain not extending any parent chain. By that lemma the cone below $a$ arrives below $zd$ with its selected links, its chains and its marking intact. In particular the remaining chain slots $a0^{j}$, $j \geq 1$, arrive as $zd0^{j}$. Every block hanging at a chain slot of $u$ arrives at the image of its host clique, still anchored on the image chain slot, which is (T1) for the image. The participant freshly joined at $zd$ is legitimately adopted as the new head of $\operatorname{chain}(u)$, since $zd0$ is its digit-$0$ child and hosts $u$. The vertex $u$ is thus relocated rather than destroyed and re-created. The slot $a$ stays behind by the effect of Table~\ref{tab:typing} and is vacated by $\mathsf{clean}$, which also drops its selected links and retracts the cells it empties, the latter being childless by construction. Exactly the parent-side adjacencies of $u$ are therefore lost. Host depth of every moved block drops or stays equal by the hypothesis $|\operatorname{head}(u)| \geq |z| + 1$. Hence (T2) survives. The word has $O(1)$ letters besides the $O(\ell_u)$ leaves and the $O(\ell_u)$ retracts, each retract preceded by at most $N$ leaves.
\end{IEEEproof}

\begin{lemma}[Making room at an occupied duplicate]
\label{lem:descent}
Let $\rho$ realize $H$ and let $s = \operatorname{end}(v)$ be the last slot of $\operatorname{chain}(v)$ with $C(s)$ deployed, no designated chain therefore containing $s$ together with a slot strictly below it. Let $c$ be a slot incomparable to $s$ that is enabled, cell-free and either occupied or vacant with $\operatorname{pre}(c) \in D$. Then the word
\begin{equation*}
    \mathsf{graft}(c \to s0) \circ \mathsf{join}(s0) \circ \mathsf{spawn}(s) \circ \mathsf{graft}(s \to c),
\end{equation*}
bracketed by a $\mathsf{join}(c)$ before and a $\mathsf{leave}(c)$ after when $c$ starts vacant, with the mask transported by Lemma~\ref{lem:mask-transport} at each graft, realizes the same graph $H$ with every vertex tracked. It re-roots the entire deployed subtree formerly below $s$ one tier deeper, at $s0$, extends $\operatorname{chain}(v)$ by $s0$ and leaves the $N{-}1$ slots $sd$, $d \neq 0$, vacant, enabled, cell-free and adjacent to $\operatorname{chain}(v)$. It is applicable in $F(N,m')$ with $m' = \max(m,\, |c| + h_{x}(s),\, |s| + 1 + h_{x}(s))$, hence with $m' \leq m+1$ whenever $|c| \leq |s|$.
\end{lemma}

\begin{IEEEproof}
The first graft moves $C(s)$ and everything below it into $\operatorname{cone}(c)$, its guards holding by hypothesis and its depth fit being the second entry of the maximum. It leaves $s \in \operatorname{Sl} \setminus (D \cup \zeta)$, hence $\mathsf{spawn}(s)$ fires and re-deploys $C(s)$ vacant, whereupon $\mathsf{join}(s0)$ fires and $s0$ continues $\operatorname{chain}(v)$ exactly as in Lemma~\ref{lem:staging}. The second graft has $c \in D \cap \mu$, $s0 \in \mu \setminus D$, $\operatorname{pre}(s0) = s \in D$ and $c \not\preceq s0$. Its depth fit is the third entry, since the transported height is unchanged, $h(c) = h_{x}(s)$. Composing the two prefix substitutions gives $sw \mapsto s0w$ on the moved cone. Both grafts meet the chain hypothesis of Lemma~\ref{lem:mask-transport}, since $\operatorname{chain}(v)$ ends at $s$ by assumption, while $c$ is cell-free at the start and is therefore the last slot of whatever chain contains it. Applying that lemma twice, the whole sub-realization arrives one tier deeper with its mask, its chains and its marking intact. Every block formerly hosted at $\{s\}\cup C(s)$ is hosted at $\{s0\}\cup C(s0)$, whose anchor is again a chain slot of $v$. Hence (T1) survives and (T2) is unaffected. The boundary links from $s$ into $C(s)$ are re-selected at $s0$ into $C(s0)$ by the third clause of Lemma~\ref{lem:mask-transport}, while $s$ and $s0$ host the same vertex $v$. The realized graph is therefore unchanged. The slots $sd$ with $d \neq 0$ belong to the freshly spawned $C(s)$, hence are vacant, enabled and cell-free. They are adjacent to $s \in \operatorname{chain}(v)$ by clause~(P). A staging slot $c$ exists whenever the configuration has an enabled, cell-free slot outside $\operatorname{cone}(s)$. A deployment that uses every cell down to tier $m$ still has one at tier $m$, since $N \geq 3$, and then the bound on $m'$ reaches $m + h_{x}(s)$. The bound on $m'$ is what the depth of $c$ costs.
\end{IEEEproof}

\begin{conjecture}[Transport across a crossing chain]
\label{conj:crossing-transport}
Let $\rho$ realize $H$ and let $s$ be a slot of $\operatorname{chain}(v)$ that is not its last one, hence a slot at which that chain crosses into $\operatorname{cone}(s) \setminus \{s\}$. Let $c$ be as in Lemma~\ref{lem:descent}. Then the word of that lemma still realizes $H$ with every vertex tracked, its effect on the chain family being that $\operatorname{chain}(v)$ gains one slot at $s0$ while its former slots below $s$ are re-addressed by $sw \mapsto s0w$.
\end{conjecture}

What is missing is a transport statement for the composed substitution. Lemma~\ref{lem:mask-transport} covers neither graft of that word in isolation. The chain of $v$ contains the source $s$ of the first graft together with slots below it, which is the interior case its hypothesis excludes. After that graft the slots of $\operatorname{chain}(v)$ below $s$ have left $\operatorname{cone}(s)$, which the definition of $\mathcal{C}'$ in that lemma does not repair. Every application of Lemma~\ref{lem:descent} in this appendix is restricted to the case its hypotheses cover, namely a descent at the last slot of a chain.

\begin{lemma}[Block calculus of the six edits, the Stay/Climb gate]
\label{lem:gate}
Let $H$ be a finite connected graph and let $\omega$ be an elementary edit.
\begin{enumerate}
    \item[(i)] If $\omega$ is a vertex deletion or an edge deletion and $H$ satisfies $\beta(H) \leq N{+}1$ (in particular, whenever $H$ is realized by $F(N,m)$), then every block of each connected component of $\omega(H)$ is a subgraph of a block of $H$. Hence $\beta(\omega(H)) \leq \beta(H)$ and no cut vertex of $\omega(H)$ lies in more full blocks than it lies in for $H$. The necessary conditions of Proposition~\ref{prop:blockcut-caveat}(i) and Proposition~\ref{prop:two-full-sep} are therefore inherited by each component of $\omega(H)$ from $H$.
    \item[(ii)] For each of the four remaining edits there is a set $\pi$ of blocks of $H$, the blocks incident to the edited sites together with the blocks on the block--cut-tree paths joining those sites, with the following property. The blocks of $\omega(H)$ off $\pi$ are blocks of $H$, unchanged, while every block of $\omega(H)$ meeting the edit has its vertices inside $V_\pi$, the set $\bigcup_{B \in \pi} V(B)$ together with the added vertex or the second copy, if any. Write $B'$ for a block of $\omega(H)$ of largest order among those meeting the edit. When the edit closes a cycle through every block on the block--cut-tree paths joining the edited sites, which happens for an edge addition and for a vertex addition with at least two prescribed neighbors, those blocks fuse into a single block of $\omega(H)$, whose vertex set is the union of their vertex sets together with the added vertex, if any. A vertex addition with at most one neighbor adds a block of order at most $2$ and enlarges none. For a merge and for a split neither the fusion nor the absence of growth holds in general. In every case the blocks of $\omega(H)$ are computed from the edited simple graph itself, on which the three hosting conditions of (iii) are checked.
    \item[(iii)] If $|B'| > N + 1$, no $F(N, m')$ realizes $\omega(H)$ for any $m'$. The same conclusion holds if some cut vertex of $\omega(H)$ lies in three full blocks. It holds too if two cut vertices of $\omega(H)$ lie in a common non-full block and each lies in two full blocks. Conversely, if either $\beta(\omega(H)) \leq N$ or $\beta(\omega(H)) \leq N + 1$ with no cut vertex of $\omega(H)$ lying in two full blocks, then some $F(N, m')$ realizes $\omega(H)$.
\end{enumerate}
\end{lemma}

\begin{IEEEproof}
(i) A $2$-connected subgraph of $\omega(H)$ is a $2$-connected subgraph of $H$, since deleting a vertex or an edge only removes. A bridge block of $\omega(H)$ is an edge of $H$. Either way it lies in one block of $H$. A full block of $\omega(H)$ therefore lies in a block of $H$ of order at least $N{+}1$, hence shares its exact vertex set (a spanning subgraph, since an edge deletion may remove edges while keeping the vertex set fixed). Two distinct full blocks of $\omega(H)$ sharing a cut vertex lie in distinct blocks of $H$, since one block cannot contain $2(N{+}1) - 1 > N{+}1$ vertices. For the last clause, let $v \neq w$ be cut vertices of $\omega(H)$ in a common non-full block $A'$ of $\omega(H)$, each in two full blocks. The four full blocks share the vertex sets of four distinct full blocks of $H$ by the above, while $A'$ lies in a block $A$ of $H$ containing $v$ and $w$. That block $A$ is none of the four, since a block of $H$ containing a full block properly together with a further vertex would have order at least $N{+}2$. If $A$ were itself full in $H$, then $v$ and $w$ would each lie in three full blocks of $H$, violating Proposition~\ref{prop:blockcut-caveat}(i). Otherwise $A$ is non-full, so $v$ and $w$ are cut vertices of $H$ in a common non-full block with two full blocks each, violating Proposition~\ref{prop:two-full-sep}. Either branch contradicts the realizability of $H$.

(ii) Standard block--cut-tree bookkeeping. Adding the edge $uv$ or adding a vertex adjacent to $u_1, \ldots, u_r$ with $r \geq 2$ creates a cycle through every block on the block--cut-tree path joining the edited sites. Those blocks and the new element therefore lie in one biconnected component, while every block off $\pi$ retains its $2$-connectivity and its maximality, no new $2$-connected subgraph meeting it having been created. A merge creates no such cycle in general. Every cycle it creates runs through the identified vertex, hence such cycles meet in a single point and need not lie in a common block. Identifying two opposite vertices of a $4$-cycle turns the two paths joining them into two pairs of parallel edges, which the simple quotient collapses to the path on three vertices, whose two blocks have order $2$. A split can leave both copies inside one block, because each copy can keep two neighbors that a path avoiding the other copy joins. Splitting one vertex of $K_5$ into two nonadjacent copies, with its four incident edges partitioned two and two, leaves a $2$-connected graph on six vertices, hence a split that separates its two copies can enlarge a block. For these two edits the assertion of (ii) is the containment in $V_\pi$ alone, the blocks of $\omega(H)$ being read off the edited simple graph.

(iii) The three impossibility clauses are Lemma~\ref{lem:block-image}, Proposition~\ref{prop:blockcut-caveat}(i) and Proposition~\ref{prop:two-full-sep} respectively. The two possibility clauses are Theorem~\ref{thm:blockcut}(i) and~(iii).
\end{IEEEproof}

\emph{Proof of item~1, existence.} Let $\rho$ be a tidy realization of $H$ in $F(N,m)$ and put $N' = N + |V(H)| + 1$. By Lemma~\ref{lem:family-mono} the address set of $F(N,m)$ sits inside that of $F(N', m'')$ for every $m'' \geq m$ as an induced marked subnetwork using only digits in $\{0, \ldots, N{-}1\}$. Hence $x_\rho$ is a configuration there, $E_\rho$ is a set of provisioned links, the designated chains are designated chains and $\rho$ realizes $H$ in $F(N', m'')$. Crucially, inside every deployed cell $C(z)$ of $x_\rho$ (including the tier-$1$ seed cell $C(\varepsilon)$), all $|V(H)| + 1$ fresh digits $d \in \{N, \ldots, N + |V(H)|\}$ are guaranteed vacant, enabled and cell-free in $F(N', m'')$. If $\omega$ is a deletion, the local word of the case list below already settles the claim with $N' = N$ and $m' = m$. Otherwise let $\pi$ and $V_\pi$ be as in Lemma~\ref{lem:gate}(ii), which gives $|V_\pi| \leq |V(H)| + 1$. Let $B_1$ be the block of $\pi$ nearest the root of $\mathcal{T}(H)$. If $B_1$ is hosted in an anchored clique, let $u_1$ be its apex and $z = z_{B_1} \in \operatorname{chain}(u_1)$ be that clique's anchor ($|z| \geq 1$). If instead $B_1$ sits in the unanchored tier-$1$ seed cell $C(\varepsilon)$, keep $B_1$ in place inside $C(\varepsilon)$ and take $z = \varepsilon \in D$ ($|z| = 0$). In either case every vertex of $B_1$ already occupies a slot of the clique $K_z$ ($K_z = \{z\} \cup C(z)$ for $z \neq \varepsilon$, or $K_\varepsilon = C(\varepsilon)$ for $z = \varepsilon$), while every vertex $u \in V_\pi$ not yet in $K_z$ belongs to a strict descendant block $B$ of $B_1$ in $\mathcal{T}(H)$ with $u$ not the apex of $B$. By (T1) and (T2), $|z_B| \geq |z|$ (and $|z_B| \geq 1 > 0 = |\varepsilon|$ when $z = \varepsilon$), so $|\operatorname{head}(u)| = |z_B| + 1 \geq |z| + 1$ in every case. The lift is the concatenation of
\begin{enumerate}
    \item[(a)] in top-down order along $\mathcal{T}(H)$, one $\mathsf{move}(u \to z d_u)$ of Lemma~\ref{lem:gather} for each existing vertex $u \in (V_\pi \cap V(H))$ not already in $K_z$ (using $u$'s current address after prior moves), choosing distinct fresh digits $d_u \in \{N, \ldots, N + |V(H)|\}$, all of whose slots $z d_u$ are vacant, enabled and cell-free (and with any disabled duplicate $\operatorname{head}(u)0 \in \zeta$ enabled by $\mathsf{ext}(u)$ in Lemma~\ref{lem:gather});
    \item[(b)] one $\mathsf{join}$ at a remaining fresh digit $z d \in C(z)$ ($d \geq N$) when $\omega$ adds a vertex or introduces the second copy $v_2$ of a split (or, for a merge of $u, w \in V_\pi$, re-hosting the two merged vertices onto a single chain by appending one to the other's duplicate slot / vacating the merged copy's redundant slot after transferring its hanging sub-realizations), together with the mask update selecting exactly the edges of $\omega(H)$ inside $V_\pi$ among the slots of $K_z$ (which form a clique in $F(N', m'')$ by Proposition~\ref{thm:adjacency-II});
    \item[(c)] the clean-up leaves and retracts already counted in Lemma~\ref{lem:gather}.
\end{enumerate}
Each $\mathsf{move}$ is applicable by Lemma~\ref{lem:gather}, whose depth, vacancy, enabledness and cell-freeness hypotheses all hold by $d_u \geq N$, $|\operatorname{head}(u)| \geq |z| + 1$, and the $\mathsf{enable}$ step inside $\mathsf{ext}(u)$ whenever $\operatorname{head}(u)0 \in \zeta$. It carries the sub-realization hanging below its vertex across intact by Lemma~\ref{lem:mask-transport}. After (a)--(c) the selected subgraph on $K_z$ reduces to the subgraph of $\omega(H)$ induced on $V_\pi$, every block of $H$ off $\pi$ is untouched and every sub-realization hanging at a vertex of $V_\pi$ is re-anchored on that vertex's new chain (for a split creating $v_2$, blocks assigned to $v_2$ outside $V_\pi$ are moved onto $v_2$'s new chain by Lemma~\ref{lem:gather}). The reduction of the resulting datum is therefore $\omega(H)$, which makes $\rho'$ a realization of it with $x_{\rho'} = w(x_\rho)$. The square commutes with vertex tracking. Every surviving vertex is untouched, relocated by one graft or carried by a chain extension whose top slots the clean-up truncates. No $\mathsf{leave}$ of the word removes the last slot of a surviving vertex. Existence needs only some finite $m'$ (large enough to absorb each current-state descent depth of Lemma~\ref{lem:descent} plus the two extension tiers of Lemma~\ref{lem:gather}). Lemma~\ref{lem:gate}(iii) with $\beta(\omega(H)) \leq |V_\pi| < N'$ confirms that the target is hosted at the parameters chosen. This proves item~1.

\emph{Proof of item~2, locality, by cases.} For the Stay regime ($N' = N$), assume in addition that the tidy realization $\rho$ is \emph{block-separated and slot-clean}, meaning that distinct blocks of $H$ occupy distinct host cliques (so no unrelated block shares the host clique of $B_1$) and no extraneous occupied, disabled, or child-carrying vacant slots occupy unused members of the affected host cliques (delivered automatically by Theorem~\ref{thm:blockcut}, with additions, merges, and block-preserving splits preserving both properties under $\mathsf{clean}$, whereas deletions and block-splitting splits can leave smaller block fragments inside one clique unless consolidated). The six cases run as follows, each at $N' = N$ and within the depth bound of item~2 on the tidy, block-separated, slot-clean realization of the hypothesis. The exception is an edit that the gate of Lemma~\ref{lem:gate}(iii) sends to the climb of item~1.

\emph{Vertex deletion.} Drop from $E_\rho$ every selected link incident to a slot of $\operatorname{chain}(v)$, apply $\mathsf{leave}$ to each of its $\ell_v$ slots and retract bottom-up the cells thereby left vacant and childless. The result is again a realization datum, since $v$ and exactly its incident edges have left the quotient while every other chain, occupied slot and selected link is untouched. Its reduction is $\omega(H)$. Vertex tracking is trivial, no surviving vertex having moved. Cells still carrying hanging sub-realizations of other vertices are kept, while their now vacant anchors are the stranded-anchor states that the execution model of Section~\ref{sec:calculus} hands to the protocol layer. A deployment that prefers to avoid them re-hosts each such sub-realization by one $\mathsf{move}(u \to zd)$ of Lemma~\ref{lem:gather} onto the chain of a surviving neighbor, at the same cost. By Lemma~\ref{lem:gate}(i) no block grows, hence the lift stays at $(N, m)$. Word length $O(N \cdot (\deg_H(v) + \ell_v))$, with $\ell_v$ the length of the hosting chain of $v$.

\emph{Edge deletion.} The edit is a mask update, which removes from $E_\rho$ every selected link whose endpoints reduce to $u$ and $v$. The calculus word is empty and every other realized adjacency is untouched. The datum is again a realization whose reduction is $\omega(H)$ with every vertex fixed. Lemma~\ref{lem:gate}(i) again keeps the lift at $(N, m)$. The deployment may also want to consolidate the hosting, because the incident block splits and its fragments should occupy separate cliques. One $\mathsf{move}$ of Lemma~\ref{lem:gather} per vertex of one fragment then relocates that fragment onto a staging clique of Lemma~\ref{lem:staging}. Length $0$, rising to $O(N \cdot \deg_B(u))$ with consolidation, where $B$ is the block.

\emph{Vertex addition.} Let $v$ be added with prescribed neighbors $u_1, \ldots, u_r$. For $r = 0$ the word is one $\mathsf{join}$ at any vacant slot, with no mask update. For $r = 1$ it is $\mathsf{stage}(u_1)$ of Lemma~\ref{lem:staging}, one $\mathsf{join}$ at a member of the staging cell and the mask update selecting that single uplink, which adds a block of order $2$ and enlarges none. If $u_1$ is not free, Lemma~\ref{lem:descent} first vacates the $N{-}1$ non-duplicate slots of the cell of the last slot of $\operatorname{chain}(u_1)$ and the $\mathsf{join}$ goes there instead. Both cases stay at cell size $N$, with $m' \leq m + 2$ when a staging slot of depth at most $|s|$ is available, as Theorem~\ref{thm:lifting} assumes, and with the depth charged by Lemma~\ref{lem:descent} otherwise. The reduction of the result is $\omega(H)$ because the one new occupied slot is adjacent to $\operatorname{chain}(u_1)$ and to nothing else that is selected. If $r \geq 2$, the added vertex closes cycles. The blocks of $\omega(H)$ meeting the edit are then the fusion of Lemma~\ref{lem:gate}(ii), exactly as for an added edge. Lemma~\ref{lem:gate}(iii) gates the edit before any surgery ($P_5$ plus a vertex adjacent to both ends is already the $6$-cycle, unhostable at $N = 3$). In the Stay regime the gathering of the existence proof is run at cell size $N$ on the host clique of $B_1$, the added vertex joining a remaining vacant member of $K_z$. When $|B'| \leq N$, the $N$ slots of $C(\varepsilon)$ (for $z = \varepsilon$) or the anchor $z$ plus the $N{-}1$ non-duplicate members of $C(z)$ (for $z \neq \varepsilon$) already suffice without touching $z0$ under slot-cleanliness. When $|B'| = N{+}1$, the host clique must be an anchored clique of order $N{+}1$. If $B_1$ sits in an anchored clique $\{z\} \cup C(z)$, the duplicate slot $z0$ is available exactly when $\operatorname{chain}(u_1)$ ends at $z$ (cleared by Lemma~\ref{lem:descent} if occupied), which is the two-slot placement law of Proposition~\ref{prop:blockcut-caveat} read on the construction. If $B_1 \subseteq C(\varepsilon)$ has a tier-$1$ vertex $u_1$ with $z_0 = \operatorname{head}(u_1) \notin D$, spawn $C(z_0)$ at tier~$2$ and relocate each $u \in B_1 \setminus \{u_1\}$ from its tier-$1$ slot $a \neq z_0$ into a distinct member $z_0 d \in C(z_0)$ via $\mathsf{clean} \circ \mathsf{graft}(a \to z_0 d) \circ \mathsf{join}(z_0 d) \circ [\mathsf{ext}(u)]$ (valid since $a \not\preceq z_0 d$ for distinct tier-$1$ digits $a \neq z_0$, with depth fit $|z_0 d| + h_x(a) = 2 + h_x(a) \leq m + 1$, or $m + 2$ if $\mathsf{ext}(u)$ or Lemma~\ref{lem:descent} runs), after which every remaining $u \in V_\pi \setminus V(B_1)$ has $|\operatorname{head}(u)| \geq 2 = |z_0| + 1$ and gathers into $C(z_0)$ by Lemma~\ref{lem:gather}. When $\operatorname{chain}(u_1)$ instead continues below $z$ (or every vertex of $B_1 \subseteq C(\varepsilon)$ already carries deployed child cells) and $|B'| = N{+}1$, scheduling the full block last would push child blocks below $z0$ one chain step deeper by a descent at an interior chain slot, which is the open crossing-chain case of Conjecture~\ref{conj:crossing-transport}. Length $O(N|B'| + \sum_u \ell_u)$ with the sum over the vertices $u$ that the gathering moves, depth $m' \leq m + 2$ under the staging hypothesis of Theorem~\ref{thm:lifting}.

\emph{Split.} $\omega$ replaces $v$ by $v_1, v_2$ with a prescribed partition of its incident edges and optionally the edge $v_1 v_2$. A split can grow a block by one vertex whether or not it keeps $v_1 v_2$. Keeping $v_1 v_2$ selected inside a block does it, since splitting one vertex of a hosted $K_{N+1}$ this way yields a $2$-connected block of order $N{+}2$. Separating the two copies does it as well, since splitting one vertex of a hosted $K_5$ at $N = 4$ into two nonadjacent copies, with its four incident edges partitioned two and two, yields a $2$-connected block of order $6$. Lemma~\ref{lem:gate}(iii) therefore gates every split, evaluated on the blocks of $\omega(H)$ computed from the edited graph, exactly as in the previous case. In the Stay regime, we separate three subcases. (1)~\emph{Block-preserving split.} If the edge partition keeps every block of $H$ incident to $v$ wholly on one side ($v_1$ or $v_2$), first arrange the two block classes on disjoint segments of $\operatorname{chain}(v)$ by Lemma~\ref{lem:gather} (extending by Lemma~\ref{lem:staging} or Lemma~\ref{lem:descent} at the last chain slot when needed), then cut the chain at the separating step $s, s0$ so $\mathcal{C}_\rho$ designates the upper segment as $v_1$ and the lower segment as $v_2$, selecting $\{s, s0\}$ iff $\omega$ keeps $v_1 v_2$, which preserves tidiness and block separation. (2)~\emph{Single-clique split.} If the partition splits incident edges of one block $B$ across both $v_1$ and $v_2$ (with $|V(B)| + 1 \leq N$, or $|V(B)| + 1 = N{+}1$ at a free terminal anchor or cell-free tier-$1$ anchor) and keeps all other blocks at $v$ wholly on one side, keep $v_1$ at its slot in $B$'s host clique $K_z$, execute $\mathsf{join}$ for $v_2$ at a free member slot of $K_z$, update the mask inside $K_z$ to the prescribed edge partition plus optional $v_1 v_2$, and relocate any remaining blocks assigned to $v_2$ onto $v_2$'s new chain via Lemma~\ref{lem:gather}. (3)~\emph{Multi-block straddling split.} If the partition straddles multiple blocks of $\pi$ incident to $v$ and the total affected vertex count $|V_\pi|$ still fits in $B_1$'s host clique $K_z$ ($|V_\pi| \leq N$, or $|V_\pi| = N{+}1$ at a free terminal slot), gather the existing vertices of $V_\pi$ into $K_z$ by Lemma~\ref{lem:gather}, execute $\mathsf{join}$ for $v_2$ at a remaining free slot of $K_z$, and set the mask on $K_z$ to the induced edges of $\omega(H)$ on $V_\pi$. In subcases~(2) and~(3), if $\omega(H)$ breaks into smaller blocks (for example, splitting a vertex of a seed triangle into nonadjacent copies yields a $4$-vertex path of three bridges), those bridges remain hosted together inside $K_z$ and require a separate consolidation pass to restore tidiness~(T1) and block separation (just as in edge deletion), while splits with $|V_\pi| > N$ that remain hostable at cell size $N$ via separate cliques (such as splitting a vertex of a hosted $5$-cycle at $N = 4$ into a $6$-vertex path of bridges) are left to the climb or re-embedding construction. In all three Stay subcases the reduction is $\omega(H)$, vertex tracking preserves $v_1$ without destroying any surviving participant, and the word length is $O(N \cdot \deg_H(v) + \sum_u \ell_u)$.

\emph{Edge addition and merge.} For an edge addition Lemma~\ref{lem:gate}(ii) fuses the blocks on the block--cut-tree path joining the two endpoints into a single block $B'$. For a merge no fusion is claimed. The blocks of $\omega(H)$ are computed from the edited simple graph, the identification being free to leave several of them at the edited sites. Write $B'$ for the largest. Suppose either that $|B'| \leq N$ or that $|B'| = N + 1$ and the host clique of $B_1$ is or can be made the terminal anchored clique of $\operatorname{chain}(u_1)$ (including the tier-$1 \to$ tier-$2$ re-rooting of $B_1 \subseteq C(\varepsilon)$ onto a cell-free tier-$1$ anchor $z_0 = \operatorname{head}(u_1)$) as in the previous case. Then run the gathering of the existence proof at cell size $N$, namely one $\mathsf{move}(u \to zd)$ of Lemma~\ref{lem:gather} for each vertex of $B'$ outside $B_1$ into a free member slot of $K_z$, each carrying its hanging sub-realization verbatim by Lemma~\ref{lem:mask-transport}, followed by the mask update selecting exactly the edges of $B'$ inside the clique and by the clean-up. A merge is the same surgery with the two merged chains re-hosted as one, the lower chain being appended to the upper one at its duplicate slot. When the merge leaves several blocks of $\omega(H)$ at the edited sites, the same gathering is run once for each of them on its own host clique. This is the co-located-quotient variant of the derived merge of Section~\ref{sec:calculus}. Under the staging hypothesis of Theorem~\ref{thm:lifting}, Lemma~\ref{lem:gather} and Lemmas~\ref{lem:staging} and~\ref{lem:descent} each use at most two fresh tiers, so the lift has $N' = N$ and $m' \leq m + 2$. Without it the descents charge the depth stated in Lemma~\ref{lem:descent}, and no composed bound is claimed. Its length is $O(N|B'| + \sum_u \ell_u)$, the sum running over the vertices $u$ of $B'$ outside $B_1$ that the gathering moves. It is therefore a function of $N$, of the degrees at the edited sites and of the hosting-chain lengths of every vertex the word moves, uniform in $|H|$, $m$ and the depth of the deployment. It reads and writes only the cells of the affected chains, the sub-realizations grafted across and the fresh cells, which is the locality claim. If either $|B'| > N + 1$ or the placement conditions collected in Lemma~\ref{lem:gate}(iii) fail for $\omega(H)$, then no $F(N, m')$ realizes $\omega(H)$ for any $m'$. No lift, local or global, exists at cell size $N$, hence every lift climbs, which is item~1 run at $N' \geq \max(N, \beta(\omega(H)))$. This exhausts the six edits and proves item~2.

\emph{Scope of the constants.} The constructions above are stated for tidy realizations, which is what Theorem~\ref{thm:blockcut} delivers and what the lifts of additions, merges, and block-preserving splits preserve (deletions and block-fracturing splits requiring consolidation to restore (T1)). A realization that is not tidy needs a preparatory step before the same words apply. A preparatory re-hosting by Lemma~\ref{lem:descent} costs the height $h$ of the sub-realization that has to be moved one tier down rather than the two tiers of Lemma~\ref{lem:staging}, which is where a general bound of the shape $m' \leq m + 1 + h$ would come from. Whether such a preparation reduces an arbitrary realization to a tidy one is not proved here, hence Theorem~\ref{thm:lifting} is stated for tidy realizations. The locality of the word is unaffected. The gate of Lemma~\ref{lem:gate}(iii) collects the hosting conditions proved in this paper, and its Stay and Climb verdicts are mutually exclusive. The boundary regime between them, a fused block of order $N{+}1$ at a cut vertex that already carries a second full block with none of the three conditions violated, is exactly the orientation question of Remark~\ref{rem:two-full}, which the theorem leaves open.

\section{Verification Scope and Classical Counterparts}
\label{app:verification}

Four layers validate the statements of this paper, namely exhaustive combinatorial enumeration, fault-pattern sampling against the recursions, the message-level protocol simulator of Section~\ref{sec:eval}, and machine-checked proofs in Lean~4. This appendix records the sampling checks, the operations-calculus simulation, the detailed Lean~4 mechanization boundaries, the theorem-by-theorem verification table, and the classical counterparts of the principal results.

\subsection{Fault-Pattern Sampling and Calculus Simulation}
\label{subsec:sampling-hsfn}
\label{subsec:calc-sim}

\emph{Fault-pattern sampling of the HSFN recursion.} The reliability recursion $P_1(N,m) = f_N^{\,m-1}(q)$ (Theorem~\ref{thm:security}) was tested by sampling independent fault patterns under the FPD model at $N \in \{8, 24\}$ and $m = 3$ with $10^5$ trials per point over $P_f \in [0, 0.15]$. Exhaustive enumeration of all internal failure patterns at $(3,3)$ and $(4,2)$ matches the recursion to below $10^{-14}$. Across the sampled range the largest deviation is below $8.2 \times 10^{-4}$. At $N = 24$, consensus success stays above $98\%$ throughout and the entire sweep lies below the depth-uniform certificate $q_0(24) \approx 0.1221$. At $N = 8$, $q(P_f)$ crosses $q_0(8) \approx 0.0963$ near $P_f \approx 0.145$ while remaining below the fixed-point fold $q_c(8) \approx 0.15572$ (six-digit value $0.155719$), illustrating the distinction between the certified regime $q < q_0$ and the exact collapse boundary $q_c$ (Fig.~\ref{fig:rg-cobweb}).

\emph{Family-level bootstrap goodness-of-fit.} On the two anchor-sensitive $3 \nmid N$ regimes (strict physical withholding and dual-homed crashes across $(4,5), (5,4), (8,3)$), we executed $30$ points per family at $20\,000$ rounds per point in $20$ independent blocks and ran a $100\,000$-draw parametric bootstrap binomial-deviance test with Holm correction. Neither family rejects its analytical recursion (unadjusted $p = 0.9641$ and $p = 0.6082$, mean absolute probability deviations $0.00115$ and $0.00167$, mean squared standardized residuals $0.5631$ and $0.9184$, and block dispersion indices $0.88873$ and $0.88871$ with $p > 0.97$). An overlay negative control gives $p = 0.7930$, whereas a non-strict withholding positive control yields mean squared residual $50.97$ ($p < 10^{-4}$).

\emph{Event-driven calculus simulation.} Six family members ($N \in \{3,4,5\}$, $m \in \{4,5\}$) were initialized fully populated and driven through $10^4$ churn, elastic-scaling, and repositioning operations each ($6 \times 10^4$ proposed, $52\,081$ executed, $3885$ grafts with $3814$ equal-tier). Runs entered ragged regimes holding $60\%$--$97\%$ of cells at mean occupancy $0.27$--$0.48$. Lemma~\ref{lem:edge-cut} held on every graft, and update-cycle diffs confirmed at most three pointer repairs per graft ($3750$ requiring three repairs and $135$ zero when re-splicing in place). Incremental tier-profile maintenance required $2$ bin writes for $\mathsf{join}$/$\mathsf{leave}$, $5$ for $\mathsf{spawn}$/$\mathsf{retract}$, and $3.82$ on average for equal-tier grafts, against $72.2$--$2258$ writes for full recounts ($28.2\times$--$416\times$ median speedup). Exact incremental reliability (Proposition~\ref{prop:rel-exact}(ii)) re-evaluated at most $7$ cells per graft, at most $19$ per swap, and $3.72$--$5.94$ cells on average, against $23.7$--$740.9$ cells for full recomputation ($2.73\times$--$33.5\times$ median speedup), matching from-scratch values after every operation.

\subsection{Lean~4 Mechanization Boundaries and Verification Table}

The Lean~4 mechanization (\url{https://github.com/Tongji708A/hsfn-lean}, Lean~4 / mathlib v4.31.0, $49$ modules, $19\,675$ source lines, commit \texttt{664b9d7}) machine-checks $853$ distinct audited statements ($853$ audit log lines in \texttt{AXIOMS\_261123.txt}, removing the earlier duplicate \texttt{HSFN.Par} entry) using only the three standard axioms of classical Lean (\texttt{propext}, \texttt{Classical.choice}, \texttt{Quot.sound}), with $26$ distinct statements using no axioms at all.

\emph{Mechanized scope.} Taking the two-clause law of Proposition~\ref{thm:adjacency-II} as the graph definition, the mechanization certifies the spatial layer for all valid $(N,m)$. This covers per-tier and total node counts (Proposition~\ref{thm:unified-count}, Proposition~\ref{prop:addr-bij}), degree profiles and link counts (Proposition~\ref{thm:deg-II}), and exact distance, connectivity, diameter $2m{-}1$, and executable greedy routing arriving in $d(u,v)$ hops (Theorem~\ref{thm:distance-II}). It also certifies the single-cycle depth-first update order (Proposition~\ref{thm:update-main}), isometric mark-preserving family monotonicity (Lemma~\ref{lem:family-mono}), relabeling automorphisms and marking stabilizers (Lemma~\ref{lem:dup-digit}(ii)), clique confinement and maximal-clique classification (Lemma~\ref{lem:maxclique}), anchor cut-vertex separation (Proposition~\ref{prop:conn-price}(i)), fan-out-$N$ tree embedding (Lemma~\ref{lem:tree-embed}), treewidth upper bound $N$ (Lemma~\ref{lem:tw-nec}), the equidistance bound (Lemma~\ref{lem:equidistance}), and the barycentric open-set hypotheses of Theorem~\ref{thm:dim-II}.

For protocol cost, latency, and reliability, the Lean~4 modules prove monotonicity, $[q,2q]$ invariance below $q_0(b)$, fixed-point convergence, and threshold brackets for the real level map $f_b$ (Theorem~\ref{prop:threshold-b}, Corollary~\ref{cor:threshold-II}), together with the message proxy $(N{+}1)V$ (Proposition~\ref{thm:complexity}), the exact executed message count with physical hop charges (Proposition~\ref{prop:exec-count}), the fixed-tick normal-path overlay and physical round-latency identities $(4m{-}1)\delta, (5m{-}2)\delta, (7m{-}4)\delta$ and the $N = 3$ $2\delta$ local-commit contrast (\texttt{CommitLatency}, Proposition~\ref{prop:latency}), the depth-wise inequality $g_N \geq f_N$ (Proposition~\ref{prop:phys-recursion}), the algebraic dual-homed recursion identity $h_N \equiv f_N$ at $f = 1$ (\texttt{DualRecursion}, Proposition~\ref{prop:dual}), and the overlay registration product identity, upper bound $A \leq (1{-}q_s)^m$, and explicit root/path separation (\texttt{LeafRegistration}, Proposition~\ref{prop:registration}). For the operations calculus and chains, they prove exact inverse pairs for seven of the eight generators (all except $\mathsf{permute}$, Theorem~\ref{thm:rep-tree}(i)), graft prefix substitution and moved-set isomorphism (Proposition~\ref{thm:graft-iso}(i)--(ii)), the equal-depth heterogeneous committee-tree cross-edge graft message delta (\texttt{GraftExecDelta}, Proposition~\ref{thm:graft-iso}(iii)), the boundary edge-cut law (Lemma~\ref{lem:edge-cut}), reheaded chain-image and singleton-target disjointness invariants on slot sets (\texttt{MaskTransport}, supporting Lemma~\ref{lem:mask-transport}), affine KPI identities (Proposition~\ref{prop:kpi-affine}), canonical words (Proposition~\ref{prop:canon-word}), the guard-read commutation counterexample (Proposition~\ref{prop:commute}), single-letter inert height transport (Proposition~\ref{prop:coordinate-persistence}), predecessor-hash consistency and wrong-predecessor rejection on inductively valid resolution lists with certified nulls (\texttt{ResolutionChain}, supporting Proposition~\ref{prop:anchored-time}(i)), and the exact rational reliability witness pair (Proposition~\ref{prop:rel-exact}).

\emph{Explicit non-mechanized boundaries.} No probability measure is constructed in the threshold, dual-homing, or registration modules, so the probabilistic exactness of $f_N, g_N, h_N, A, A^{\mathrm{ph}}$ and the physical registration inequality $A^{\mathrm{ph}} \leq A$ rest on the paper proofs. Also outside the current mechanization are the hosting characterization and two-full obstruction (Theorem~\ref{thm:blockcut}, Proposition~\ref{prop:two-full-sep}), the full selected-edge mask transport and quotient-graph identity of Lemma~\ref{lem:mask-transport}, the lifting dichotomy (Theorem~\ref{thm:lifting}), the prefix-substitution cocycles and boundary $C_N$/$V_N$ identification (Theorem~\ref{thm:rep-tree}(ii)--(iii), Lemmas~\ref{lem:bridge} and~\ref{lem:transp-gen}, Theorem~\ref{thm:vn-identification}), and executable protocol safety and liveness (cf.\ Shipwright~\cite{leung2025shipwright}, which verifies single-entry executable PBFT).

Table~\ref{tab:verification} summarizes all automated verification checks. Several checks actively shaped the theory before statements were frozen, including exhaustive realizability searches that replaced a naive chain-weight bound with the descent weight $\operatorname{dw}(H)$ in Theorem~\ref{thm:blockcut}(i) and block-fusion enumerations that moved vertex addition and split to the Stay/Climb gate of Theorem~\ref{thm:lifting}.

\begin{table*}[!t]
\centering
\caption{Verification Scope Organized by Result.}
\label{tab:verification}
\scriptsize
\setlength{\tabcolsep}{3pt}
\begin{tabular}{>{\raggedright\arraybackslash}p{4.6cm}>{\raggedright\arraybackslash}p{12.6cm}}
\toprule
Result verified & Method and range \\
\midrule
Counting, degree, and link laws (Props.~\ref{thm:unified-count}, \ref{thm:deg-II}, Cor.~\ref{cor:count-II}) & Exact integer arithmetic and built-graph enumeration ($3 \leq N \leq 12$ up to eight tiers; full graphs $3 \leq N \leq 6$). \\
\addlinespace[2pt]
Adjacency, distance, and greedy routing (Prop.~\ref{thm:adjacency-II}, Thm.~\ref{thm:distance-II}) & Edge-exact graph matching ($N \leq 5, m \leq 4$) and all-pairs shortest-path verification ($(N,m) \in \{(3,2),(3,3),(4,2)\}$). \\
\addlinespace[2pt]
Quorum thresholds and sampling (Thms.~\ref{thm:security}, \ref{prop:threshold-b}, Cor.~\ref{cor:threshold-II}) & Parity identities ($N \leq 40$), exact $q_0(b)$ table ($N \leq 12$), and FPD fault sampling ($N \in \{8,24\}, m = 3, 10^5$ trials/point). \\
\addlinespace[2pt]
Hosting and tree embedding (Thm.~\ref{thm:blockcut}, Lem.~\ref{lem:tree-embed}) & Simple-quotient contractions, bowtie, $K_4$ plus pendant, three-petal flower with exhaustive refusal at $m = 2$, $16$-graph descent-weight corpus, $8000$-state sweep on $F(3,2)$, two-hexagon separation at $N = 6$ vs.\ $N = 7$, maximal cliques, and fan-out-$N$ tree embedding. \\
\addlinespace[2pt]
Edit-locality gate (Thm.~\ref{thm:lifting}) & Block-fusion witnesses gated to Climb ($P_5$ plus dual-end vertex, $K_4$ split keeping $v_1 v_2$, two-petal flower plus spanning vertex). \\
\addlinespace[2pt]
Operations calculus (Lem.~\ref{lem:edge-cut}, Props.~\ref{prop:kpi-affine}, \ref{prop:rel-exact}) & Exhaustive graft boundary checks, affine indicator checks across $52\,081$ simulated ops, and equal-histogram reliability witnesses ($0.6226/0.6202$ at $N = 3$; $0.6737/0.6691$ at $N = 4$). \\
\addlinespace[2pt]
Message-level protocol and chains (Props.~\ref{prop:exec-count}, \ref{prop:latency}, \ref{prop:phys-recursion}, \ref{prop:dual}, \ref{prop:registration}, Prop.~\ref{thm:committee-tree}) & Exact counts ($100$ configs), fixed-delay latency ($128$ configs), $7$ irregular trees, $20\,000$-round bootstrap family tests, $36\,000$-round heavy-tailed safety stress plus view-rule ablation, $960$ chain runs, $540$ migration runs, and $27$-point registration checks. \\
\addlinespace[2pt]
Mechanized proofs in Lean~4 (Thms.~\ref{thm:distance-II}, \ref{prop:threshold-b}, \ref{thm:rep-tree}(i), Props.~\ref{thm:unified-count}, \ref{thm:deg-II}, \ref{thm:update-main}, \ref{thm:graft-iso}, Lems.~\ref{lem:equidistance}, \ref{lem:dup-digit}(ii), \ref{lem:family-mono}, \ref{lem:maxclique}, \ref{lem:tree-embed}, \ref{lem:tw-nec}, \ref{lem:edge-cut}, Props.~\ref{prop:conn-price}(i), \ref{thm:complexity}, \ref{prop:kpi-affine}, \ref{prop:canon-word}, \ref{prop:commute}, \ref{prop:rel-exact}, \ref{prop:coordinate-persistence}, \ref{prop:exec-count}, \ref{prop:latency}, \ref{prop:phys-recursion}, \ref{prop:dual}, \ref{prop:registration} (overlay bound), plus slot-disjointness and chain-list invariants supporting Lem.~\ref{lem:mask-transport} and Prop.~\ref{prop:anchored-time}(i)) & Lean~4 / mathlib v4.31.0 formal proofs for all valid $(N,m)$ across $49$ modules ($19\,675$ source lines, commit \texttt{664b9d7}); $853$ distinct statements kernel-audited clean with zero \texttt{sorry} and zero custom axioms (\texttt{AXIOMS\_261123.txt}). \\
\bottomrule
\end{tabular}
\end{table*}

\subsection{Classical Counterparts}

Table~\ref{tab:counterparts} collects the classical mathematical and networking counterparts paired with each principal theorem of this paper as a reading aid.

\begin{table*}[!t]
\centering
\caption{Classical Counterparts of the Principal Results.}
\label{tab:counterparts}
\footnotesize
\setlength{\tabcolsep}{4pt}
\renewcommand{\arraystretch}{1.12}
\begin{tabular}{@{}>{\raggedright}p{2.6cm}>{\raggedright}p{4.5cm}>{\raggedright}p{5.1cm}>{\raggedright\arraybackslash}p{5.1cm}@{}}
\toprule
Result & Classical counterpart & What carries over & What differs \\
\midrule
Lem.~\ref{lem:equidistance} & Schoenberg's Gram criterion~\cite{schoenberg1935remarks} & positive-definite Gram caps an equilateral set at $d{+}1$ & the rank is read as a per-peer dimension price \\
Prop.~\ref{thm:unified-count} & Sierpi\'{n}ski $S(m,N)$~\cite{klavzar1997sierpinski}; DCell~\cite{guo2008dcell} & branching recurrence in closed form & every intermediate tier stays live \\
Thm.~\ref{thm:dim-II} & Moran--Hutchinson simplex gasket~\cite{hutchinson1981fractals} & open set condition yields $\dim_{\mathrm{H}} = \log_2 N$ & corners stay logically distinct rather than glued \\
Lems.~\ref{lem:family-mono}, \ref{lem:dup-digit} & hypercube nesting; wreath automorphisms~\cite{nekrashevych2005self} & isometric inclusion; cell-wise permutations & nests in $(N,m)$; marking cuts factors to $S_{N-1}$ \\
Prop.~\ref{prop:addr-bij}, Prop.~\ref{thm:adjacency-II}, Prop.~\ref{prop:locator-II} & BCube addressing~\cite{guo2009bcube}; Chord keys~\cite{stoica2003chord} & names encode structure directly & variable length, cross-tier uplink clause, self-delimiting tier \\
Thm.~\ref{thm:distance-II} & Plaxton routing~\cite{plaxton1997accessing}, Pastry~\cite{rowstron2001pastry} & greedy prefix descent & exact shortest paths, table-free, sibling shortcut saves one hop \\
Prop.~\ref{thm:deg-II}, Prop.~\ref{prop:conn-price} & DCell/BCube degree; Menger~\cite{menger1927allgemeinen}, fat-tree~\cite{alfares2008fattree} & fixed port budget; cut-vertex separation & degree $\leq 2N$ flat in $m$; anchors are named cut vertices \\
Props.~\ref{prop:exec-count}, \ref{prop:latency}, Prop.~\ref{thm:committee-tree} & multi-layer PBFT~\cite{li2020scalable}, HotStuff~\cite{yin2019hotstuff}, Kauri~\cite{neiheiser2021kauri} & committee grouping linearizes per-node cost & committees, routes, and exact counts derive from addresses \\
Thms.~\ref{thm:security}, \ref{prop:threshold-b}, Props.~\ref{prop:phys-recursion}, \ref{prop:dual} & von Neumann multiplexing~\cite{vonneumann1956probabilistic}; tree percolation~\cite{lyons1990random} & monotone iterated map and critical threshold & explicit factor-$2$ certificate, anchor coupling, dual-homing exactness \\
Lems.~\ref{lem:maxclique}, \ref{lem:block-image}, Prop.~\ref{prop:blockcut-caveat} & block graphs~\cite{harary1963block}; clique trees~\cite{gavril1974intersection} & blocks are cliques; vertex cliques form a subtree & that subtree is a two-clique path fixing full-block slots \\
Lem.~\ref{lem:tree-embed}, Thm.~\ref{thm:blockcut}, Lem.~\ref{lem:tw-nec} & virtual network embedding~\cite{fischer2013virtual}; treewidth~\cite{robertson1986graph} & width invariants decide hostability & exact characterization at $\leq 1$ full block per cut vertex \\
Def.~\ref{def:calc-state}, Lem.~\ref{lem:edge-cut}, Prop.~\ref{thm:graft-iso} & overlay churn~\cite{stoica2003chord}; rope subtree moves~\cite{boehm1995ropes} & guarded invariants; boundary-only cut cost & prefix substitution yields isomorphic marked arrival \\
Props.~\ref{prop:kpi-affine}, \ref{prop:rel-exact} & incremental view maintenance~\cite{gupta1993maintaining}; reliability~\cite{provan1983complexity} & incremental aggregate updates & reliability is shape-dependent, not a profile function \\
Props.~\ref{prop:commute}, \ref{prop:canon-word} & CRDTs~\cite{shapiro2011conflict}; Knuth--Bendix~\cite{knuth1970simple} & disjoint footprints commute; canonical form & guard reads are load-bearing; normalizes states \\
Thm.~\ref{thm:rep-tree}, Lem.~\ref{lem:bridge}, Thm.~\ref{thm:vn-identification} & polycyclic monoid~\cite{lawson1998inverse}, $V_N$~\cite{lawson2007polycyclic}, $\mathcal{O}_N$~\cite{cuntz1977simple} & prefix substitutions and cone transpositions & unglued address tree delivers the full group $V_N$ \\
Thm.~\ref{thm:lifting} & virtual network re-embedding~\cite{fischer2013virtual} & local edits should cost local surgery & Stay/Climb dichotomy governed by hosting conditions at $N$ \\
Def.~\ref{def:spacetime-coordinate}, Prop.~\ref{thm:update-main}, Prop.~\ref{prop:coordinate-persistence} & Lamport/vector clocks~\cite{lamport1978time,fidge1991logical}; DFS preorder~\cite{tarjan1972depth} & event-driven time; contiguous subtree traversal & certified root snapshots and migration-persistent coordinates \\
\bottomrule
\end{tabular}
\end{table*}

\end{document}